\documentclass[sigconf]{acmart}

\renewcommand\footnotetextcopyrightpermission[1]{} 

\makeatletter
\renewcommand\@formatdoi[1]{\ignorespaces}
\makeatother

\usepackage{amsmath,amssymb,amsfonts}
\usepackage{algorithm}
\usepackage{graphicx}
\usepackage{textcomp}
\usepackage{xcolor}

\newtheorem{Definition}{\textbf{Definition}}

\newtheorem{Proposition}{Proposition}

\newtheorem{Proof}{Proof}
\newtheorem{Example}{Example}
\newtheorem{Problem}{Problem}
\usepackage{booktabs}
\usepackage[T1]{fontenc}
\usepackage[justification=centering]{caption}  
\usepackage{balance}

\newcommand{\stitle}[1]{\noindent{\bf #1}}
\usepackage{subfigure}
\usepackage[shortlabels]{enumitem}
\usepackage{multirow}
\usepackage[noend]{algpseudocode}
\usepackage{algorithmicx,algorithm}
\usepackage{marginnote}
\usepackage{multicol}

\usepackage{times}
\usepackage{bbm}
\usepackage{epstopdf}
\usepackage{url}

\def\BibTeX{{\rm B\kern-.05em{\sc i\kern-.025em b}\kern-.08em
    T\kern-.1667em\lower.7ex\hbox{E}\kern-.125emX}}

\begin{document}

\title{Towards Truss-Based Temporal Community Search}

\author{Huihui Yang, Chunxue Zhu, Longlong Lin, Pingpeng Yuan}
\authornote{Both authors contributed equally to this research.}
\email{hh\_yang@hust.edu.cn, cxzhu@hust.edu.cn, longlonglin@swu.edu.cn, ppyuan@hust.edu.cn}
\affiliation{%
  \institution{School of Computer Science and Technology, Huazhong University of Science and Technology}
  \city{Wuhan}
  \country{CA}
  \postcode{430074}
}




\begin{abstract}
Identifying communities from temporal networks facilitates the understanding of potential dynamic relationships among entities, which has already received extensive applications. However, existing methods primarily rely on lower-order connectivity (e.g., temporal edges) to capture the structural and temporal cohesiveness of the community, often neglecting higher-order temporal connectivity, which leads to sub-optimal results. To overcome this dilemma, we propose a novel temporal community model named maximal-$\delta$-truss (\textit{MDT}). This model emphasizes maximal temporal support, ensuring all edges are connected by a sequence of triangles with elegant temporal properties. To search the \textit{MDT} containing the user-initiated query node $q$ (\textit{q-MDT}), we first design a powerful local search framework with some effective pruning strategies. This approach aims to identify the solution from the small temporal subgraph which is expanded from $q$. To further improve the performance on large graphs, we build the temporal trussness index (\textit{TT}-index) for all edges. Leveraging the \textit{TT}-index allows us to efficiently search high-probability target subgraphs instead of performing a full search across the entire input graph. Empirical results on nine real-world  networks and seven competitors demonstrate the superiority of our solutions in terms of efficiency, effectiveness, and scalability.
\end{abstract}




\maketitle

\section{Introduction} \label{sec:introduction}
Modeling the data as graphs to mine the implicit relationships among entities has become an important means of data analysis. Community mining is one of the most important tools to understand the underlying structure of the graphs. Generally, community mining can be  divided into {community detection} \cite{Fortunato2009Community, DBLP:conf/aaai/LinLJ23,newman2004fast,DBLP:journals/eswa/HeLYLJW24,DBLP:journals/corr/abs-2406-07357} 
and {community search} \cite{barbieri2015efficient,chen2020finding,cui2014local}. The former aims to identify $\emph{all}$ communities that meet the specified constraints from the perspective of global criteria. The latter tends to find $\emph{a}$ specific community containing the user-initiated query vertex. Therefore, {community search} is more user-friendly, which can be applied in personalized recommendations, infectivity analysis, and so on \cite{fang2020survey}.

Nowadays, community search has been receiving sustained and widespread attention, and many models have been proposed in the literature  \cite{chang2019cohesive}.  
For example, 
Yin et al. \cite{yin2017local} developed local diffusion algorithms for finding clusters of nodes with the minimum motif conductance based on higher-order subgraph structures (e.g., $k$-clique). However, most research still concentrates on static graphs and ignores the rich temporal interaction information of real-world networks\cite{bogdanov2011mining, DBLP:journals/jcst/ZhuLYJ22, DBLP:journals/corr/abs-2302-08740}. For instance, in social networks, people exchange messages at various times, and in collaboration networks, researchers collaborate to publish papers in different years.
Therefore, the communities identified by existing static community search methods cannot adequately capture temporal relationships, resulting in suboptimal solutions in practical applications.

With growing interest in temporal graphs, there has been some exploration of community mining on temporal graphs. For example, Li et al. \cite{DBLP:conf/icde/LiSQYD18} developed persistent $k$-core communities on temporal networks. Qin et al. \cite{qin2019mining} attempted to mine periodic cliques in temporal networks. Chu et al. \cite{OL:chu2019online} explored the bursting community by extending the static density model. Clearly, these studies capture the temporal information by extending existing lower-order community models (e.g., $k$-core), which pay more attention to the lower-order relationship of nodes and edges. 
In fact, people tend to focus more on the tightness of the interaction rather than the specific interaction time. For instance, in social network analysis, it was observed that the strength of relationships between users, such as the frequency of communication and the depth of engagement, holds greater significance than the exact timestamps of their interactions.

Motivated by these observations, we intend to explore methods for searching higher-order temporal community. Unlike temporal edges, each higher-order temporal connectivity pattern (a.k.a., temporal motifs \cite{DBLP:conf/wsdm/ParanjapeBL17}) is a small temporal subgraph. The communities identified through higher-order structures may become more specific and meaningful, but the computational complexity will also increase accordingly. 
Recently, Fu et al. \cite{LMega:fu2020local} proposed the \textit{L-MEGA} model to investigate the higher-order graph clustering on temporal networks, though it suffers from low efficiency (Section \ref{sec:experiments}). Our study focuses on the truss model, which is a simple type of higher-order structure. While more complex motifs may yield more meaningful community structures, the associated computational overhead can be prohibitive, particularly in practical applications. Therefore, we prioritize efficiency and applicability by employing the truss model.

Given that the well-known $k$-truss model effectively captures higher-order structural cohesiveness with near-linear time complexity \cite{akbas2017truss,DBLP:conf/sigmod/HuangCQTY14,DBLP:journals/pvldb/WangC12}, we aim to search for higher-order temporal communities based on this model.
An intuitive approach to enable community search with the $k$-truss model is to convert temporal networks into static networks. Unfortunately, this transformation scales the graph to the square of its original size, resulting in prohibitively high space\&time overhead for massive networks (see Section \ref{sec:pre}). In this paper, we define a novel temporal community model and propose efficient algorithms to address these challenges.
Our main contributions are summarized as follows:

\stitle{\underline{Novel Model.}} We propose a novel temporal community model named maximal-$\delta$-truss (\textit{MDT}), which is based on the temporal support (i.e., the number of edges participating in temporal triangles). The \textit{MDT} model can seamlessly capture both higher-order structural cohesiveness and the intensity of temporal interactions. 

\stitle{\underline{Efficient Algorithms.}} To search for the specific \textit{MDT} containing the user-initiated query node, we first develop a local search method with powerful pruning strategies to identify the solution within the small temporal subgraph expanded from the query node. Then, combining the elegant properties of \textit{MDT}, we construct the Temporal Trussness index (\textit{TT}-index) for edges. Equipped with the \textit{TT}-index, we can efficiently search for highly probable target subgraphs rather than performing a full search of the expanded subgraphs. In this way, we accelerate the search process.

\stitle{\underline{Extensive Experiments.}} We employ nine real-world temporal networks and seven competitors to evaluate the efficiency, scalability, and effectiveness of our solutions. These results indicate that our methods are more efficient and effective than the baselines. For example, our methods can process massive temporal networks (e.g., the million-node DBLP dataset) in a few minutes, whereas some baselines cannot obtain the result within two days. Additionally, our case studies also demonstrate that our model can mine more meaningful higher-order temporal communities that the competitors cannot identify. Our source codes and datasets are available at https://anonymous.4open.science/r/MDT-2C28.

\section{Preliminaries} \label{sec:pre}
A temporal graph $\mathcal{G}=(V,\mathcal{E},\mathcal{T})$ records all interactions and their temporal information between the nodes in $V$ through temporal edges in $\mathcal{E}$ during the time interval $\mathcal{T}$. Here, a temporal edge $e{(u,v,t)} \in \mathcal{E}$ denotes an interaction between $u$ and $v$ at time $t$.
When disregarding the temporal aspect of $\mathcal{G}$, we define its static graph $\bar{\mathcal G}=(V,\bar {\mathcal E})$ where $\bar {\mathcal E}=\{\bar e{(u,v)}|\exists e{(u,v,t)}\in{\mathcal{G}}\}$. For simplicity, we also denote static edges in $\bar{\mathcal{G}}$ as $\bar e$. We define a function $\odot$ to covert temporal graphs (or edges) into static graphs (or edges), specifically $\odot(\mathcal{G})=\bar{\mathcal{G}}$ and $\odot(e{(u,v,t)})=e{(u,v)}$. For a temporal graph $\mathcal{G}$, given a vertex set $S\subset{V}$, we can derive its induced temporal subgraph $\mathcal{G_S}=(S,\mathcal{E}_S,\mathcal{T}_{S})$ where $\mathcal{E}_S = \{e{(u,v,t)}|u,v\in{S},e{(u,v,t)}\in {\mathcal{E}}\}$. 

\begin{figure*}
	\centering
	\subfigure[Temporal graph $\mathcal{G}$]{ 
		\label{fig:basic_example:a} 
		\includegraphics[width=0.23\textwidth]{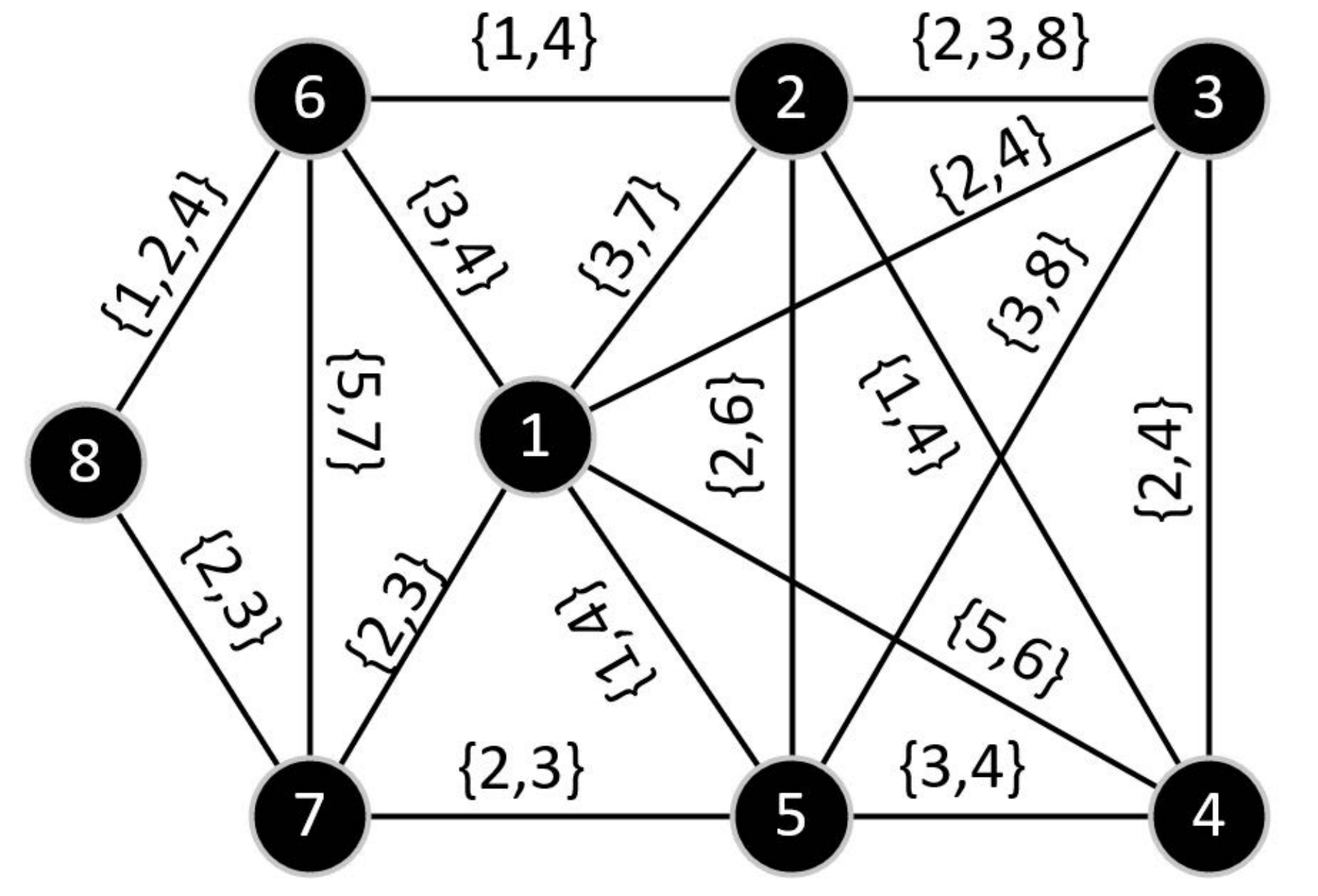} 
	} 
	\subfigure[Static graph $\bar{ \mathcal{G}}$ of $\mathcal{G}$]{ 
		\label{fig:basic_example:b} 
		\includegraphics[width=0.23\textwidth]{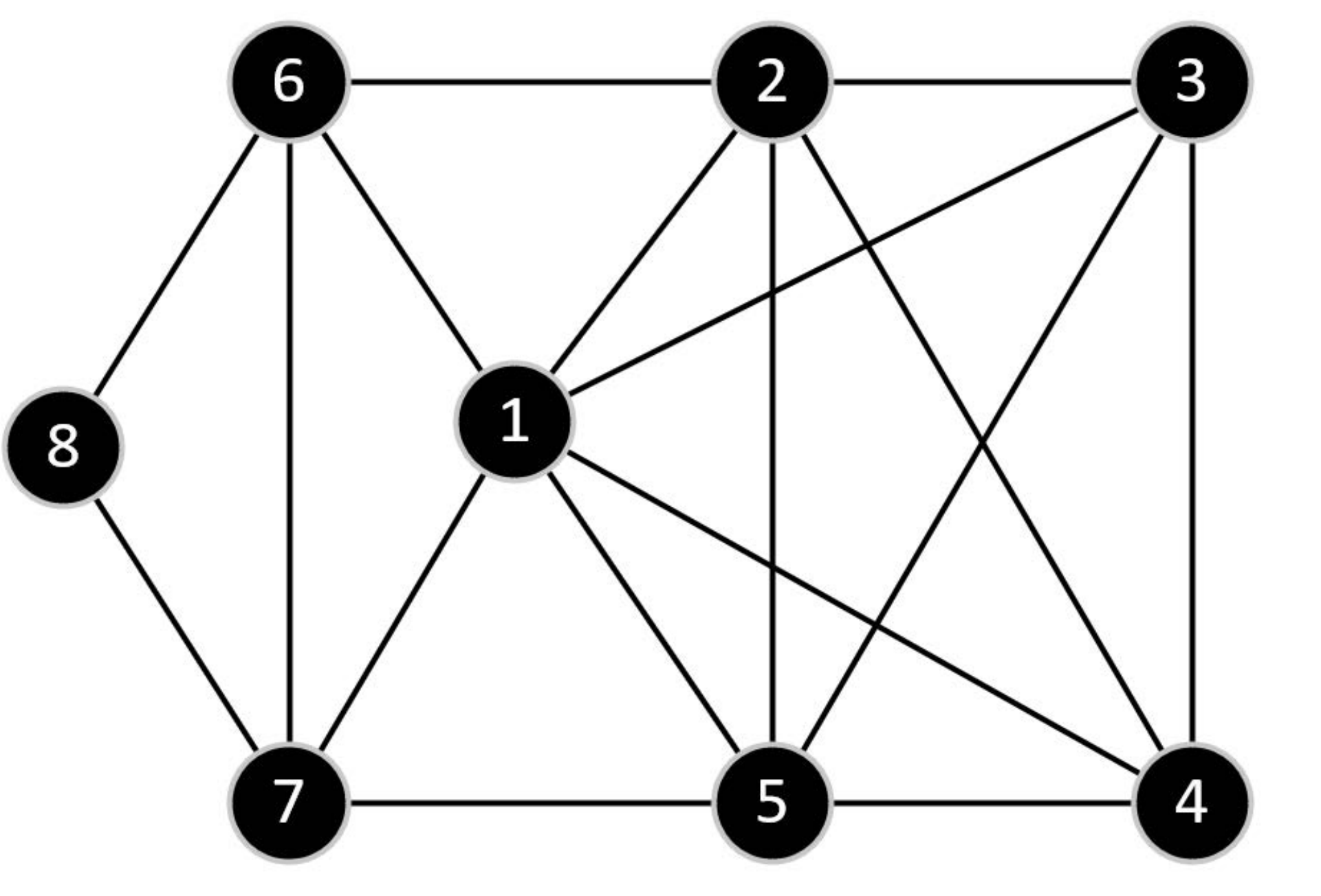} 
	} 
	\subfigure[Temporal subgraph $\mathcal{G}_{S_{1}}$]{ 
		\label{fig:basic_example:c}
		\includegraphics[width=0.23\textwidth]{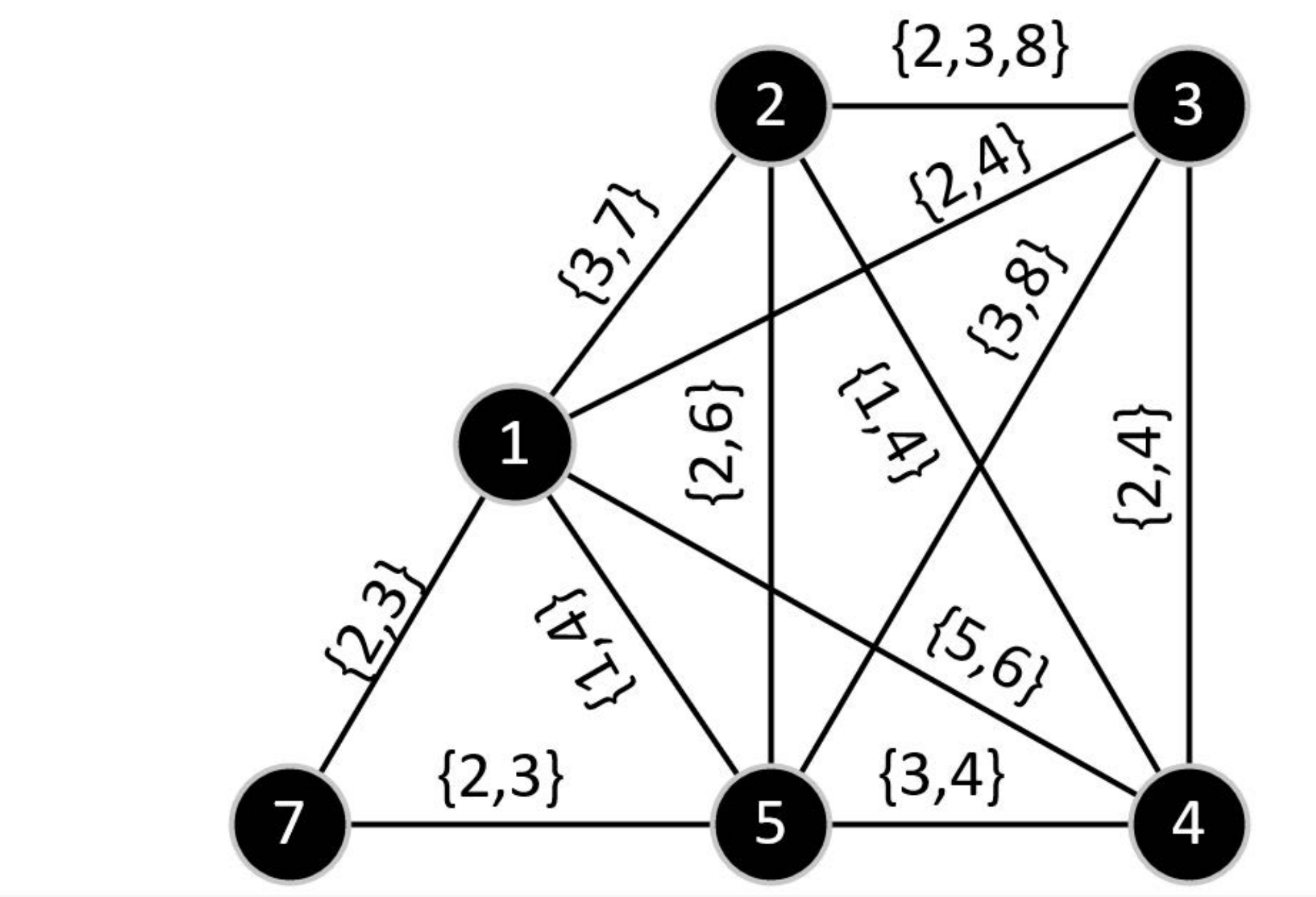} 
	} 
	\subfigure[Temporal subgraph $\mathcal{G}_{S_{2}}$]{ 
		\label{fig:basic_example:d}
		\includegraphics[width=0.23\textwidth]{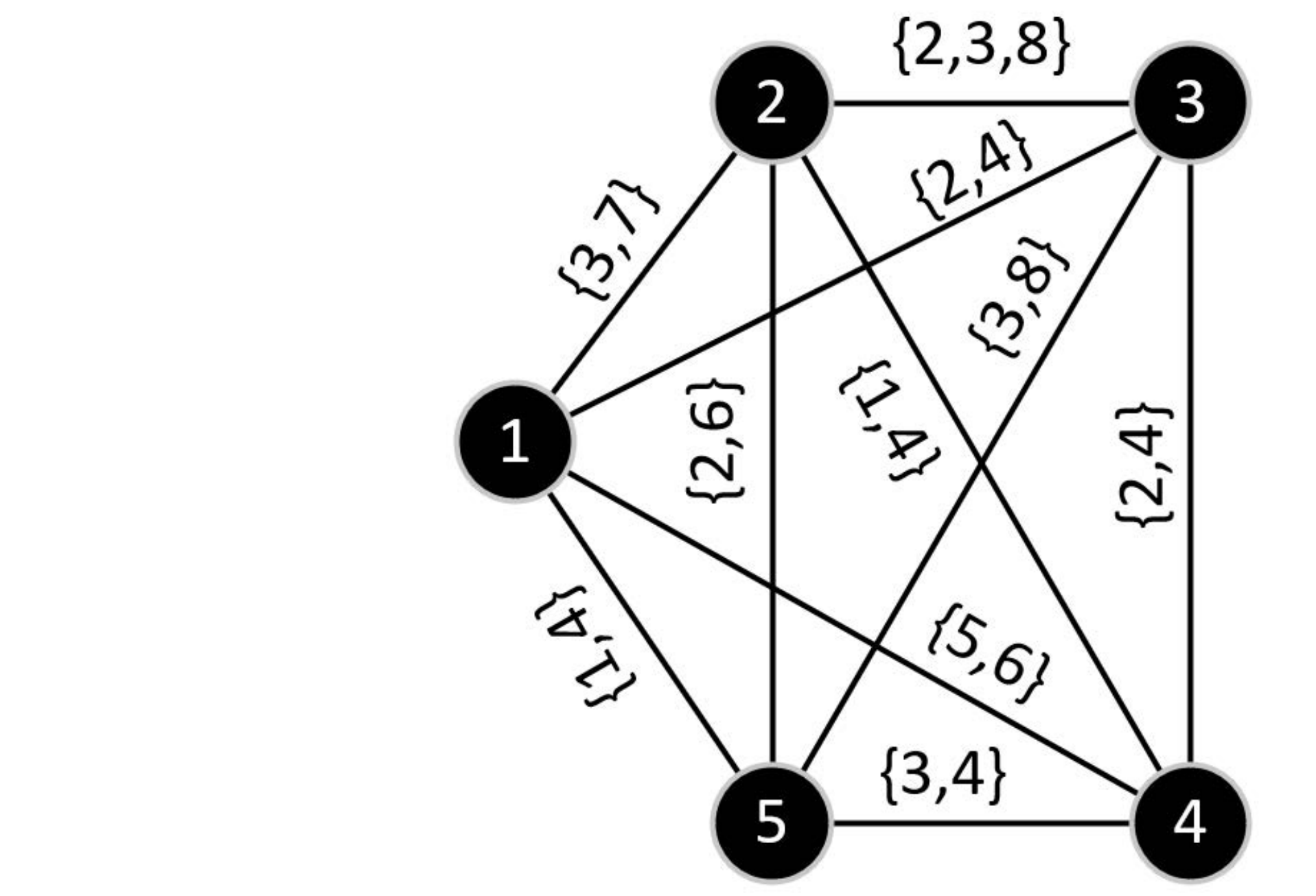} 
	} 
        \vspace{-0.3cm}
	\caption{Temporal (sub)graph and static subgraph}
	\label{fig:basic_example}
    \vspace{-0.3cm}
\end{figure*}

\begin{Example}
\figurename\ref{fig:basic_example:a} shows a temporal graph $\mathcal{G}$ with 8 nodes and 36 temporal edges, and $\mathcal{T}=[1,8]$. \figurename\ref{fig:basic_example:b} is the static graph of $\mathcal{G}$. The temporal subgraphs induced by the vertex sets $S_1=\{1,2,3,4,5,7\}$ and $S_2=\{1,2,3,4,5\}$ are depicted in \figurename\ref{fig:basic_example:c} and \figurename\ref{fig:basic_example:d}, respectively.
\end{Example}

\begin{Definition}[Temporal Triangle] \label{def:temp_triangle}
In a temporal graph $\mathcal{G}=(V,\mathcal{E},\mathcal{T})$, given three temporal edges $e_1(u,v,t_1)$, $e_2(u,w,t_2)$, $e_3(v,w,t_3) \in \mathcal{E}$, if their static edges $\bar e_1(u,v)$, $\bar e_2(u,w)$, $\bar e_3(v,w)$ can form a triangle $\bar \triangle_{uvw}$, then we say $\triangle_{e_1e_2e_3}$ is a temporal triangle. 
\end{Definition}

Similarly, the function $\odot$ converts a temporal triangle into its static counterpart, denoted as $\odot (\triangle)=\bar \triangle$. 
Let $\Delta(\triangle)=\max\{|t_i-t_j||1\leq i,j\leq 3\}$ denote the time span of $\triangle_{e_1e_2e_3}$ where $e_1$, $e_2$, and $e_3$ correspond to timestamps $t_1$, $t_2$, and $t_3$, respectively. Various time spans indicate different levels of interaction tightness, and individuals have varying criteria for assessing this tightness based on the time span. In a temporal graph, for a personalized parameter $\delta$ specified by users, the temporal property of a static triangle is characterized by the number of temporal triangles whose time span is no greater than $\delta$. This is expressed by the formula $N(\bar \triangle,\delta)$=$|\{\triangle|\odot (\triangle)=\bar \triangle, \Delta(\triangle) \leq \delta\}|$. Sometimes $N(\bar \triangle,\delta)$ is abbreviated as $N(\bar \triangle)$.

\begin{Definition}[$\delta$-Temporal Support]
\label{def:tsup} Given an integer $\delta$, the $\delta$-temporal support of a static edge $\bar e(u,v)$ in the temporal graph $\mathcal {G}$ is denoted as $TSup_{\mathcal G}(\bar e,\delta)$=$|\{\triangle_{e_1e_2e_3}|\exists e_i \in \triangle_{e_1e_2e_3}(i \in \{1, 2, 3\}),  \odot (e_i)=\bar e,\Delta( \triangle_{e_1e_2e_3})\leq \delta\}|$. That is, it is the number of temporal triangles in the temporal graph $\mathcal{G}$ that involve a temporal edge $e$ satisfying $\odot (e)=\bar e$ and have a time span not exceeding $\delta$.
\end{Definition}

The temporal support of an edge can be rewritten as $TSup(\bar e,\delta)=\sum_{\bar e\in \bar \triangle}N(\bar \triangle,\delta)$. When the context is clear, we abbreviate $TSup_{\mathcal{G}}(\bar e,\delta)$ to $TSup(\bar e)$. It is straightforward to state the following proposition.

\begin{Proposition} \label{pro:g_l_sup}
Given a temporal graph $\mathcal{G}$ and its subgraph $\mathcal S$, for any $\bar{e} \in \mathcal S$, we have $TSup_{\mathcal{G}}(\bar{e})\geq$ $TSup_{\mathcal S}(\bar{e})$.
\end{Proposition}

Two triagnles $\triangle_{i}$ and $\triangle_{i+1}$ are considered connected if they share at least one common edge or vertex. Formally, this is expressed as $\triangle_i \cap \triangle_{i+1} \neq \emptyset$. Based on this definition, we define higher-order connectivity as follows:
\begin{Definition}[Higher-order Connectivity] \label{def:higher-order connct}
Given a parameter $\delta$, two triangles $\triangle_1, \triangle_n$ are considered to be higher-order connected if the following two conditions are satisfied: (1) $\triangle_1$ and $\triangle_n$ can be connected by a sequence of triangles $\triangle_{1}$, ..., $\triangle_{i}$, ..., $\triangle_{n}$, where $n\geq 2$ and each pair of consecutive triangles $\triangle_{i}$ and $\triangle_{i+1}$ in this sequence must satisfy $\triangle_i \cap \triangle_{i+1} \neq \emptyset$. (2) $\forall 1 \leq i \leq n$, $N(\bar {\triangle_i}) \neq 0$.

\end{Definition}


\begin{Definition}[$(k,\delta)$-truss]\label{def:k-delta-truss}	\label{def:problem1}
Given a temporal graph $\mathcal{G}$ and integers $\delta$ and $k$, a temporal subgraph $\mathcal S$ is called a \textit{$(k,\delta)$-truss} if it satisfies the following conditions:
(1) Cohesive: for every edge $\bar e \in \mathcal{S}$, $TSup_{\mathcal{S}}(\bar e,\delta) \geq k$. 
(2) Connective: for every pair of edges $\bar e_1,\bar e_2 \in {\mathcal{S}}$, there exist triangles $\triangle_x$ and $\triangle_y$ such that $\bar e_1 \in \triangle_x,\bar e_2 \in \triangle_y$, and $\triangle_x=\triangle_y$ or $\triangle_x$ and $\triangle_y$ are higher-order connected. 
(3) Maximal subgraph: there does not exist a subgraph $\mathcal S' \subset \mathcal G$ such that $\mathcal S\subset{\mathcal S'}$ and $\mathcal S'$ satisfies condition (1) and (2).
\end{Definition}

Conceptually, a $(k,\delta)$-truss not only remains the cohesiveness of k-truss model, but also captures the tightness of time. Specifically, condition (1) ensures that the subgraph is densely connected under the structural constraints of the truss model, while the restriction $\delta$ on temporal triangles guarantees that the nodes are tightly connected. Condition (2) further enhances this by ensuring that all edges in $(k,\delta)$-truss are strongly connected through a series of powerful and stable triangles. 
Generally, a larger $k$ indicates that edges are connected through tighter temporal triangles, resulting in greater structural density and temporal compactness. Such communities are more desirable to users. Nevertheless, choosing an appropriate parameter $k$ is challenging. Small $k$ values may yield a large number of solutions, while large $k$ values might result in no matching communities. However, $k$ is closely related to the user-specified $\delta$, which helps in finding the maximal-$\delta$-truss with the maximal $k^*$.

\begin{Definition}[Maximal-$\delta$-truss, MDT]
\label{def:problem2}
Given a temporal graph $\mathcal{G}$ and an integer $\delta$, a temporal subgraph $\mathcal S$ is a maximal-$\delta$-truss of $\mathcal{G}$ if $\mathcal S$ is a $(k,\delta)$-truss such that $k$ is maximal, i.e. there is no other $(k',\delta)$-truss with $k'>k$.
\end{Definition}


\begin{Example}
\figurename\ref{fig:basic_example:c} is an $(8,3)$-truss of \figurename\ref{fig:basic_example:a}, where each edge participates in at least eight temporal triangles with $\Delta(\triangle)\leq 3$.  \figurename\ref{fig:basic_example:d} is a $(12,3)$-truss. Since no other $(k,3)$-truss exists for $k>12$, \figurename\ref{fig:basic_example:d} is the maximal-$\delta$-truss.
\end{Example}

Users are typically more interested in communities that include the target  nodes. So we aim to mine a \textit{MDT} for a given query node.

\begin{Problem}[Maximal-$\delta$-truss mining]\label{def:problem}
For a temporal graph $\mathcal{G}$, a query node $q$, and an integer $\delta$, our objective is to find a maximal-$\delta$-truss containing $q$. This is denoted as q-MDT.
\end{Problem}

\stitle{Theoretical challenges.} 
Maximal-$\delta$-truss extends the properties of $k$-truss while maintaining the elegant properties of $k$-truss, such as $(k,\delta)$-truss $\subseteq (k-1,\delta)$-truss. Based on the idea from \textit{decompose} \cite{DBLP:conf/sigmod/HuangCQTY14}, a naive global search method (\textit{GS}) for the \textit{q-MDT} problem is to iteratively remove the edges with the lowest temporal support. 
Specifically, we need to ascertain the temporal support of all edges to guide the selection process for edge deletion. According to Definition \ref{def:temp_triangle}, a temporal triangle is a closed sequence of three temporal edges. Therefore, for a triangle $\bar \triangle_{uvw}=\{\bar e_1(u,v), \bar e_2(u,w), \bar e_3(v,w)\}$, we can count the number of the temporal triangle $\triangle_{e_1e_2e_3}$ with $\Delta(\triangle_{e_1e_2e_3})\leq \delta$ by checking the time span for all permutations of the temporal edges sequence. Then we compute $N(\triangle,\delta)$ for all triangles and accumulate $N(\triangle)$ to determine the temporal support for edges. Subsequently, we iteratively delete the edge $\bar e$ with the lowest temporal support and update the temporal support for edges sharing a triangle with $\bar e$. This continues until only one edge induced by $q$ remains, and then \textit{decompose} returns the maximal temporal subgraph $\mathcal S$ with the lowest temporal support and maximal $k^*$. Finally, \textit{GS} selects the connected components of $\mathcal S$ containing the query node $q$ by checking the connectivity of edges induced by $q$. This yields the final solution to our problem.

\textit{GS} is not efficient since it needs to compute the temporal support for all edges in $\mathcal{G}$ and then greedily deletes the edges. Some edges may not meet the given temporal support and are irrelevant to the solution. For example, edges not connected to those induced by $q$ through a sequence of static triangles (against condition (1) in Definition \ref{def:higher-order connct}) are not part of the solution, but \textit{GS} still computes their temporal support. In addition, it is time consuming for \textit{GS} as it begins its search from the entire graph. To address these inefficientcies, we propose a local search strategy in Section \ref{sec:local}.

\section{Local Search Strategy of \textit{q-MDT}} \label{sec:local}
One straightforward way to calculate the temporal support is to count the valid permutations of temporal edges. 
However, it is costly since there exist a large number of unnecessary operations. Here, we first calculate the temporal support by extracting edge timestamps and applying a sliding window technique. Subsequently, we locally explore the expanded temporal subgraph around the query node $q$ to find the solution of \textit{q-MDT}.


\subsection{Sliding Windows for Temporal Support} \label{sec:temp_trian_count}
 
\begin{figure}[H]
    \centering
    \vspace{-0.4cm}
    \includegraphics[width=0.45\textwidth]{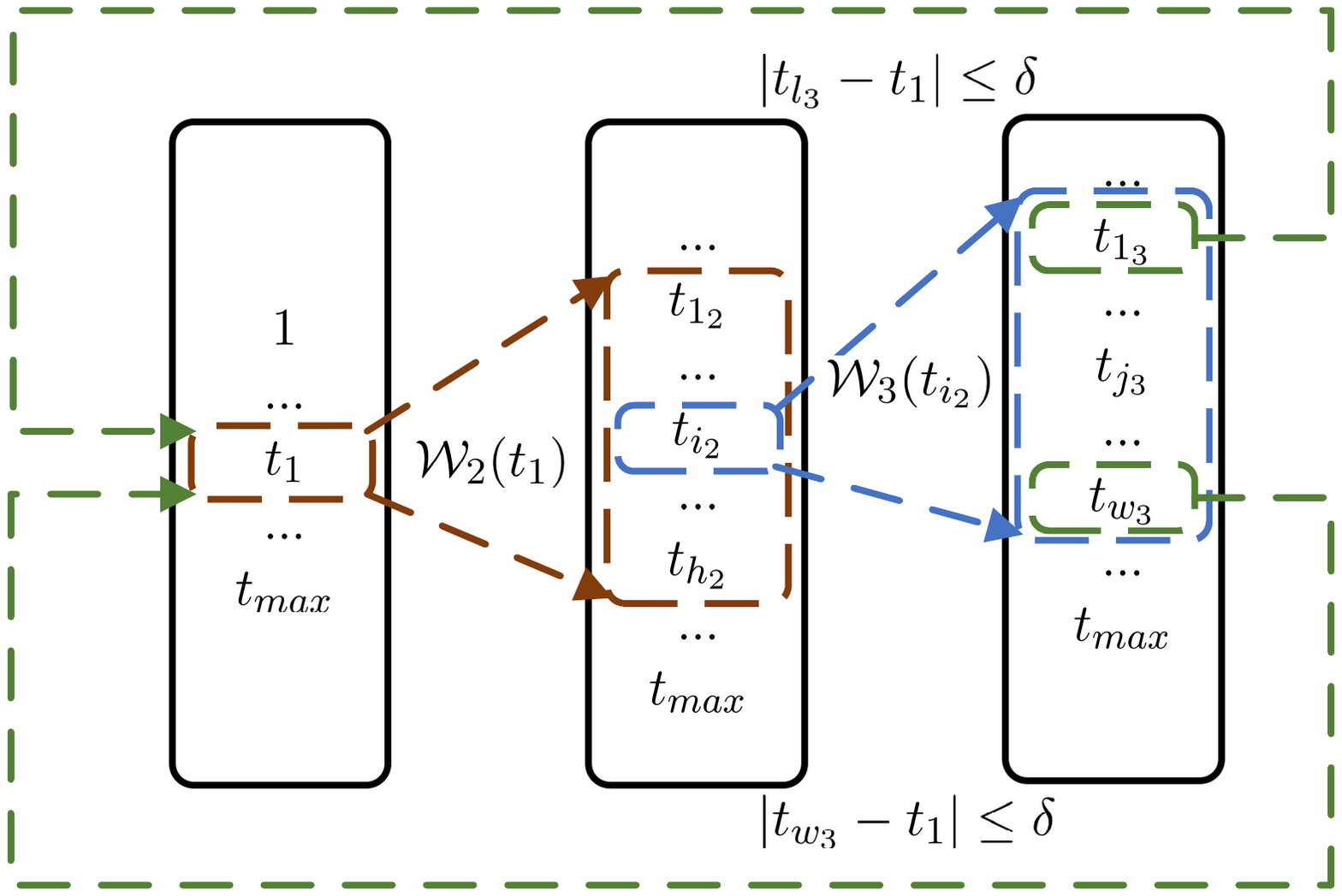} 
    \vspace{-0.6cm}
    \caption{Sliding window for counting $N(\triangle,\delta)$}
    \label{fig:cdt} 
    \vspace{-0.2cm}
\end{figure}

Computing $N(\triangle,\delta)$ by listing permutations of all temporal edges is expensive and unnecessary. However, if the time span of two temporal edges $e_1(u,v,t_1), e_2(u,w,t_2)$ exceeds $\delta$, then any permutations that contain these two edges definitely cannot contribute to the temporal support. Inspired by this, we propose a new strategy using sliding windows to count the $\delta$-temporal triangles, as illustrated in \figurename \ref{fig:cdt}. 

For a static triangle $\triangle=\{\bar e_1,\bar e_2,\bar e_3\}$, three sliding windows, each corresponding to one of the static edges, are used to count the number of temporal triangles that satisfy $\Delta(\tilde \triangle)\leq \delta$.
We first extract three timestamp lists $T(u,v)$, $T(u,w)$, $T(v,w)$ for edges in $\triangle_{uvw}$. Here, each list $T(x,y)$ contains the timestamps of all temporal edges in $\mathcal{G}=(V,\mathcal{E},\mathcal{T})$ corresponding to the static edge $(x,y)$, i.e., $T(x,y)=\{t|e(x,y,t)\in \mathcal E\}$.
Specifically, we sort these lists in ascending order by their size and name them as $T_1$, $T_2$, $T_3$ respectively, such that $|T_1|\leq |T_2| \leq |T_3|$. For each timestamp $t_1$ in $T_1$, the algorithm maintains a window $\mathcal{W}_2(t_1)={t_{1_2},...,t_{i_2},...,t_{h_2}}$ consisting of timestamps from $T_2$, such that every $t_{i_2} \in \mathcal{W}_2(t_1)$ satisfies $|t_1-t_{i_2}|\leq{\delta}$. Similarly, $\forall t_{i_2} \in \mathcal{W}_2$, the algorithm records the window $\mathcal{W}_3(t_{i_2})=\{t_{1_3},...,t_{j_3},...,t_{w_3}\}$ for $T_3$, where $|t_{i_2}-t_{j_3}|\leq \delta$ and $|t_1-t_{j_3}|\leq \delta$ for $\forall t_{j_3} \in \mathcal{W}_3(t_{i_2})$. 

Using sliding windows, we avoid the need to compute over the full $|T_1| \times |T_2| \times |T_3|$ space. Instead, for each timestamp $t_1$ in $T_1$, the number of temporal triangles with a time span not greater than $\delta$ is given by $\sum_{\substack{t_{i_2} \in \mathcal{W}_2(t_1)}} |\mathcal{W}_3(t_{i_2})|$. 
Therefore, we can quickly obtain the number of temporal triangles in $\mathcal{G}$ induced by $\triangle_{uvw}$. 
By accumulating $N(\triangle,\delta)$ for each triangle containing the edge $\bar e$, we can get the temporal support for $\bar e$.

\subsection{Local Exploration for Candidate Subgraph} \label{subsec:local}
Condition (2) in Definition \ref{def:problem1} requires that all edges in a $(k,\delta)$-truss must be connected, thus according to Definition \ref{def:higher-order connct}, all edges must be connected by a series of triangles where $N(\triangle,\delta) \neq 0$. Consequently, if an edge cannot be connected to any edge induced by $q$ through a set of triangles, or if these triangles cannot induce a temporal triangle with a time span no greater than $\delta$, then these edges must not appear in \textit{q-MDT}. 
Inspired by this, we locally explore the candidate temporal subgraph containing the \textit{q-MDT} to avoid processing irrelevant edges.

Suppose $\mathcal{S^*}$ is a \textit{q-MDT}, denoted as $(k^*,\delta)$-truss. For each edge $\bar{e} \in \mathcal{S^*}$, we have $TSup_{\mathcal{S^*}}(\bar{e})\geq{k^*}$. If an edge $\bar e_0$ with $TSup_{\mathcal{G}}(\bar{e}_0)<{k^*}$, it cannot appear in a \textit{q-MDT} according to Proposition \ref{pro:g_l_sup}. Consequently, searching for a \textit{q-MDT} in $\mathcal{G}$ is equivalent to searching it in the temporal subgraph $\mathcal{H}$, where each edge $\bar{e}$ satisfies $TSup_{\mathcal{G}}(\bar{e})\geq{k^*}$. 
Since $k^*$ is unknown, a naive approach is to calculate the temporal support for all edges induced by $q$ and choose the smallest value $k_l$ as the threshold to extract the subgraph where $TSup_{\mathcal G}(\bar e)\geq k_l$. However, $k_l$ is usually small, sometimes even 0, leading to an overly large subgraph and high computational overhead. 

We leverage the upper bound of $k^*$ and design a local search algorithm (\textit{LS}) that performs a binary search on the temporal graph to get the expanded temporal subgraph. The key idea of \textit{LS} is to find a small expanded temporal subgraph that contains the query node \textit{q}, and then explore the exact solution in this subgraph. Given parameters $\delta$ and the original $k$, \textit{LS} adopts a "compute-while-expanding" strategy to derive the expanded temporal subgraph $\mathcal H$. Initially, an edge $\bar e(q,v)$ induced by $q$ is selected. We then check if $TSup_{\mathcal{G}}(\bar{e}(q,v))\geq{k}$, where $k$ is an adjustable threshold. If the condition is satisfied, $\bar e(q,v)$ is considered as the first "expandable" edge. For each "expandable" edge $\bar e$, it computes the temporal support of the edges that are in the same triangle with $\bar e$, and picks up the new "expandable" edges for iterative expansion. The expanded temporal subgraph is obtained by iteratively expanding from the "expandable" edges $\bar e(q,v)$.


\begin{Proposition} \label{obr:km_vary}
If an exact solution cannot be found in $\mathcal{H}$ constructed with threshold $k$, the subsequent threshold $k'$ for the next round must be smaller than $k$, i.e. $k'< k$.
\end{Proposition}

\begin{Proposition} \label{obr:expan_graph}
For two temporal subgraphs $\mathcal{H}_1$ and $\mathcal{H}_2$ expanded with thresholds $k_1$ and $k_2$ respectively, if $k_1\leq {k_2}$, then $\mathcal{H}_2 \subseteq{\mathcal H}_1$.
\end{Proposition}

\textit{LS} computes the temporal support of an edge only when it is accessed, thereby avoiding unnecessary computations and explorations of infeasible edges. For each expansion, the algorithm iteratively adjusts the threshold $k$ for temporal support and adds the edges whose temporal support is not less than $k$ to the expanded subgraph $\mathcal{H}$. This ensures that the expanded subgraph is neither too large nor too small. When computing the temporal support for an edge, if its current value is no less than $k$ then we can immediately stop access of the triangles it participates in. So both "compute-while-expanding" and "early stop" strategies improve the performance of exploration.

For the detailed algorithm description and pseudocode, please refer to Appendix~\ref{appendix-code-LS}. Additionally, we provide an example in Appendix~\ref{appendix-example-local} and analyze its complexity in Appendix~\ref{appendix-complexity}.

%



\section{Fast \textit{q-MDT} Search with Temporal Trussness} \label{sec:index}

Compared with the baseline algorithm \textit{GS}, the local exploration strategy shows better performance.  \textit{LS} still incurs  some unnecessary visits because it needs to search some candidate subgraphs before the target communities are found. Moreover, the search space of the local method is related to the query node and the parameter $\delta$. So, it accesses a much larger search space when the graph is large or the parameter is big. In the cases, it generally leads to longer response time. 
One reason is that $LS$ method does not know where the possible \textit{q-MDT}s are. If we know in advance the maximum value of $k$ (i.e. temporal trussness) in which each edge can participate in the $(k,\delta)$-truss, then we can search the highly possible target subgraphs instead of full search of the expanded subgrphs. By this way, we can prune much search space. Moreover, many queries may share the same parameter $\delta$. Especially it is true when the timespan of a graph is not large. Due to the above reasons, we propose a method using the temporal trussness to efficiently find solutions for \textit{q-MDT} query.

\subsection{Temporal Trussness (TT)}
Each edge may appear in multiple triangles. However, A subgraph has several edges which have different values of $\delta$-temporal support. 
Here, we first define the temporal trussness of a subgraph, then define the temporal trussness of an edge.

\begin{Definition}[Temporal Trussness] \label{def:tau}
Given an integer $\delta$, the temporal trussness of a temporal subgraph $\mathcal{H}$ is the minimum temporal support of the edges in it, i.e.  $\tau(\mathcal{S},\delta)=\min\{TSup_{\mathcal{S}}(\bar{e}) | \bar{e} \in { { \mathcal{S}}}\}$. The temporal trussness of $\bar e \in  {\mathcal G}$ is the maximal trussness of the temporal subgraphs that $\bar e$ participates in, where $\tau_{\mathcal{G}}(\bar{e},\delta)=\max\{\tau(\mathcal S,\delta)|\bar{e}\in {\mathcal{S}}, \mathcal{S}\subset{\mathcal{G}}\}$.
\end{Definition}
When it does not cause confusion, we abbreviate $\tau(\mathcal{S},\delta)$ and $\tau_{\mathcal{G}}(\bar{e},\delta)$ as $\tau(\mathcal{S})$ and $\tau(\bar{e})$, respectively. 

If we want to know the temporal trussness of an edge, we first need to compute the $N(\triangle)$ of each triangle containing the edge so that we can get its temporal support and thus the temporal trussness.
Suppose there are $|\triangle_{\mathcal{G}}|$ triangles in $\mathcal G$, we need $|\triangle_{\mathcal{G}}|\times t_{max}\times \delta ^ 2$ to get $N(\triangle)$ of all triangles, so this method takes $|\triangle_{\mathcal{G}}|\times (t_{max})^2\times \delta ^ 2$ to compute the temporal support for the edges in $\mathcal{G}$ under different $\delta$. It is very expensive if the temporal graph is large and spans a large time interval. 
Fortunately, although this way is a bit unsatisfactory, it still provides us with a new idea. When we set $\delta=t_{max}$ to compute the corresponding temporal support, as all temporal edges span no greater than $t_{max}$, so all the triangles in $\mathcal{G}$ satisfy that $N(\triangle)>0$ and all the temporal triangles in $\mathcal{G}$ satisfy that $\Delta(\tilde \triangle)\leq t_{max}$. For each triangle $\triangle$ $\subset$ $\mathcal{G}$, $N(\triangle)$ is the number of the permutations of the temporal edges that lay on the edge in $\triangle$. Inspired on this observing, before introducing a novel algorithm to calculate the temporal support, we import a novel definition called $\delta$-slice which is a set of subgraphs whose time span is $\delta$.

\begin{Definition}[$\delta$-slice]  \label{def:delta_slice}
For a temporal graph $\mathcal{G}$ and an integer $\delta$, a $\delta$-slice is a sequence of temporal subgraphs of $\mathcal G$ in which the time span of each subgraph is $\delta$, denoted as $\mathcal G^\delta$=$\{\mathcal{G}_{[1,\delta+1]}$, $\mathcal{G}_{[2,\delta+2]}$, ..., $\mathcal{G}_{[t_{max}-\delta,t_{max}]}\}$.
	\end{Definition}
	 
The range of $\delta$ is [$0$, $t_{max}$]. We shorten $\mathcal{G}_{[i,i+\delta]}$ to 
$\mathcal{G}_{i}^{\delta}$. When no misunderstanding occurs, $\mathcal{G}_{i}^{\delta}$ is further abbreviated to $\mathcal{G}_{i}$. 
Since the temporal graph can be decomposed into a list of temporal subgraphs, each of which starts at a specific timestamp and spans a specific interval $\delta$, we need to compute the temporal trussness of each temporal subgraph in the list. Computing the temporal trussness of all edges is time-consuming if a temporal graph has a large time span. In the following, we will discuss how to speed up the computation.

For each $\mathcal{G}_j$ in $\delta$-slice, it is easy to know the number of temporal triangles $N_{\mathcal{G}_j}(\triangle_{uvw})$=$|T_{\mathcal{G}_j}(u,v)|\times$ $|T_{\mathcal{G}_j}(u,w)|\times|T_{\mathcal{G}_j}(v,w)|$, where $T_{\mathcal{G}_j}(u,v)$ is the list of timestamps in edge $\bar e(u,v)$. By Definition \ref{def:tsup}, we get the following propositions and prove them in Appendix~\ref{appendix-proofs}.


\begin{Proposition}     \label{pro:trian_tsup}
Let $\mathcal{G}^\delta$ be a $\delta$-slice of  $\mathcal{G}$. For a triangle $\triangle \subseteq \mathcal{G}$, its 
corresponding $N_{\mathcal{G}_{[i, i+\delta+1]}}(\triangle,\delta)$=  $N_{\mathcal{G}_i^\delta}(\triangle,\delta)$+$N_{\mathcal{G}_{i+1}^\delta}(\triangle,\delta)$-$ N_{\mathcal{G}_{i+1}^{\delta-1}}(\triangle,\delta-1)$.
\end{Proposition}

\begin{Proposition}  \label{pro:trian_up}
Given a temporal graph $\mathcal{G}$ and an integer $\delta$, for a triangle $\triangle \subseteq \mathcal{G}$, its  
$N_{\mathcal{G}}(\triangle,\delta)=N_{\mathcal{G}}(\triangle,\delta-1)+\sum_{i=1}^{t_{max}-\delta}(\Phi_{N_{\mathcal{G}_i^\delta}(\triangle,\delta)}-\Phi_{N_{\mathcal{G}_{i+1}^{\delta-1}}(\triangle,\delta-1)})$, where $\Phi_{N_{\mathcal{G}_{i}^\delta}(\triangle,\delta)}=N_{\mathcal{G}_{i}^\delta}(\triangle,\delta)-N_{\mathcal{G}_{i}^{\delta-1}}(\triangle,\delta-1)$.
\end{Proposition}

\begin{Proposition}  \label{pro:edge_tsup}
Given a temporal graph $\mathcal{G}$, an integer $\delta$ and the corresponding $\delta$-slice, $(\delta-1)$-slice, the temporal support of an edge $\bar{e}\in \mathcal{G}$ is given by: $TSup_{\mathcal{G}}(\bar{e},\delta)=TSup_{\mathcal{G}}(\bar{e},\delta-1)+\sum_{j=1}^{n}\sum_{i=1}^{t_{max}-\delta}(\Phi_{N_{\mathcal{G}_{i}^\delta}(\triangle_j,\delta)}-\Phi_{N_{\mathcal{G}_{i+1}^{\delta-1}}(\triangle_j,\delta-1)})$, where $\bar{e} \in \triangle_j$.
\end{Proposition}

\subsection{Bottom-up Incremental Computing} \label{subsec:index_build} 
\begin{figure}
    \centering		
    \includegraphics[width=0.95\linewidth]{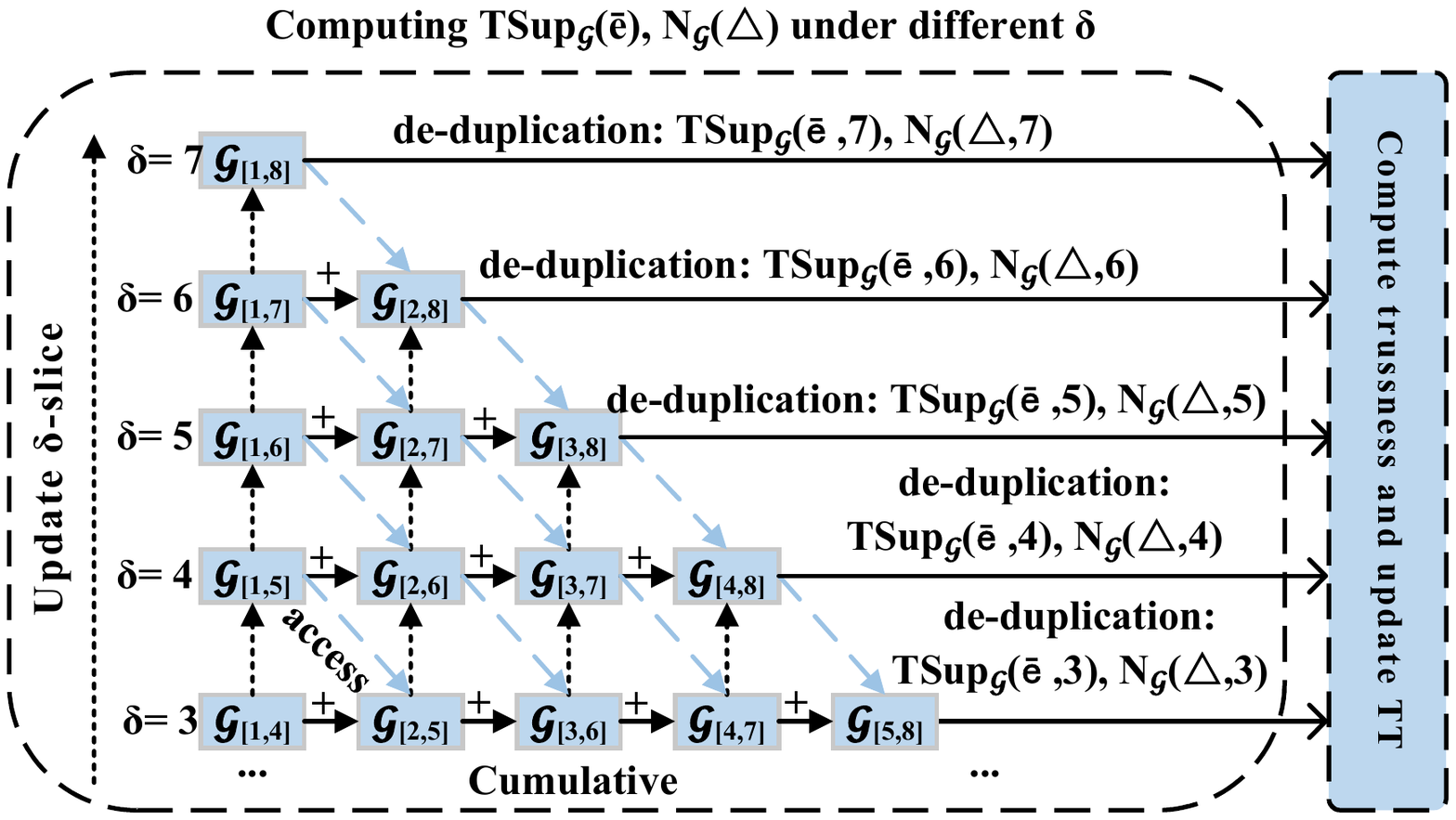}
    \vspace{-0.7cm}
    \caption{The construction framework of the \textit{TT}-index}
    \label{fig:index_framework} 
    \vspace{-0.4cm}
\end{figure}

The temporal support of any edge under $\delta>t_{max}$ is same as that under $\delta=t_{max}$, since the time span of any two different temporal edges in $\mathcal{G}$ is not larger than $t_{max}$. So, based on previous propositions, we proposed the bottom-up algorithm (Algorithm \ref{alg:index_construct}) to compute the temporal support to obtain the trussness of the edges under different $\delta$. In the algorithm, we increase $\delta$ from $0$ to $t_{max}$ and incrementally compute temporal support and trussness. 
The algorithm includes two stages (\figurename\ref{fig:index_framework}).
The trussness of each edge $\bar e \in \mathcal G$ will be saved in a list of pairs $(\delta,\tau(\bar e))$, such as $\{(0,\tau_0),...,(i,\tau_i),...,(t_{max},\tau_{t_{max}})\}$. Since it is a list of edge temporal trussness, we named it as \textit{TT}-index.

\begin{algorithm}[b]
\caption{Bottom-up Computing of Temporal Trussness} \label{alg:index_construct}

\begin{flushleft}
		\hspace*{0.02in} {\bf Input:}
A temporal graph $\mathcal{G}$\\
		\hspace*{0.02in} {\bf Output:}  The \textit{TT}-index for $\mathcal{G}$
	\end{flushleft}
	\begin{algorithmic}[1]
\State     Initial $\Phi[\triangle]=0$;
	\For{$\delta$ $\gets{0}$ upto $t_{max}$}\State CountAllTSup($\mathcal{G},\mathcal{H},\delta, \Phi$);	
 \State decompose($\mathcal{G}$,$\delta$,\textit{TT});
	\EndFor
	\State \Return{\textit{TT}};
 \end{algorithmic}
\end{algorithm}

\begin{algorithm}[t]
\caption{CountAllTSup} 	\label{alg:count_all_sup}

\begin{flushleft}
		\hspace*{0.02in} {\bf Input:}
$\mathcal{G}$, $\delta$, $\mathcal G^\delta$, $\mathcal G^{\delta-1}$ and $\Phi$ \\
		\hspace*{0.02in} {\bf Output:} $\forall \triangle \subseteq \mathcal{G},N_{\mathcal{G}}(\triangle,\delta), \forall \bar e \in \mathcal{G}, TSup_{\mathcal{G}}(\bar e,\delta)$
	\end{flushleft}
	\begin{algorithmic}[1]
	\State	Let $\mathcal{P}[\triangle]$ record the value of $\Phi_{N_{\mathcal{G}_{1}^\delta}(\triangle,\delta)}$  ;
	\State	Initial $\Phi'=\Phi$;
		\For {$t \gets{1}$ upto $t_{max}-\delta$}
			\For{$\bar{e}(u,v) \in \mathcal G_{[t+\delta,t+\delta]}$}	\State $\mathcal{G}_{t}^\delta=\mathcal{G}_{t}^{\delta-1}\cup e(u,v,t)$;
			\EndFor
			\For{$\bar{e}(u,v)$ in $ \mathcal G_{[t+\delta,t+\delta]}$}
				\For{$w$ in $\mathcal {D}_{\mathcal G_{t}^\delta}(u) \cap \mathcal {D}_{\mathcal G_{t}^\delta}(v) $}
	\State$s=|T_{\mathcal{G}_{t}^\delta}(u,v)|*|T_{\mathcal{G}_{t}^\delta}(u,w)|*|T_{\mathcal{G}_{t}^\delta}(v,w)|$;
\State $\Phi[\triangle_{uvw}]+=s-N_{\mathcal{G}_{t}^{\delta-1}}(\triangle_{uvw})$;
					\If{$t=1$}
						\State $\mathcal{P}[\triangle_{uvw}]=\Phi[\triangle_{uvw}]$;
					\EndIf
					\State $N_{\mathcal{G}_{t}^\delta}(\triangle_{uvw})=s$;
				\EndFor
			\EndFor
		\EndFor
		\For {$\triangle_{uvw}$ in $\Phi$}
			\If{$\Phi[\triangle_{uvw}]=0$} \State continue;
   \EndIf
			\If{\textit{TT}[$\triangle_{uvw}$] has not been recorded}
				\State $\textit{TT}[\triangle_{uvw}]\gets{\delta}$
			\EndIf
			\State $\varphi=\Phi[\triangle_{uvw}]-\Phi'[\triangle_{uvw}]$;
			\State $N_{\mathcal{G}}(\triangle_{uvw},\delta)\gets{N_{\mathcal{G}}(\triangle_{uvw},\delta-1)+\varphi}$;
			\State $TSup_{\mathcal{G}}(\bar e(u,v),\delta)\gets{TSup_{\mathcal{G}}(\bar e(u,v),\delta-1)+\varphi}$;
			\State $TSup_{\mathcal{G}}(\bar e(u,w),\delta)\gets{TSup_{\mathcal{G}}(\bar e(u,w),\delta-1)+\varphi}$;
			\State $TSup_{\mathcal{G}}(\bar e(v,w),\delta)\gets{TSup_{\mathcal{G}}(\bar e(v,w),\delta-1)+\varphi}$;
	\State $\Phi[\triangle_{uvw}]=\Phi[\triangle_{uvw}]-\mathcal{P}[\triangle_{uvw}]$;
		\EndFor
  \end{algorithmic}
\end{algorithm}

In the first stage (Left part of \figurename \ref{fig:index_framework}, Algorithm \ref{alg:count_all_sup}), the algorithm iteratively generates $\delta$-slice of $\mathcal{G}$ and computes the number of temporal triangles in each interval. Specifically, when the algorithm increases $\delta$ by 1, temporal subgraph $\mathcal{G}_{i}^{\delta-1}$ in $(\delta-1)$-slice will be expanded to include the edges in snapshot $ \mathcal G_{[i+\delta,i+\delta]}$ (Line 4-5). So, the temporal subgraph $\mathcal{G}_{i}^{\delta}$ can be derived from $\mathcal{G}_{i}^{\delta-1}$. The added edges will induce new triangles and change $N(\triangle)$ of the existing triangles, as well as the $TSup(\bar e)$ of the edges in these triangles.	
The algorithm computes $N_{\mathcal{G}_{i}^{\delta}}(\triangle,\delta)$ for all triangles and then updates $\Phi$ based on the increment (Lines 8-9). The increment of $N(\triangle)$ for triangles in $\mathcal{G}_{1}^\delta$ (denoted as $\mathcal{P}$) is saved for the next iteration (Lines 10-11). The algorithm then traverses and updates these triangles and their relevant edges. Specifically, for a triangle $\triangle$, if the increment $\Phi[\triangle]=0$, this indicates that $N(\triangle,\delta)$ remains unchanged, i.e., $N(\triangle,\delta)=N(\triangle,\delta-1)$, and thus the edges within it are unaffected (Lines 14-15). Otherwise, the algorithm updates $N(\triangle,\delta)$ for the triangles and adjusts the temporal support of the edges within those triangles (Lines 18-22). Additionally, if a triangle has an increment, it must be true that $N(\triangle)>0$. Consequently, Algorithm \ref{alg:count_all_sup} updates the \textit{TT}-index (Lines 16-17) to record the minimum value of $\delta$ at which $N(\triangle,\delta)>0$, allowing for quick determination of edge connectivity. Finally, the algorithm updates the increment to facilitate the next iteration (Line 23).

In the next stage (Right part of \figurename \ref{fig:index_framework}), the adapted decomposition algorithm \textit{decompose} \cite{DBLP:conf/sigmod/HuangCQTY14} computes the trussness and continually updates the \textit{TT}-index (Line 4 in Algorithm \ref{alg:index_construct}) until all reasonable value of $\delta$ had been processed. The temporal trussness of each edge for $\delta$ is at least as large as its temporal trussness for $\delta-1$, namely $\tau(\bar e,\delta)\geq{\tau(\bar e,\delta-1)}$. In fact, the temporal trussness of an edge may be the same for adjacent $\delta$ values. To conserve storage, only the changed pairs are saved. Specifically, a pair $(\delta_i,\tau_i)$ for an edge is stored in \textit{TT} if and only if there are no other pairs $(\delta_j,\tau_j)$ satisfy $\delta_j<\delta_i$ and $\tau_i\leq \tau_j$.

\subsection{Query Processing}
As previously mentioned, for a specific $\delta$, the temporal trussness of an edge $\tau(\bar e)$ is the maximal $k$ that participates in a $(\tau(\bar e),\delta)$-truss. For a query node $q$, identifying the $k^*$ of the \textit{q-MDT} becomes straightforward, as it is equal to the smallest trussness of the edges induced by $q$. \textit{TTS} algorithm (Algorithm \ref{alg:index_search}) outlines the strategy of using \textit{TT}-index to search for \textit{q-MDT}.
The algorithm first queries the trussness of all edges induced by $q$ from \textit{TT} and marks the max trussness as $k^*$ (Lines 2-8). Specifically, for a given $\delta$, to identify the trussness of edges in $\mathcal{G}$, the algorithm calls Procedure $FindIndex$ to access \textit{TT} and returns a pair $(d,\tau)$, where $d\leq \delta$ and $d$ is maximal. During this process, a deque $Q$ is maintained to store the edges induced by $q$. If an edge has a trussness with $\tau<k^*$, it is excluded from the \textit{q-MDT}, and the algorithm marks it as visited to avoid repeat checks (Lines 4-5). Otherwise, the algorithm continues to remove edges from $Q$ until no edge in $Q$ has a trussness less than $k^*$ (Lines 6-7). Then, the algorithm pushes the current edge into $Q$ and update $k^*$ accordingly (Line 8). 

\begin{algorithm}
    \caption{Searching \textit{q-MDT} using Temporal Trussness (TTS)}
 \label{alg:index_search}
\begin{flushleft}
		\hspace*{0.02in} {\bf Input:}
$\mathcal{G}$, $q$, $\delta$, and \textit{TT}\\
		\hspace*{0.02in} {\bf Output:} The \textit{q-MDT} solution
	\end{flushleft}
	\begin{algorithmic}[1]
\State 	Initial $vis=\emptyset$, $Q=\emptyset$, $k^*=0$, $i=0$;
	\For{$u$ in $\mathcal{D}_{\mathcal{G}}(q)$}
		\State $(k,d)=FindIndex(\mathcal{G},\bar e(q,u),\delta)$;
		\If{$k<k^*$}
		\State	$vis\gets{vis} \cup{\bar e(q,u)}$;
			\textbf{continue};
		\EndIf
		\While{$!Q.empty()$ and $\tau(Q.front())<k$}
			\State $Q.pop\_front()$; 
 $vis\gets{vis} \cup{\bar e(q,u)}$;
		\EndWhile
		\State $Q.push\_front(\bar e)$; 
		$k^*=max(k^*,k)$;
	\EndFor
	\While{$!Q.empty()$}
		\State $\bar{e}=Q.pop\_front(); Q_s=\emptyset$\;
		\If{$\bar e(q,u)$ is in $vis$}
			\State \textbf{continue};
		\EndIf
		\State $\mathcal{S}_{i++}=\emptyset$; $Q_s.push(\bar e(u,v))$;
		\While{$!Q_s.empty()$}
		\State	$\bar e(u,v)=Q_s.pop();$~$\mathcal S_i.add(\{e(u,v,t)\in {\mathcal G}\})$;
			\For{$w \in \mathcal D(u) \cap \mathcal D(v)$}
				\If{$TT[\triangle_{uvw}]>\delta$}\State \textbf{continue};
                \If{$FindIndex(\bar e(u,w),\delta).second\geq k^*$ and $FindIndex(\bar e(v,w),\delta).second\geq k^*$}
					\If{$\bar e_1(u,w)$ is not in $vis$}
				\State		$Q_s.push(\bar e_1)$; $vis\gets{vis\cup{\bar e_1}}$;
					\EndIf
					\If{$\bar e_2(v,w)$ is not in $vis$}
					\State	$Q_s.push(\bar e_2)$; $vis\gets{vis\cup{\bar e_2}}$;
					\EndIf
				\EndIf
    \EndIf
			\EndFor
		\EndWhile
	\EndWhile
	\State \Return ${\mathcal S_1,\mathcal S_2,..., \mathcal S_i}$;
 \end{algorithmic}
\end{algorithm}		
	
After getting the value of $k^*$, the search process starts with the edges in $Q$ induced by $q$ whose trussness are no less than $k^*$. For each edge $\bar e(u,v) \in Q$, the algorithm accesses the edges that are in the same triangle with $\bar e(u,v)$. For the triangle $\triangle \in \{\triangle_{uvw}|\bar e(u,w),\bar e(v,w) \in  \mathcal G\}$, to judge whether $\triangle$ can be a bridge to connect two edges, the algorithm only needs to access the \textit{TT}-index to obtain $TT[\triangle]$ (the minimal $\delta$ makes it satisfy $N(\triangle)>0$), instead of recalculating the $N(\triangle,\delta)$ of it. In more details, $TT[\triangle]>\delta$ indicates that the triangle $\triangle$ cannot satisfy condition (2) in Definition \ref{def:k-delta-truss} under the current $\delta$, so that we cannot find the connected edges through $\triangle$ (Lines 17-18). Otherwise, the algorithm continues to judge if the edges within this triangle are valid (Line 19). According to the Definition \ref{def:problem2}, if the trussness of both $bar e(u,w)$ and $bar (v,w)$ is not less than $k^*$, this indicates that both edges can be contained in \textit{q-MDT}. In this way, we may be able to search some other edges in the final solution through these edges. Therefore, Algorithm \ref{alg:index_search} pushes them into queue $Q_s$ and updates their states as visited (Lines 16-20). When $Q_s$ is empty, the current temporal subgraph $\mathcal S_i$ is a \textit{q-MDT}. And when $Q$ is empty, all qualified \textit{q-MDT}s have been found and the algorithm returns. An example is provided in Appendix \ref{appendix-example-TTS} to illustrate the query process.

	



\section{Experimental Evaluation}
\label{sec:experiments}
We conducted experiments on nine real-world datasets and selected seven state-of-the-art models (\textit{MAPPR}, \textit{k-truss}, \textit{OL}, \textit{PCore}, \textit{DCCS}, \textit{FirmCore}, \textit{L-MEGA}) as benchmarks to evaluate the efficiency, effectiveness, and scalability of the proposed methods. \textit{GS} adopts the global strategy with a \textit{Sliding Window} procedure for \textit{q-MDT} search. \textit{LS} and \textit{TTS} are our optimized methods which search the \textit{q-MDT} with local strategy and temporal trussness index (\textit{TT}-index), respectively. We set \textit{$\delta$=8} unless specified otherwise. Further details about the experimental setup are provided in Appendix~\ref{appendix-settings}.


\subsection{Efficiency Evaluation}

\textbf{Exp-1: Storage and Construction Time for Building \textit{TT}.} 
Table \ref{tab:exp_index_building} shows the storage requirements and construction time for \textit{TT}. $\textit{TT}[\bar e]$ and $\textit{TT}[\triangle]$ are the storage of the temporal trussness for each edge and triangle, respectively. Regarding construction time, in all datasets, our algorithm builds the \textit{TT}-index within one day for all datasets. It is usually necessary to sacrifice a certain amount of space to achieve this result. For example, $\textit{TT}[\triangle]$ requires up to about five times the storage of the raw temporal graphs (\textit{LS}) in the $Thiers$ dataset. However, the storage does not increase in all datasets. The size of \textit{TT}[$\triangle$] is even smaller than that of the input graph in the $Rmin$, $Twitter$ and $Facebook$ datasets. This discrepancy arises due to \textit{TT}[$\triangle$]'s strong dependence on both the time span and the structure of the input graphs. A longer time span results in a greater number of $\triangle$ stored in \textit{TT}[$\triangle$], consequently enlarging the storage requirements. 
For the same reasons, $\textit{TT}[\bar e]$ consumes between 0.25 to 98 times the space occupied by the original temporal graphs in \textit{LS}. 
Theoretically, on the $Rmin$ dataset, $\textit{TT}[\bar e]$ would take up 5579 times more space than the original graph, but in practice it is only about 98 times. This demonstrate the effectiveness of our non-domain strategy.

\begin{table} 
	\centering
	\caption{Storage (in MB) and Construction Time (in Sec.)}
	\vspace{-0.2cm}
	\scalebox{1}{
		\begin{tabular}{c|r|rrr}
			\toprule
			\multirow{2}{*}{Dataset} & \multirow{2}{*}{LS} & \multicolumn{3}{c}{TTS}\\
            \cline{3-5}
             &  &  \vspace{0.2em} $\textit{TT}[\bar e]$ &$\textit{TT}[\triangle]$ &  Index Time \\
			\midrule
			Rmin & 0.854  & 83.3 & 0.237   & 43174  \\ 
			Primary &0.348  & 0.885 & 0.810 & 23    \\
			Lyon &2.81   & 4.11 &  12.42 & 1248    \\
			Thiers & 3.98  &16.1  &  20.9  &  5570  \\ 	 
			Twitter & 6.69 &1.7 & 1.17  & 19    \\
			Facebook & 7.15 &16.4  & 1.62 & 1407 \\
			Enron & 10  &  50.5  & 10.11  &   5026 \\ 
			Lkml & 4.04  & 51.4  & 14.22  &  3386  \\ 
			
			DBLP & 191.82  & 210.38 & 404.21  & 53826 \\ 
			\bottomrule		
	    \end{tabular}}
	\label{tab:exp_index_building}
	\vspace{-0.7cm}
\end{table}

\begin{table*}
	\centering
	\caption{The Running Time of Temporal Methods (ms). $*$ indicates the corresponding method cannot obtain results within two days.}
	\vspace{-0.2cm}
	\scalebox{1}{
		\begin{tabular}{c|rrrrrrrr}
			\toprule
			Dataset & OL & PCore & DCCS  & FirmCore & L-MEGA & GS &LS &TTS \\
			\midrule
			Rmin & 47.121  &769.308  & *   & 4112.281   &2.420     & 0.577   &0.469  & 0.003\\ 
			Primary &69.515  & 11.463 & 4.094 & 0.959   &16.0203     &0.924  &0.352 &0.005 \\
			Lyon &1635.670   & 38524.712 &19.335 & 3.587   &787.731    &50.560  &48.514 &0.831 \\
			Thiers & 2077.790  &10603.183  &175.428  & 16.084  &290.311   &56.754 &26.501 & 0.379\\ 	 
			Twitter & 353.517 &66.362   &8.323   &26.036       &*   &7.887 &0.444 &0.056 \\
			Facebook & 51.213 &28967.205  &3796.355  & 2413.352   & *    &4.435 & 0.206 &0.018 \\
			Enron & 267.906  &  153909.163  &491.361   &2073.825   &*     &27.515 &4.489 &0.283 \\ 
			Lkml & 3130.130  & 159965.765  & 1638.877  &44.882      &*      &79.022  &16.547 &0.839 \\ 
			DBLP & 2435.710  & 185.330 & 5236.058  &8080.973       &*    &446.099 &339.240 &40.660 \\ 
            \midrule
			AVG.RANK & 6 & 7 &5  &4  &8 &3  &2  & 1 \\
			\bottomrule		
	\end{tabular}}
	\label{tab:running_time_for_various_method}
\end{table*}

\begin{figure*}
 \setlength{\abovecaptionskip}{2mm}
  \setlength{\belowcaptionskip}{-2mm}
	\centering
	\subfigure[Facebook (vary $\delta$)]{ 
		\label{fig:parameter-subfig2:a} 
		\includegraphics[width=0.23\linewidth]{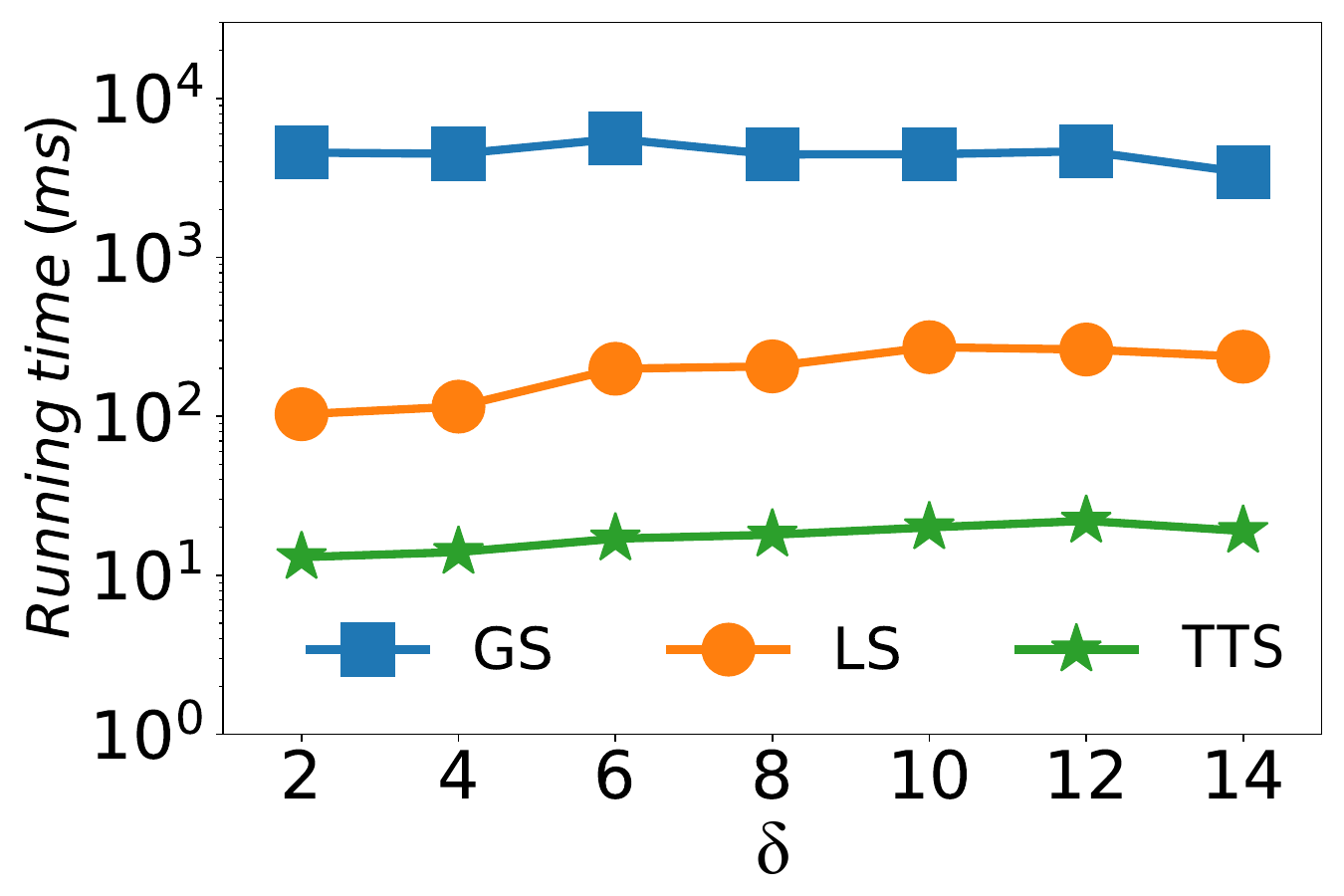} 
	} 
	\subfigure[Enron (vary $\delta$)]{ 
		\label{fig:parameter-subfig2:b} 
		\includegraphics[width=0.23\linewidth]{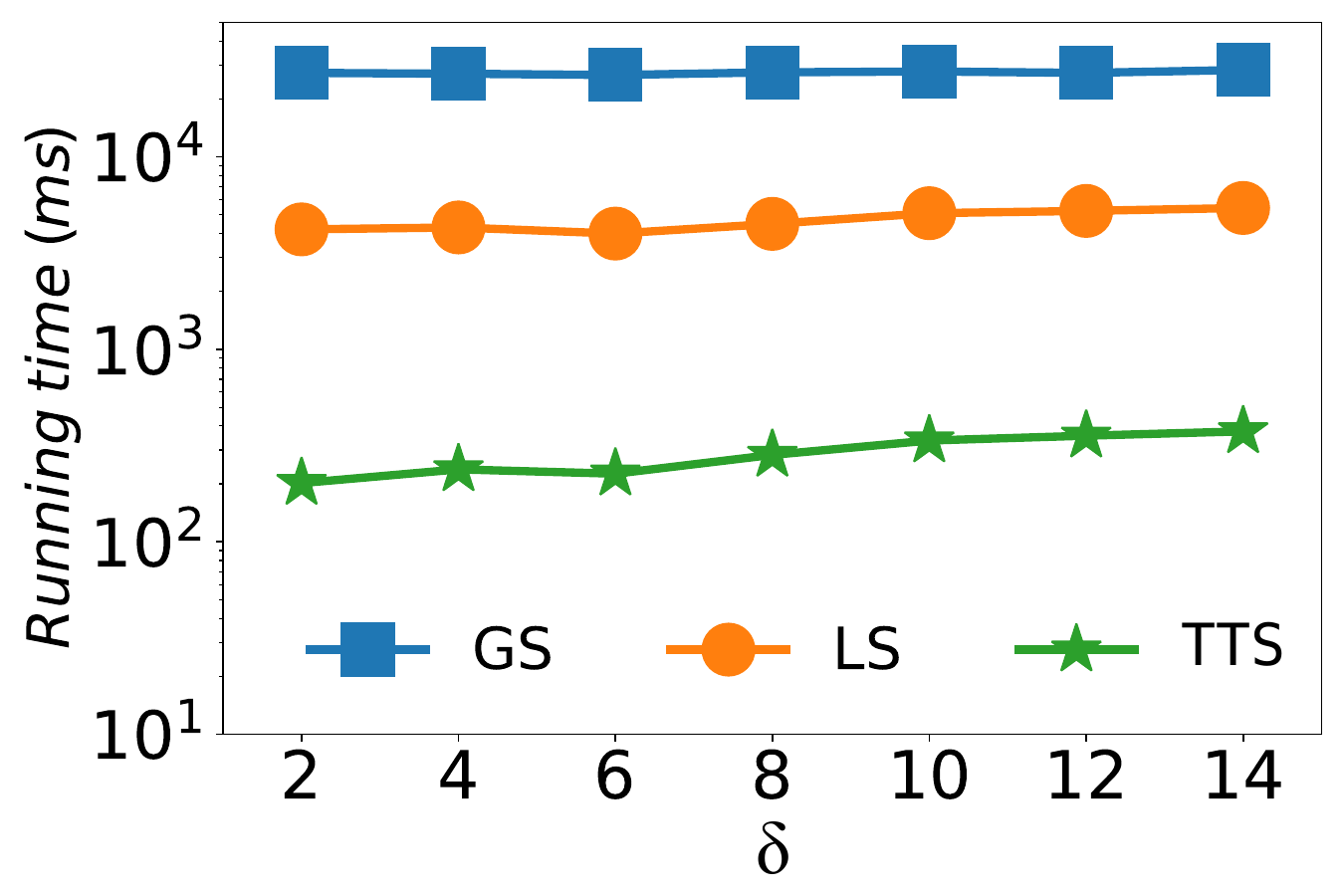} 
	} 
	\subfigure[Lkml (vary $\delta$)]{ 
		\label{fig:parameter-subfig2:c}
		\includegraphics[width=0.23\linewidth]{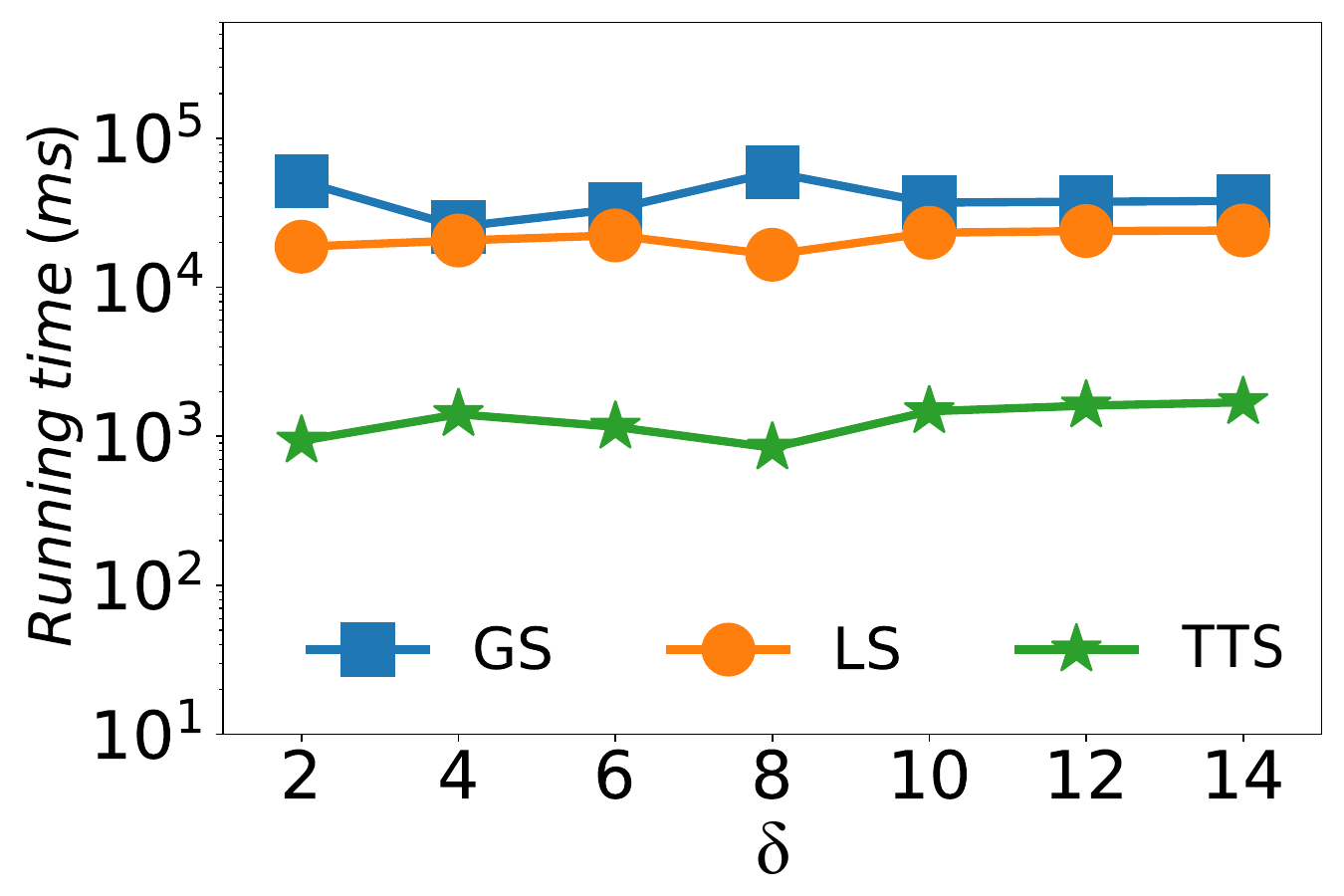} 
	} 
	\subfigure[DBLP (vary $\delta$)]{ 
		\label{fig:parameter-subfig2:d} 
		\includegraphics[width=0.23\linewidth]{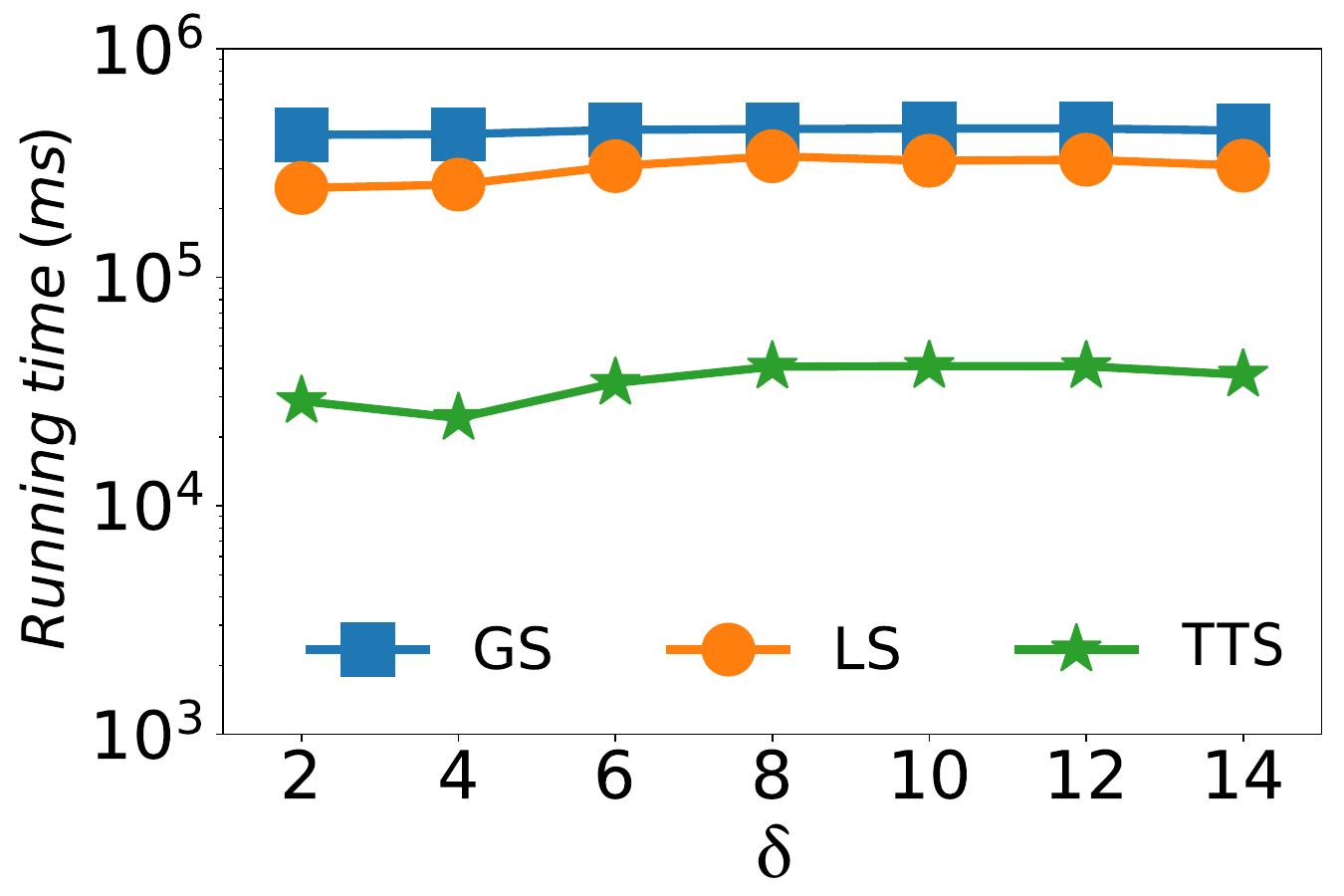} 
	} 
        \vspace{-0.3cm}
	\caption{ Running Time of Different Methods with Varying  $\delta$.} 
    \label{fig:parameter_effect}
    
\end{figure*}

\begin{figure*}
    \centering
	\subfigure[Facebook (vary $d_{\mathcal{G}}(q)$)]{ 
		\label{deg-fig:deg-subfig2:a} 
		\includegraphics[width=0.23\linewidth]{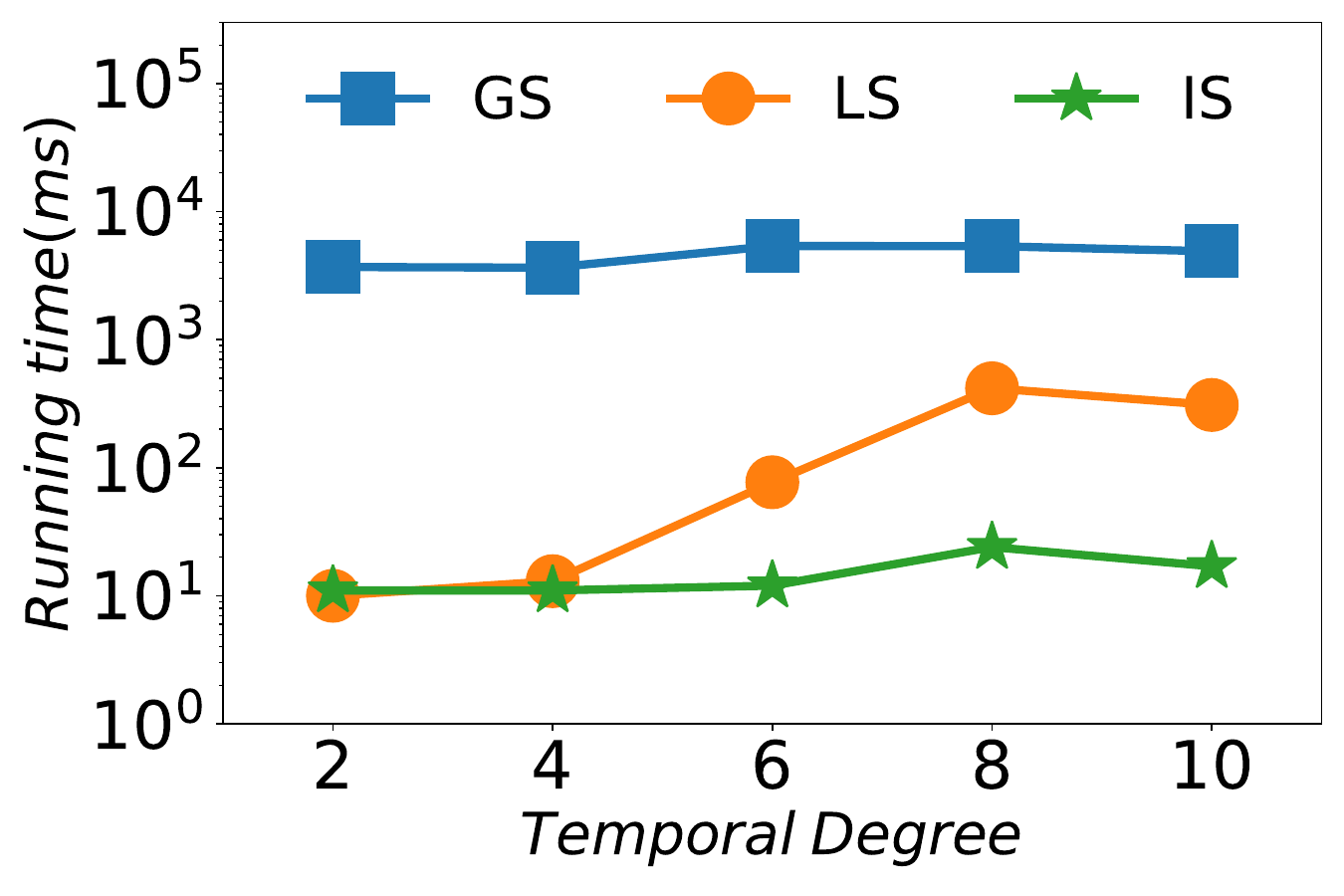} 
	} 
	\subfigure[Enron (vary $d_{\mathcal{G}}(q)$)]{ 
		\label{fig:deg-subfig2:b} 
		\includegraphics[width=0.23\linewidth]{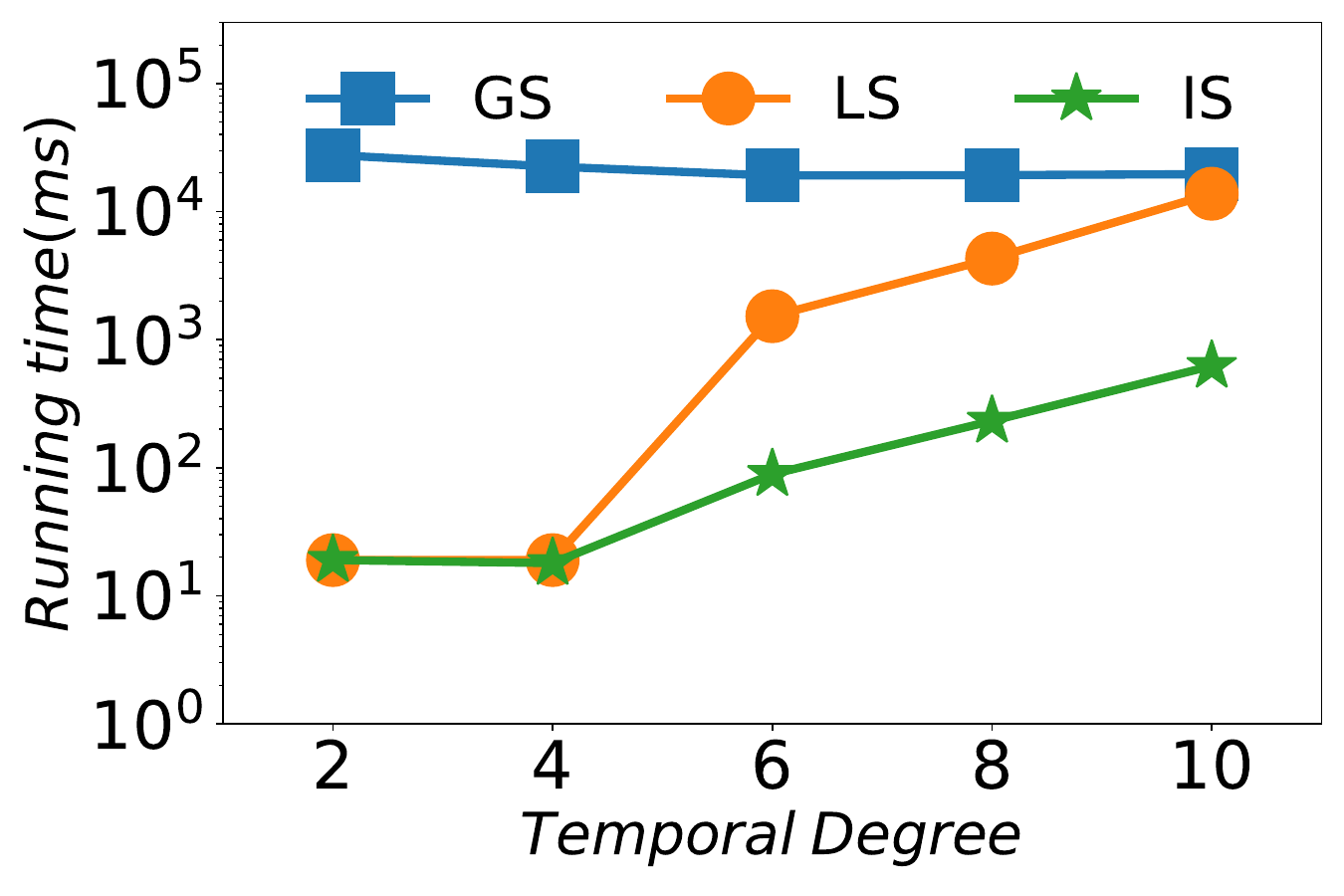} 
	} 
	\subfigure[Lkml (vary $d_{\mathcal{G}}(q)$)]{ 
		\label{fig:deg-subfig2:c}
		\includegraphics[width=0.23\linewidth]{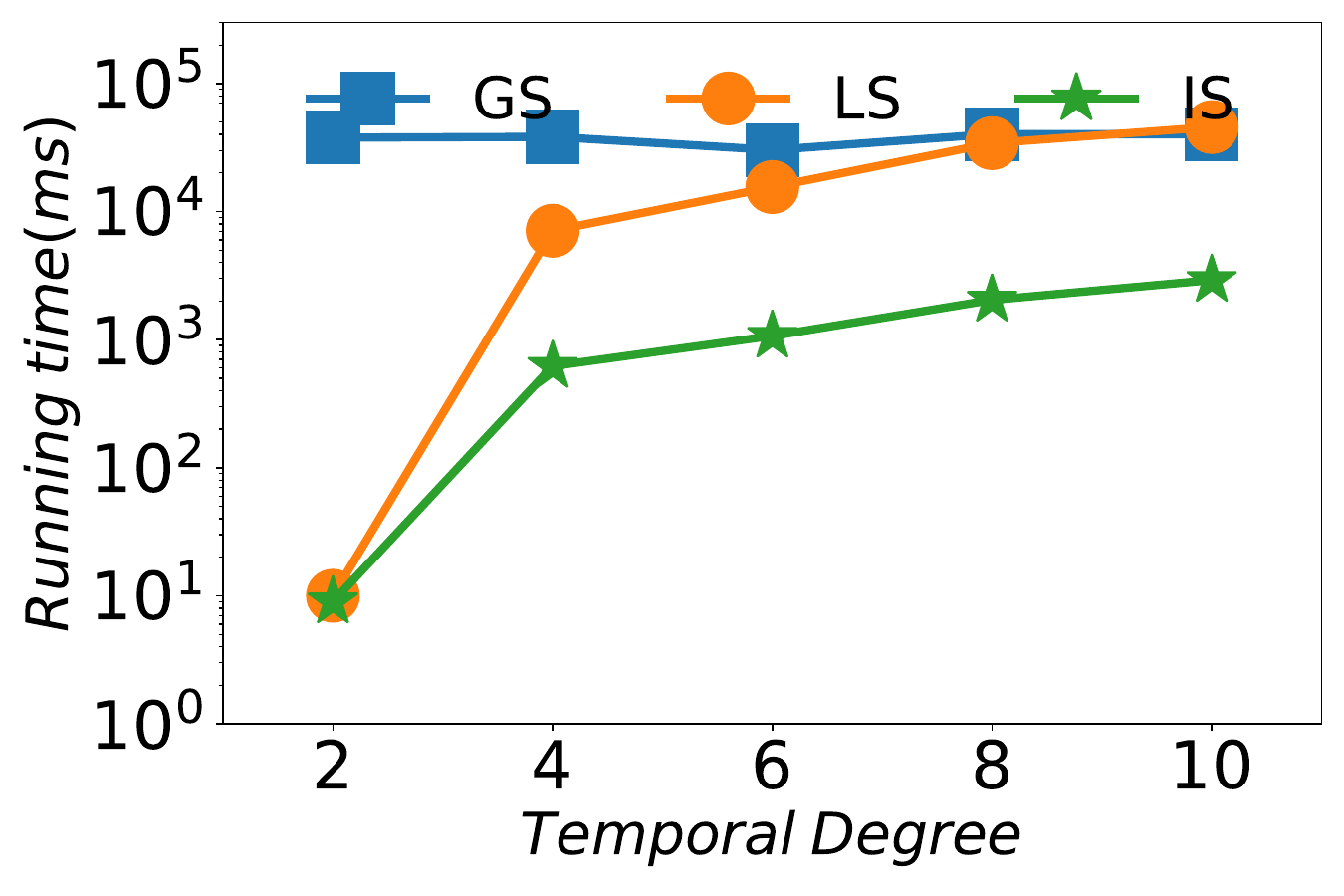} 
	} 
	\subfigure[DBLP (vary $d_{\mathcal{G}}(q)$)]{ 
		\label{fig:deg-subfig2:d} 
		\includegraphics[width=0.23\linewidth]{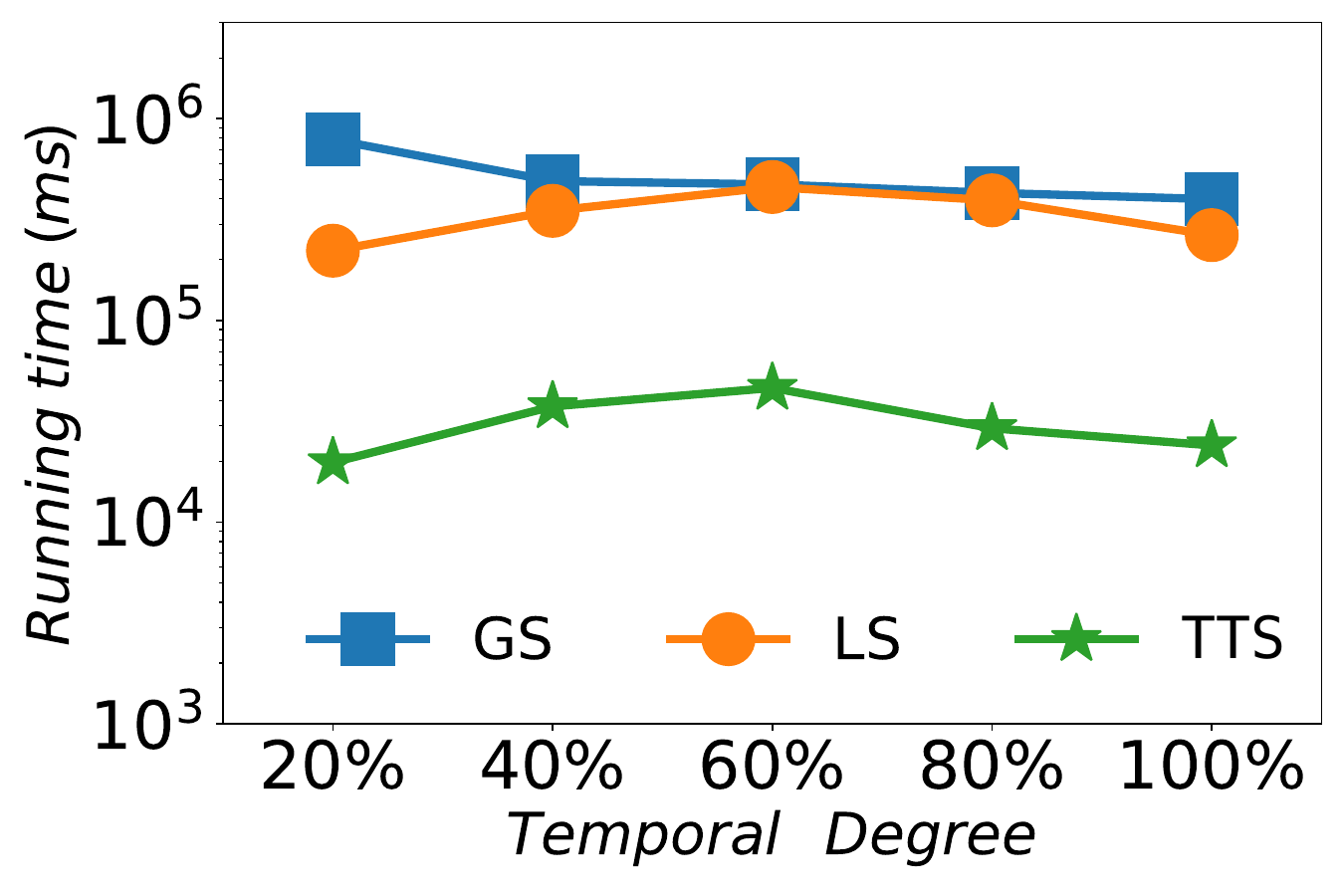} 
	} 
        \vspace{-0.3cm}
	\caption{Running Time with Varying Temporal Degree of Query Nodes.}
	\label{fig:deg_effect}
\end{figure*}

\textbf{Exp-2: The Running Time of Different Methods.} Table \ref{tab:running_time_for_various_method} shows the running time of eight methods on nine datasets, where the last row displays the ranking based on the average rank. \textit{TTS} and \textit{LS} rank top and second, respectively. Especially, \textit{TTS} takes less than one second on eight datasets and under a minute on $DBLP$. \textit{GS} is slightly slower than \textit{LS} due to its unnecessary calculations. \textit{FirmCore} and \textit{DCCS} are the fourth and fifth, respectively, because they spend much time finding the basic core model and integrating them as solutions.  \textit{OL} and \textit{PCore} perform poorly since both require many iterations to return optimal results. \textit{L-MEGA} returns the solutions on four small datasets within a reasonable time but fails to return a solution within two days for the other five datasets. This demonstrates that our optimal strategies are effective in practice.

{\textbf{Exp-3: Running Time of Different Methods with Varying $\delta$}}. 
In this experiment, we selected four datasets and three different search strategies to report the effect of parameter $\delta$ on the running time, as shown in \figurename\ref{fig:parameter_effect}. 
The search strategy \textit{GS} always requires the longest time to return a solution across all datasets, highlighting the efficacy of the optimization strategies employed in \textit{LS}.
\textit{TTS} has the best performance since it only executes a simple search strategy based on the \textit{TT}-index, which minimizes redundant accesses. Compared to \textit{GS}, \textit{TTS} achieves a speedup of at least two orders of magnitude across most datasets, indicating its capability to efficiently handle a high volume of queries.


\textbf{Exp-4: Running Time with Varying Temporal Degree of Query Nodes}. We selected query nodes based on varying temporal degrees to analyze their impact on running time. For each dataset, nodes were sorted by ascending temporal degree and divided into five equally sized buckets. From each bucket, 100 nodes were randomly chosen as query seeds for searching the \textit{q-MDT} using three methods. The results (see Fig. \ref{fig:deg_effect}) indicate that \textit{GS} is insensitive to the temporal degree of query nodes due to its nature as a global algorithm, which needs to search the entire input graph.
In contrast, both \textit{LS} and \textit{TTS} show increased running times as the temporal degree of query nodes rises. One reason is that nodes with higher temporal degrees tend to participate in larger \textit{q-MDTs}, necessitating longer search times. \textit{TTS} particularly benefits in these scenarios compared to \textit{LS}, as it focuses on searching a restricted set of nodes. Therefore, \textit{TTS} is more practical, especially when users prioritize communities where initial nodes have extensive interactions with others.

\begin{table*}
	\centering
	\caption{Effectiveness of Temporal Methods. $*$ denotes the corresponding model cannot be returned in two days.}
	\label{tab:effec_various_method}
	\vspace{-0.2cm}
	\scalebox{1}{
		\begin{tabular}{c|crrrrrrrr}
			\toprule
			\multicolumn{2}{c}{}
			&k-truss &MAPPR & OL & PCore & DCCS  & FirmCore & L-MEGA & q-MDT \\
			\midrule
			\multirow{9}*{   \textit{HTD}   }
			&Rmin & 0.00007   &  0.00089   &  0.00498  &  0.00176   & *  & 0.00414 & 0.00021     &  0.01106   \\ 
			&Primary & 0.00479   &  0.03768   & 0.13751   & 0.0461    &  0.15874    & 0.0221    & 0.0669  & 0.1609\\
			&Lyon & 0.02038   &  0.09283   & 0.14888   &  0.78574   & 0.09654     & 0.1   & 0.00021 &  0.1426\\
			&Thiers & 0.00925   &  0.04642   & 0.12805   &  0.09655   &  0.08434    & 0.04516    & 0.01963 & 0.13751 \\ 	 
			&Twitter &  0.0008  &  0.001   & 0.34056   & 0.00341    & 0.11696     & 0.08879    & *  &0.36365  \\
			&Facebook & 0.01016   &  0.00001   & 0.01508   &0.15081     & 0.1  & 0.00001    &*  & 0.07368  \\
			&Enron & 0.00025   & 0.00685    & 0.01804   & 0.00241    & 0.00001     & 0.00418    &*  &0.14095   \\ 
			&Lkml & 0.06299   & 0.18989    & 0.26226   & 0.13408    & 0.01125     & 0.16105    &* & 0.52042  \\ 
			&DBLP & 0.39787   & 0.57478    & 0.6401   & 0.51182    & 0.22405     & 0.54407    &*  &0.80436   \\ 
                \cline{2-10}
			&AVG.RANK &7    & 5    & 2   &    3 &  4    &  6   &8  & 1 \\
			\midrule
			\multirow{9}*{   \textit{HTC}   }
			&Rmin & 1   & 0.73225    &0.90037    &0.67877  &*   &0.67787  &0.7211      & 0.57939    \\ 
			&Primary &0.79225    & 0.56849    & 0.97105   & 0.52811    & 0.80813  & 0.86898 &  0.54241    & 0.7802\\
			&Lyon & 1   &  0.83441   &0.95451    &0.90039  & 0.99767  & 0.97431 & 0.75084 &0.91382 \\
			&Thiers & 1   &  0.78531   & 0.75383   & 0.98391    & 0.9391  & 0.9966 & 0.68487 &0.66646  \\ 	 
			&Twitter &0.74328    & 0.70028    & 0.96348   & 0.77779    &0.92273   &0.9314  & *     &0.65159  \\
			&Facebook & 0.6932   & 0.676    & 0.69151   & 0.69329    &0.25791   &0  &*      &0.66891  \\
			&Enron &0.82007    & 0.68761    & 0.98123   & 0.96308    & 0.8645  &0.89774  & *   & 0.60762 \\ 
			&Lkml &  0.99698  & 0.87261    &0.99189    & 0.93365    &0.90057   & 0.90263 & *     & 0.81753 \\ 
			&DBLP & 0.51533   & 0.66011    & 0.77391   & 0.81835    & 0.8868  &0.03288  &* & 0.64509 \\ 
                \cline{2-10}
			&AVG.RANK & 7   & 3    & 8   &5     &6      &4     &1  &2  \\   
			\bottomrule		
	\end{tabular}}
	
        \vspace{-0.1cm}
\end{table*}

\begin{figure*}
	\centering
	\subfigure[PCore]{ 
		\label{fig:case:PCore} 
		\includegraphics[width=0.27\linewidth]{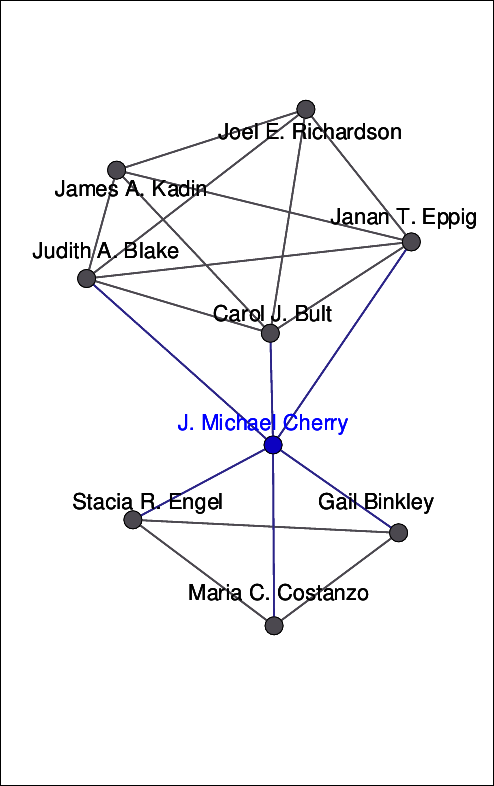} 
	} 
 \hspace{0.7cm}
	\subfigure[OL]{ 
		\label{fig:case:ol} 
		\includegraphics[width=0.27\linewidth]{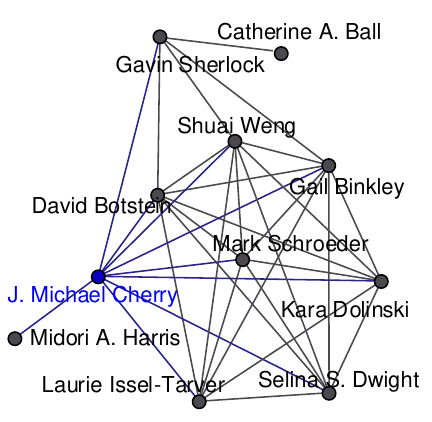}} \hspace{0.7cm}
	\subfigure[Our model (q-MDT)]{ 
		\label{fig:case:our} 
		\includegraphics[width=0.3\linewidth]{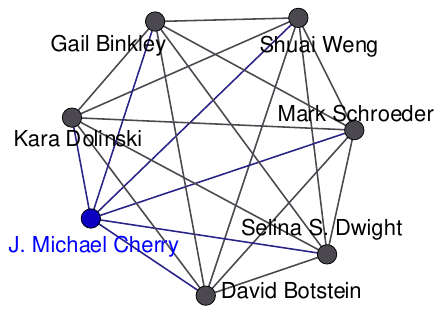} 
	} \vspace{-0.3cm}
	\caption{Case studies on DBLP.  The blue vertex (i.e., J. Michael Cherry) is the query vertex.} 	\vspace{-0.3cm}
	\label{fig:case_study}

\end{figure*}

\subsection{Effectiveness Evaluation}
Currently, there are no established metrics for assessing the attributes of higher-order temporal communities. Existing well-known metrics such as \textit{EDB} (Edge Density Burstiness) \cite{Linshixiong9351686,OL:chu2019online,DBLP:journals/jcst/ZhuLYJ22} and \textit{TC} (Temporal Conductance) \cite{zhang2022significant,Linshixiong9351686,DBLP:journals/jcst/ZhuLYJ22} focus on temporal edges and are therefore suitable for evaluating low-order temporal communities. To address this gap, we propose two new metrics, \textit{HTD} and \textit{HTC}, which incorporate temporal triangles instead of temporal edges. In \textit{HTD} and \textit{HTC}, temporal triangles must satisfy $\Delta(\triangle)\leq \delta^*$. Here, we let 

\begin{equation}
HTD=\sqrt[3]{\frac{|\{\triangle|\Delta(\triangle)\leq \delta^*, \odot (\triangle) \subset \mathcal{G}_S\}|}{|S||S-1||S-2||T_{S}|^3}}
\end{equation}

\begin{equation}
HTC=\frac{|HTcut(S,V\backslash S)|}{min\{HTvol(S),HTvol(V\backslash S\}}
\end{equation}

\begin{equation}
HTcut(S,V\backslash S)=\{\triangle_{uvw}|\Delta(\triangle)\leq \delta^*, \exists u,v,w\in S\}
\end{equation}

\begin{equation}
HTvol(S)=\{\triangle_{uvw}|u \in S,\Delta(\triangle)\leq \delta^*\}
\end{equation}

To ensure fairness, we first compute the average gap of the temporal edges to estimate $\delta^*$ as proposed by Li et al. \cite{DBLP:conf/icde/LiSQYD18}. Subsequently, this estimate is utilized in the calculation of temporal triangles. This approach allows us to assess the density of higher-order temporal triangles effectively while preserving the integrity of experimental results. A higher value of \textit{HTD} signifies a denser temporal community. \textit{HTC} measures the degree distribution within the solution, where a smaller \textit{HTC} indicates more intra-solution connections and fewer external ties.

\textbf{Exp-5: Effectiveness of Temporal Models.}
For \textit{HTD} metric, our model obtains the best grade on six datasets (Table \ref{tab:effec_various_method}). Following closely is \textit{OL}, which defines bursting communities as clique-like subgraphs with higher triangle density. Other temporal models, such as \textit{PCore}, \textit{DCCS}, and \textit{FirmCore}, show better performance than the static \textit{k-truss} because the latter does not consider temporal information of subgraphs. In contrast, \textit{L-MEGA} underperforms as it only optimizes the higher-order conductance at each timestamp. In terms of \textit{HTC} metric, our model and \textit{L-MEGA} exhibit similar performance on datasets where \textit{L-MEGA} runs successfully. The reason is that \textit{L-MEGA} optimizes its solutions based on triangle conductance \cite{LMega:fu2020local}, which is highly relevant with \textit{HTC}. In contrast, \textit{MAPPR}, which also utilizes triangle conductance to identify communities, performs less effectively than \textit{L-MEGA}. This is because \textit{MAPPR} is a completely static method, but \textit{L-MEGA} uses the temporal information to optimize its solution. In general, \textit{PCore} outperforms other core models such as \textit{DCCS} and \textit{FirmCore} due to its strong persistence correlation with $\delta^*$. It's notable that the \textit{HTC} of \textit{k-truss} on datasets like $Rmin$, $Lyon$ and $Thiers$ reaches $1$, indicating that their solutions are indistinguishable with the other parts. This occurs because these datasets exhibit dense structures when temporal information is disregarded, and returned almost the entire graph. This experiment underscores our model's capability to discover higher-quality communities compared to other models.

\textbf{Exp-6: Case Study on DBLP.}
In this section, we present results for \textit{k-truss}, \textit{PCore}, \textit{OL}, and \textit{q-MDT}. Due to space limitations, we omit similar results obtained by other models. The community identified by \textit{k-truss} includes over 1,000 authors spanning diverse research domains like computational biology, chemical reactions, and data mining. This community is too large to visualize effectively in a figure, which is why no visualization is provided. Notably, \textit{k-truss} only captures the structural cohesiveness of temporal graphs but does not consider the temporal information.
\textit{PCore} returns a community with nine authors, divided into two groups connected by J. Michael Cherry (see \figurename\ref{fig:case:PCore}). The upper part consists of authors primarily affiliated with the Jackson Laboratory, whereas the lower group consists of individuals from Stanford University. Consequently, the community exhibits loose connections and lacks cohesion around J. Michael Cherry, resulting in a less dense structure.
\textit{OL} returns a community with 11 authors as shown in Fig.\ref{fig:case:ol}. Compared with \textit{PCore}, \textit{OL} provides a more meaningful solution. However, it includes two authors, Catherine A. Ball and Midori A. Harris, who each collaborate with only one other author within this community. In fact, Catherine A. Ball shares research interests in Gene Expression Profiling and Bioinformatics with Gavin Sherlock, who frequently collaborates with J. Michael Cherry. 
Our model (\textit{q-MDT}) identifies a community consisting of seven authors, all specializing in genetics as J. Michael Cherry (see \figurename\ref{fig:case:our}). They were colleagues of J. Michael Cherry at Stanford University from 1990 to 2013. David Botstein and J. Michael Cherry co-founded the Saccharomyces Genome Database, a significant international resource connecting genomic sequences with biological functions. This community formation is attributed to their shared research interests and enduring collaboration with J. Michael Cherry over the years. In conclusion, our model \textit{q-MDT} can find more practical and meaningful communities in real-world scenarios.

Additionally, the detailed information about scalability experiments is available in Appendix\ref{appendix-scalability}.

\vspace{-0.3cm}
\section{Conclusion}
In this paper, we introduce a novel higher-order temporal community model called maximal-$\delta$-truss (\textit{MDT}), in which all edges are connected by a sequence of triangles with well-defined temporal properties. The \textit{MDT} model captures both the structural and temporal information of subgraphs through these temporal triangles within a constrained time span. To find a \textit{MDT} around a specific query node $q$ (\textit{q-MDT}), we propose a local strategy that combines an expanding algorithm to incrementally explore potential subgraphs, followed by searching \textit{q-MDT} in the expanded temporal subgraph. Additionally, we develop the \textit{TT}-index to expedite queries, facilitating efficient processing of large-scale graph queries. Empirical results on nine real-world networks, compared with seven competitors, demonstrate the efficiency, effectiveness, and scalability of our solutions.

\bibliographystyle{ACM-Reference-Format}
\bibliography{temporaltruss}


\begin{thebibliography}{52}


\ifx \showCODEN    \undefined \def \showCODEN     #1{\unskip}     \fi
\ifx \showDOI      \undefined \def \showDOI       #1{#1}\fi
\ifx \showISBNx    \undefined \def \showISBNx     #1{\unskip}     \fi
\ifx \showISBNxiii \undefined \def \showISBNxiii  #1{\unskip}     \fi
\ifx \showISSN     \undefined \def \showISSN      #1{\unskip}     \fi
\ifx \showLCCN     \undefined \def \showLCCN      #1{\unskip}     \fi
\ifx \shownote     \undefined \def \shownote      #1{#1}          \fi
\ifx \showarticletitle \undefined \def \showarticletitle #1{#1}   \fi
\ifx \showURL      \undefined \def \showURL       {\relax}        \fi
\providecommand\bibfield[2]{#2}
\providecommand\bibinfo[2]{#2}
\providecommand\natexlab[1]{#1}
\providecommand\showeprint[2][]{arXiv:#2}

\bibitem[Akbas and Zhao(2017)]%
        {akbas2017truss}
\bibfield{author}{\bibinfo{person}{Esra Akbas} {and} \bibinfo{person}{Peixiang Zhao}.} \bibinfo{year}{2017}\natexlab{}.
\newblock \showarticletitle{Truss-based community search: a truss-equivalence based indexing approach}.
\newblock \bibinfo{journal}{\emph{PVLDB}} \bibinfo{volume}{10}, \bibinfo{number}{11} (\bibinfo{year}{2017}), \bibinfo{pages}{1298--1309}.
\newblock


\bibitem[Andersen and Chellapilla(2009)]%
        {andersen2009finding}
\bibfield{author}{\bibinfo{person}{Reid Andersen} {and} \bibinfo{person}{Kumar Chellapilla}.} \bibinfo{year}{2009}\natexlab{}.
\newblock \showarticletitle{Finding dense subgraphs with size bounds}. In \bibinfo{booktitle}{\emph{International workshop on algorithms and models for the web-graph}}. \bibinfo{pages}{25--37}.
\newblock


\bibitem[Barbieri et~al\mbox{.}(2015)]%
        {barbieri2015efficient}
\bibfield{author}{\bibinfo{person}{Nicola Barbieri}, \bibinfo{person}{Francesco Bonchi}, \bibinfo{person}{Edoardo Galimberti}, {and} \bibinfo{person}{Francesco Gullo}.} \bibinfo{year}{2015}\natexlab{}.
\newblock \showarticletitle{Efficient and effective community search}.
\newblock \bibinfo{journal}{\emph{DMKD}} \bibinfo{volume}{29}, \bibinfo{number}{5} (\bibinfo{year}{2015}), \bibinfo{pages}{1406--1433}.
\newblock


\bibitem[Bogdanov et~al\mbox{.}(2011)]%
        {bogdanov2011mining}
\bibfield{author}{\bibinfo{person}{Petko Bogdanov}, \bibinfo{person}{Misael Mongiov{\`\i}}, {and} \bibinfo{person}{Ambuj~K Singh}.} \bibinfo{year}{2011}\natexlab{}.
\newblock \showarticletitle{Mining heavy subgraphs in time-evolving networks}. In \bibinfo{booktitle}{\emph{ICDM}}. \bibinfo{pages}{81--90}.
\newblock


\bibitem[Chang and Qin(2019)]%
        {chang2019cohesive}
\bibfield{author}{\bibinfo{person}{Lijun Chang} {and} \bibinfo{person}{Lu Qin}.} \bibinfo{year}{2019}\natexlab{}.
\newblock \showarticletitle{Cohesive subgraph computation over large sparse graphs}. In \bibinfo{booktitle}{\emph{ICDE}}. \bibinfo{pages}{2068--2071}.
\newblock


\bibitem[Chang et~al\mbox{.}(2013)]%
        {chang2013efficiently}
\bibfield{author}{\bibinfo{person}{Lijun Chang}, \bibinfo{person}{Jeffrey~Xu Yu}, \bibinfo{person}{Lu Qin}, \bibinfo{person}{Xuemin Lin}, \bibinfo{person}{Chengfei Liu}, {and} \bibinfo{person}{Weifa Liang}.} \bibinfo{year}{2013}\natexlab{}.
\newblock \showarticletitle{Efficiently computing k-edge connected components via graph decomposition}. In \bibinfo{booktitle}{\emph{SIGMOD}}. \bibinfo{pages}{205--216}.
\newblock


\bibitem[Chen et~al\mbox{.}(2020)]%
        {chen2020finding}
\bibfield{author}{\bibinfo{person}{Lu Chen}, \bibinfo{person}{Chengfei Liu}, \bibinfo{person}{Rui Zhou}, \bibinfo{person}{Jiajie Xu}, \bibinfo{person}{Jeffrey~Xu Yu}, {and} \bibinfo{person}{Jianxin Li}.} \bibinfo{year}{2020}\natexlab{}.
\newblock \showarticletitle{Finding effective geo-social group for impromptu activities with diverse demands}. In \bibinfo{booktitle}{\emph{SIGKDD}}. \bibinfo{pages}{698--708}.
\newblock


\bibitem[Cheng et~al\mbox{.}(2011)]%
        {DBLP:conf/icde/ChengKCO11}
\bibfield{author}{\bibinfo{person}{James Cheng}, \bibinfo{person}{Yiping Ke}, \bibinfo{person}{Shumo Chu}, {and} \bibinfo{person}{M.~Tamer {\"{O}}zsu}.} \bibinfo{year}{2011}\natexlab{}.
\newblock \showarticletitle{Efficient core decomposition in massive networks}. In \bibinfo{booktitle}{\emph{ICDE}}. \bibinfo{pages}{51--62}.
\newblock


\bibitem[Chu et~al\mbox{.}(2019)]%
        {OL:chu2019online}
\bibfield{author}{\bibinfo{person}{Lingyang Chu}, \bibinfo{person}{Yanyan Zhang}, \bibinfo{person}{Yu Yang}, \bibinfo{person}{Lanjun Wang}, {and} \bibinfo{person}{Jian Pei}.} \bibinfo{year}{2019}\natexlab{}.
\newblock \showarticletitle{Online density bursting subgraph detection from temporal graphs}.
\newblock \bibinfo{journal}{\emph{PVLDB}} \bibinfo{volume}{12}, \bibinfo{number}{13} (\bibinfo{year}{2019}), \bibinfo{pages}{2353--2365}.
\newblock


\bibitem[Cohen(2008)]%
        {cohen2008trusses}
\bibfield{author}{\bibinfo{person}{Jonathan Cohen}.} \bibinfo{year}{2008}\natexlab{}.
\newblock \showarticletitle{Trusses: Cohesive subgraphs for social network analysis}.
\newblock \bibinfo{journal}{\emph{National security agency technical report}} \bibinfo{volume}{16}, \bibinfo{number}{3.1} (\bibinfo{year}{2008}).
\newblock


\bibitem[Cui et~al\mbox{.}(2014)]%
        {cui2014local}
\bibfield{author}{\bibinfo{person}{Wanyun Cui}, \bibinfo{person}{Yanghua Xiao}, \bibinfo{person}{Haixun Wang}, {and} \bibinfo{person}{Wei Wang}.} \bibinfo{year}{2014}\natexlab{}.
\newblock \showarticletitle{Local search of communities in large graphs}. In \bibinfo{booktitle}{\emph{SIGMOD}}. \bibinfo{pages}{991--1002}.
\newblock


\bibitem[Fang et~al\mbox{.}(2020)]%
        {fang2020survey}
\bibfield{author}{\bibinfo{person}{Yixiang Fang}, \bibinfo{person}{Xin Huang}, \bibinfo{person}{Lu Qin}, \bibinfo{person}{Ying Zhang}, \bibinfo{person}{Wenjie Zhang}, \bibinfo{person}{Reynold Cheng}, {and} \bibinfo{person}{Xuemin Lin}.} \bibinfo{year}{2020}\natexlab{}.
\newblock \showarticletitle{A survey of community search over big graphs}.
\newblock \bibinfo{journal}{\emph{The VLDB Journal}} \bibinfo{volume}{29}, \bibinfo{number}{1} (\bibinfo{year}{2020}), \bibinfo{pages}{353--392}.
\newblock


\bibitem[Fortunato(2009)]%
        {Fortunato2009Community}
\bibfield{author}{\bibinfo{person}{Santo Fortunato}.} \bibinfo{year}{2009}\natexlab{}.
\newblock \showarticletitle{Community detection in graphs}.
\newblock \bibinfo{journal}{\emph{Physics Reports}} \bibinfo{volume}{486}, \bibinfo{number}{3} (\bibinfo{year}{2009}), \bibinfo{pages}{75--174}.
\newblock


\bibitem[Fu et~al\mbox{.}(2020)]%
        {LMega:fu2020local}
\bibfield{author}{\bibinfo{person}{Dongqi Fu}, \bibinfo{person}{Dawei Zhou}, {and} \bibinfo{person}{Jingrui He}.} \bibinfo{year}{2020}\natexlab{}.
\newblock \showarticletitle{Local Motif Clustering on Time-Evolving Graphs}. In \bibinfo{booktitle}{\emph{SIGKDD}}. \bibinfo{pages}{390--400}.
\newblock


\bibitem[Gao et~al\mbox{.}(2020)]%
        {PCCD_sigir_20}
\bibfield{author}{\bibinfo{person}{Zheng Gao}, \bibinfo{person}{Hongsong Li}, \bibinfo{person}{Zhuoren Jiang}, {and} \bibinfo{person}{Xiaozhong Liu}.} \bibinfo{year}{2020}\natexlab{}.
\newblock \showarticletitle{Detecting User Community in Sparse Domain via Cross-Graph Pairwise Learning}. In \bibinfo{booktitle}{\emph{SIGIR}}. \bibinfo{pages}{139--148}.
\newblock


\bibitem[Goldberg(1984)]%
        {goldberg1984finding}
\bibfield{author}{\bibinfo{person}{AV Goldberg}.} \bibinfo{year}{1984}\natexlab{}.
\newblock \showarticletitle{Finding a maximum density subgraph}.
\newblock \bibinfo{journal}{\emph{Uni. California, Berkeley}} (\bibinfo{year}{1984}).
\newblock


\bibitem[Han et~al\mbox{.}(2016)]%
        {zhongming2016ncss}
\bibfield{author}{\bibinfo{person}{Zhongming Han}, \bibinfo{person}{Xusheng Tan}, \bibinfo{person}{Yan Chen}, {and} \bibinfo{person}{Dagao Duan}.} \bibinfo{year}{2016}\natexlab{}.
\newblock \showarticletitle{NCSS: An effective and efficient complex network community detection algorithm}.
\newblock \bibinfo{journal}{\emph{Scientia Sinica Informationis}} \bibinfo{volume}{46}, \bibinfo{number}{4} (\bibinfo{year}{2016}), \bibinfo{pages}{431--444}.
\newblock


\bibitem[Hashemi et~al\mbox{.}(2022)]%
        {DBLP:conf/www/HashemiBL22}
\bibfield{author}{\bibinfo{person}{Farnoosh Hashemi}, \bibinfo{person}{Ali Behrouz}, {and} \bibinfo{person}{Laks V.~S. Lakshmanan}.} \bibinfo{year}{2022}\natexlab{}.
\newblock \showarticletitle{FirmCore Decomposition of Multilayer Networks}. In \bibinfo{booktitle}{\emph{{WWW}}}. \bibinfo{pages}{1589--1600}.
\newblock


\bibitem[He et~al\mbox{.}(2024)]%
        {DBLP:journals/eswa/HeLYLJW24}
\bibfield{author}{\bibinfo{person}{Yue He}, \bibinfo{person}{Longlong Lin}, \bibinfo{person}{Pingpeng Yuan}, \bibinfo{person}{Ronghua Li}, \bibinfo{person}{Tao Jia}, {and} \bibinfo{person}{Zeli Wang}.} \bibinfo{year}{2024}\natexlab{}.
\newblock \showarticletitle{{CCSS:} Towards conductance-based community search with size constraints}.
\newblock \bibinfo{journal}{\emph{Expert Syst. Appl.}}  \bibinfo{volume}{250} (\bibinfo{year}{2024}), \bibinfo{pages}{123915}.
\newblock


\bibitem[Hong et~al\mbox{.}(2022)]%
        {GraphReformCD_WWW22}
\bibfield{author}{\bibinfo{person}{Jiwon Hong}, \bibinfo{person}{Dong-hyuk Seo}, \bibinfo{person}{Jeewon Ahn}, {and} \bibinfo{person}{Sang-Wook Kim}.} \bibinfo{year}{2022}\natexlab{}.
\newblock \showarticletitle{GraphReformCD: Graph Reformulation for Effective Community Detection in Real-World Graphs}. In \bibinfo{booktitle}{\emph{WWW}}. \bibinfo{pages}{180--183}.
\newblock


\bibitem[Huang et~al\mbox{.}(2014)]%
        {DBLP:conf/sigmod/HuangCQTY14}
\bibfield{author}{\bibinfo{person}{Xin Huang}, \bibinfo{person}{Hong Cheng}, \bibinfo{person}{Lu Qin}, \bibinfo{person}{Wentao Tian}, {and} \bibinfo{person}{Jeffrey~Xu Yu}.} \bibinfo{year}{2014}\natexlab{}.
\newblock \showarticletitle{Querying k-truss community in large and dynamic graphs}. In \bibinfo{booktitle}{\emph{SIGMOD}}. \bibinfo{pages}{1311--1322}.
\newblock


\bibitem[Li et~al\mbox{.}(2015)]%
        {DBLP:journals/pvldb/LiQYM15}
\bibfield{author}{\bibinfo{person}{Rong{-}Hua Li}, \bibinfo{person}{Lu Qin}, \bibinfo{person}{Jeffrey~Xu Yu}, {and} \bibinfo{person}{Rui Mao}.} \bibinfo{year}{2015}\natexlab{}.
\newblock \showarticletitle{Influential Community Search in Large Networks}.
\newblock \bibinfo{journal}{\emph{{PVLDB}}} \bibinfo{volume}{8}, \bibinfo{number}{5} (\bibinfo{year}{2015}), \bibinfo{pages}{509--520}.
\newblock


\bibitem[Li et~al\mbox{.}(2018)]%
        {DBLP:conf/icde/LiSQYD18}
\bibfield{author}{\bibinfo{person}{Rong{-}Hua Li}, \bibinfo{person}{Jiao Su}, \bibinfo{person}{Lu Qin}, \bibinfo{person}{Jeffrey~Xu Yu}, {and} \bibinfo{person}{Qiangqiang Dai}.} \bibinfo{year}{2018}\natexlab{}.
\newblock \showarticletitle{Persistent Community Search in Temporal Networks}. In \bibinfo{booktitle}{\emph{ICDE}}. \bibinfo{pages}{797--808}.
\newblock


\bibitem[Li et~al\mbox{.}(2017)]%
        {li2017effective}
\bibfield{author}{\bibinfo{person}{Yuan Li}, \bibinfo{person}{Yuhai Zhao}, \bibinfo{person}{Guoren Wang}, \bibinfo{person}{Feida Zhu}, \bibinfo{person}{Yubao Wu}, {and} \bibinfo{person}{Shengle Shi}.} \bibinfo{year}{2017}\natexlab{}.
\newblock \showarticletitle{Effective k-vertex connected component detection in large-scale networks}. In \bibinfo{booktitle}{\emph{DASFAA}}. \bibinfo{pages}{404--421}.
\newblock


\bibitem[Lin et~al\mbox{.}(2024a)]%
        {DBLP:journals/corr/abs-2406-07357}
\bibfield{author}{\bibinfo{person}{Longlong Lin}, \bibinfo{person}{Tao Jia}, \bibinfo{person}{Zeli Wang}, \bibinfo{person}{Jin Zhao}, {and} \bibinfo{person}{Rong{-}Hua Li}.} \bibinfo{year}{2024}\natexlab{a}.
\newblock \showarticletitle{{PSMC:} Provable and Scalable Algorithms for Motif Conductance Based Graph Clustering}.
\newblock \bibinfo{journal}{\emph{CoRR}}  \bibinfo{volume}{abs/2406.07357} (\bibinfo{year}{2024}).
\newblock


\bibitem[Lin et~al\mbox{.}(2023)]%
        {DBLP:conf/aaai/LinLJ23}
\bibfield{author}{\bibinfo{person}{Longlong Lin}, \bibinfo{person}{Ronghua Li}, {and} \bibinfo{person}{Tao Jia}.} \bibinfo{year}{2023}\natexlab{}.
\newblock \showarticletitle{Scalable and Effective Conductance-Based Graph Clustering}. In \bibinfo{booktitle}{\emph{AAAI}}. \bibinfo{pages}{4471--4478}.
\newblock


\bibitem[Lin et~al\mbox{.}(2022a)]%
        {Linshixiong9351686}
\bibfield{author}{\bibinfo{person}{Longlong Lin}, \bibinfo{person}{Pingpeng Yuan}, \bibinfo{person}{Rong{-}Hua Li}, {and} \bibinfo{person}{Hai Jin}.} \bibinfo{year}{2022}\natexlab{a}.
\newblock \showarticletitle{Mining Diversified Top-r Lasting Cohesive Subgraphs on Temporal Networks}.
\newblock \bibinfo{journal}{\emph{{IEEE} Trans. Big Data}} \bibinfo{volume}{8}, \bibinfo{number}{6} (\bibinfo{year}{2022}), \bibinfo{pages}{1537--1549}.
\newblock


\bibitem[Lin et~al\mbox{.}(2022b)]%
        {DBLP:journals/tsmc/LinYLWLJ22}
\bibfield{author}{\bibinfo{person}{Longlong Lin}, \bibinfo{person}{Pingpeng Yuan}, \bibinfo{person}{Rong{-}Hua Li}, \bibinfo{person}{Jifei Wang}, \bibinfo{person}{Ling Liu}, {and} \bibinfo{person}{Hai Jin}.} \bibinfo{year}{2022}\natexlab{b}.
\newblock \showarticletitle{Mining Stable Quasi-Cliques on Temporal Networks}.
\newblock \bibinfo{journal}{\emph{{IEEE} Trans. Syst. Man Cybern. Syst.}} \bibinfo{volume}{52}, \bibinfo{number}{6} (\bibinfo{year}{2022}), \bibinfo{pages}{3731--3745}.
\newblock


\bibitem[Lin et~al\mbox{.}(2024b)]%
        {DBLP:journals/corr/abs-2302-08740}
\bibfield{author}{\bibinfo{person}{Longlong Lin}, \bibinfo{person}{Pingpeng Yuan}, \bibinfo{person}{Rong{-}Hua Li}, \bibinfo{person}{Chun{-}Xue Zhu}, \bibinfo{person}{Hongchao Qin}, \bibinfo{person}{Hai Jin}, {and} \bibinfo{person}{Tao Jia}.} \bibinfo{year}{2024}\natexlab{b}.
\newblock \showarticletitle{{QTCS:} Efficient Query-Centered Temporal Community Search}.
\newblock \bibinfo{journal}{\emph{Proc. {VLDB} Endow.}} \bibinfo{volume}{17}, \bibinfo{number}{6} (\bibinfo{year}{2024}), \bibinfo{pages}{1187--1199}.
\newblock


\bibitem[Liu et~al\mbox{.}(2020)]%
        {liu2020truss}
\bibfield{author}{\bibinfo{person}{Qing Liu}, \bibinfo{person}{Minjun Zhao}, \bibinfo{person}{Xin Huang}, \bibinfo{person}{Jianliang Xu}, {and} \bibinfo{person}{Yunjun Gao}.} \bibinfo{year}{2020}\natexlab{}.
\newblock \showarticletitle{Truss-based community search over large directed graphs}. In \bibinfo{booktitle}{\emph{SIGMOD}}. \bibinfo{pages}{2183--2197}.
\newblock


\bibitem[Ma et~al\mbox{.}(2017)]%
        {ma2017fast}
\bibfield{author}{\bibinfo{person}{Shuai Ma}, \bibinfo{person}{Renjun Hu}, \bibinfo{person}{Luoshu Wang}, \bibinfo{person}{Xuelian Lin}, {and} \bibinfo{person}{Jinpeng Huai}.} \bibinfo{year}{2017}\natexlab{}.
\newblock \showarticletitle{Fast computation of dense temporal subgraphs}. In \bibinfo{booktitle}{\emph{ICDE}}. \bibinfo{pages}{361--372}.
\newblock


\bibitem[Newman(2004)]%
        {newman2004fast}
\bibfield{author}{\bibinfo{person}{Mark~EJ Newman}.} \bibinfo{year}{2004}\natexlab{}.
\newblock \showarticletitle{Fast algorithm for detecting community structure in networks}.
\newblock \bibinfo{journal}{\emph{Physical review E}} \bibinfo{volume}{69}, \bibinfo{number}{6} (\bibinfo{year}{2004}), \bibinfo{pages}{066133}.
\newblock


\bibitem[Paranjape et~al\mbox{.}(2017)]%
        {DBLP:conf/wsdm/ParanjapeBL17}
\bibfield{author}{\bibinfo{person}{Ashwin Paranjape}, \bibinfo{person}{Austin~R. Benson}, {and} \bibinfo{person}{Jure Leskovec}.} \bibinfo{year}{2017}\natexlab{}.
\newblock \showarticletitle{Motifs in Temporal Networks}. In \bibinfo{booktitle}{\emph{WSDM}}. \bibinfo{pages}{601--610}.
\newblock


\bibitem[Pardalos and Xue(1994)]%
        {pardalos1994maximum}
\bibfield{author}{\bibinfo{person}{Panos~M Pardalos} {and} \bibinfo{person}{Jue Xue}.} \bibinfo{year}{1994}\natexlab{}.
\newblock \showarticletitle{The maximum clique problem}.
\newblock \bibinfo{journal}{\emph{Journal of global Optimization}} \bibinfo{volume}{4}, \bibinfo{number}{3} (\bibinfo{year}{1994}), \bibinfo{pages}{301--328}.
\newblock


\bibitem[Porter et~al\mbox{.}(2022)]%
        {WWW2022_2}
\bibfield{author}{\bibinfo{person}{Alexandra~M. Porter}, \bibinfo{person}{Baharan Mirzasoleiman}, {and} \bibinfo{person}{Jure Leskovec}.} \bibinfo{year}{2022}\natexlab{}.
\newblock \showarticletitle{Analytical Models for Motifs in Temporal Networks}. In \bibinfo{booktitle}{\emph{Companion of The Web Conference 2022, Virtual Event / Lyon, France, April 25 - 29, 2022}}, \bibfield{editor}{\bibinfo{person}{Fr{\'{e}}d{\'{e}}rique Laforest}, \bibinfo{person}{Rapha{\"{e}}l Troncy}, \bibinfo{person}{Elena Simperl}, \bibinfo{person}{Deepak Agarwal}, \bibinfo{person}{Aristides Gionis}, \bibinfo{person}{Ivan Herman}, {and} \bibinfo{person}{Lionel M{\'{e}}dini}} (Eds.). \bibinfo{publisher}{{ACM}}, \bibinfo{pages}{903--909}.
\newblock
\urldef\tempurl%
\url{https://doi.org/10.1145/3487553.3524669}
\showDOI{\tempurl}


\bibitem[Qin et~al\mbox{.}(2020)]%
        {qin2020periodic}
\bibfield{author}{\bibinfo{person}{Hongchao Qin}, \bibinfo{person}{Ronghua Li}, \bibinfo{person}{Ye Yuan}, \bibinfo{person}{Guoren Wang}, \bibinfo{person}{Weihua Yang}, {and} \bibinfo{person}{Lu Qin}.} \bibinfo{year}{2020}\natexlab{}.
\newblock \showarticletitle{Periodic communities mining in temporal networks: Concepts and algorithms}.
\newblock \bibinfo{journal}{\emph{TKDE}} (\bibinfo{year}{2020}).
\newblock


\bibitem[Qin et~al\mbox{.}(2019)]%
        {qin2019mining}
\bibfield{author}{\bibinfo{person}{Hongchao Qin}, \bibinfo{person}{Rong-Hua Li}, \bibinfo{person}{Guoren Wang}, \bibinfo{person}{Lu Qin}, \bibinfo{person}{Yurong Cheng}, {and} \bibinfo{person}{Ye Yuan}.} \bibinfo{year}{2019}\natexlab{}.
\newblock \showarticletitle{Mining periodic cliques in temporal networks}. In \bibinfo{booktitle}{\emph{ICDE}}. \bibinfo{pages}{1130--1141}.
\newblock


\bibitem[Qin et~al\mbox{.}(2022)]%
        {Qintkde2020}
\bibfield{author}{\bibinfo{person}{Hongchao Qin}, \bibinfo{person}{Rong-Hua Li}, \bibinfo{person}{Ye Yuan}, \bibinfo{person}{Guoren Wang}, \bibinfo{person}{Weihua Yang}, {and} \bibinfo{person}{Lu Qin}.} \bibinfo{year}{2022}\natexlab{}.
\newblock \showarticletitle{Periodic Communities Mining in Temporal Networks: Concepts and Algorithms}.
\newblock \bibinfo{journal}{\emph{IEEE Transactions on Knowledge and Data Engineering}} \bibinfo{volume}{34}, \bibinfo{number}{8} (\bibinfo{year}{2022}), \bibinfo{pages}{3927--3945}.
\newblock


\bibitem[Rozenshtein et~al\mbox{.}(2020)]%
        {rozenshtein2020finding}
\bibfield{author}{\bibinfo{person}{Polina Rozenshtein}, \bibinfo{person}{Francesco Bonchi}, \bibinfo{person}{Aristides Gionis}, \bibinfo{person}{Mauro Sozio}, {and} \bibinfo{person}{Nikolaj Tatti}.} \bibinfo{year}{2020}\natexlab{}.
\newblock \showarticletitle{Finding events in temporal networks: segmentation meets densest subgraph discovery}.
\newblock \bibinfo{journal}{\emph{Knowledge and Information Systems}} \bibinfo{volume}{62}, \bibinfo{number}{4} (\bibinfo{year}{2020}), \bibinfo{pages}{1611--1639}.
\newblock


\bibitem[Tang et~al\mbox{.}(2022)]%
        {TangCRC22}
\bibfield{author}{\bibinfo{person}{Yifu Tang}, \bibinfo{person}{Jianxin Li}, \bibinfo{person}{Nur Al~Hasan Haldar}, \bibinfo{person}{Ziyu Guan}, \bibinfo{person}{Jiajie Xu}, {and} \bibinfo{person}{Chengfei Liu}.} \bibinfo{year}{2022}\natexlab{}.
\newblock \showarticletitle{Reliable Community Search in Dynamic Networks}.
\newblock \bibinfo{journal}{\emph{Proc. VLDB Endow.}} \bibinfo{volume}{15}, \bibinfo{number}{11} (\bibinfo{year}{2022}), \bibinfo{pages}{2826–2838}.
\newblock


\bibitem[Tsourakakis(2015)]%
        {DBLP:conf/www/Tsourakakis15a}
\bibfield{author}{\bibinfo{person}{Charalampos~E. Tsourakakis}.} \bibinfo{year}{2015}\natexlab{}.
\newblock \showarticletitle{The K-clique Densest Subgraph Problem}. In \bibinfo{booktitle}{\emph{{WWW}}}. \bibinfo{pages}{1122--1132}.
\newblock


\bibitem[Wang and Cheng(2012)]%
        {DBLP:journals/pvldb/WangC12}
\bibfield{author}{\bibinfo{person}{Jia Wang} {and} \bibinfo{person}{James Cheng}.} \bibinfo{year}{2012}\natexlab{}.
\newblock \showarticletitle{Truss Decomposition in Massive Networks}.
\newblock \bibinfo{journal}{\emph{{PVLDB}}} \bibinfo{volume}{5}, \bibinfo{number}{9} (\bibinfo{year}{2012}), \bibinfo{pages}{812--823}.
\newblock


\bibitem[Wang et~al\mbox{.}(2020)]%
        {cikm2020}
\bibfield{author}{\bibinfo{person}{Jingjing Wang}, \bibinfo{person}{Yanhao Wang}, \bibinfo{person}{Wenjun Jiang}, \bibinfo{person}{Yuchen Li}, {and} \bibinfo{person}{Kian-Lee Tan}.} \bibinfo{year}{2020}\natexlab{}.
\newblock \showarticletitle{Efficient Sampling Algorithms for Approximate Temporal Motif Counting}. In \bibinfo{booktitle}{\emph{CIKM}}. \bibinfo{pages}{1505–1514}.
\newblock


\bibitem[Wang et~al\mbox{.}(2022)]%
        {TOIS_2021_2}
\bibfield{author}{\bibinfo{person}{Lili Wang}, \bibinfo{person}{Chenghan Huang}, \bibinfo{person}{Ying Lu}, \bibinfo{person}{Weicheng Ma}, \bibinfo{person}{Ruibo Liu}, {and} \bibinfo{person}{Soroush Vosoughi}.} \bibinfo{year}{2022}\natexlab{}.
\newblock \showarticletitle{Dynamic Structural Role Node Embedding for User Modeling in Evolving Networks}.
\newblock \bibinfo{journal}{\emph{{ACM} Trans. Inf. Syst.}} \bibinfo{volume}{40}, \bibinfo{number}{3} (\bibinfo{year}{2022}), \bibinfo{pages}{46:1--46:21}.
\newblock


\bibitem[Wu and Hao(2015)]%
        {wu2015review}
\bibfield{author}{\bibinfo{person}{Qinghua Wu} {and} \bibinfo{person}{Jin-Kao Hao}.} \bibinfo{year}{2015}\natexlab{}.
\newblock \showarticletitle{A review on algorithms for maximum clique problems}.
\newblock \bibinfo{journal}{\emph{European Journal of Operational Research}} \bibinfo{volume}{242}, \bibinfo{number}{3} (\bibinfo{year}{2015}), \bibinfo{pages}{693--709}.
\newblock


\bibitem[Yang et~al\mbox{.}(2016)]%
        {DBLP:conf/kdd/YangYWCZL16}
\bibfield{author}{\bibinfo{person}{Yi Yang}, \bibinfo{person}{Da Yan}, \bibinfo{person}{Huanhuan Wu}, \bibinfo{person}{James Cheng}, \bibinfo{person}{Shuigeng Zhou}, {and} \bibinfo{person}{John C.~S. Lui}.} \bibinfo{year}{2016}\natexlab{}.
\newblock \showarticletitle{Diversified Temporal Subgraph Pattern Mining}. In \bibinfo{booktitle}{\emph{SIGKDD}}. \bibinfo{pages}{1965--1974}.
\newblock


\bibitem[Yin et~al\mbox{.}(2017)]%
        {yin2017local}
\bibfield{author}{\bibinfo{person}{Hao Yin}, \bibinfo{person}{Austin~R Benson}, \bibinfo{person}{Jure Leskovec}, {and} \bibinfo{person}{David~F Gleich}.} \bibinfo{year}{2017}\natexlab{}.
\newblock \showarticletitle{Local higher-order graph clustering}. In \bibinfo{booktitle}{\emph{SIGKDD}}. \bibinfo{pages}{555--564}.
\newblock


\bibitem[Yu et~al\mbox{.}(2021)]%
        {YUVLDB2021}
\bibfield{author}{\bibinfo{person}{Michael Yu}, \bibinfo{person}{Dong Wen}, \bibinfo{person}{Lu Qin}, \bibinfo{person}{Ying Zhang}, \bibinfo{person}{Wenjie Zhang}, {and} \bibinfo{person}{Xuemin Lin}.} \bibinfo{year}{2021}\natexlab{}.
\newblock \showarticletitle{On Querying Historical K-Cores}.
\newblock \bibinfo{journal}{\emph{Proc. VLDB Endow.}} \bibinfo{volume}{14}, \bibinfo{number}{11} (\bibinfo{year}{2021}), \bibinfo{pages}{2033–2045}.
\newblock


\bibitem[Yuan et~al\mbox{.}(2018)]%
        {DBLP:journals/tkde/YuanQZCY18}
\bibfield{author}{\bibinfo{person}{Long Yuan}, \bibinfo{person}{Lu Qin}, \bibinfo{person}{Wenjie Zhang}, \bibinfo{person}{Lijun Chang}, {and} \bibinfo{person}{Jianye Yang}.} \bibinfo{year}{2018}\natexlab{}.
\newblock \showarticletitle{Index-Based Densest Clique Percolation Community Search in Networks}.
\newblock \bibinfo{journal}{\emph{{IEEE} Trans. Knowl. Data Eng.}} \bibinfo{volume}{30}, \bibinfo{number}{5} (\bibinfo{year}{2018}), \bibinfo{pages}{922--935}.
\newblock


\bibitem[Zhang et~al\mbox{.}(2022)]%
        {zhang2022significant}
\bibfield{author}{\bibinfo{person}{Yifei Zhang}, \bibinfo{person}{Longlong Lin}, \bibinfo{person}{Pingpeng Yuan}, {and} \bibinfo{person}{Hai Jin}.} \bibinfo{year}{2022}\natexlab{}.
\newblock \showarticletitle{Significant Engagement Community Search on Temporal Networks}. In \bibinfo{booktitle}{\emph{DASFAA}}. \bibinfo{pages}{250--258}.
\newblock


\bibitem[Zhu et~al\mbox{.}(2022)]%
        {DBLP:journals/jcst/ZhuLYJ22}
\bibfield{author}{\bibinfo{person}{Chun{-}Xue Zhu}, \bibinfo{person}{Longlong Lin}, \bibinfo{person}{Pingpeng Yuan}, {and} \bibinfo{person}{Hai Jin}.} \bibinfo{year}{2022}\natexlab{}.
\newblock \showarticletitle{Discovering Cohesive Temporal Subgraphs with Temporal Density Aware Exploration}.
\newblock \bibinfo{journal}{\emph{J. Comput. Sci. Technol.}} \bibinfo{volume}{37}, \bibinfo{number}{5} (\bibinfo{year}{2022}), \bibinfo{pages}{1068--1085}.
\newblock


\bibitem[Zhu et~al\mbox{.}(2018)]%
        {dccs:zhu2018diversified}
\bibfield{author}{\bibinfo{person}{Rong Zhu}, \bibinfo{person}{Zhaonian Zou}, {and} \bibinfo{person}{Jianzhong Li}.} \bibinfo{year}{2018}\natexlab{}.
\newblock \showarticletitle{Diversified coherent core search on multi-layer graphs}. In \bibinfo{booktitle}{\emph{ICDE}}. \bibinfo{pages}{701--712}.
\newblock


\end{thebibliography}
\balance

\appendix
\newpage
\section{Related Work}
\stitle{Static Community Mining.}
Delving deeply into graphs to capture inside relationships among entities has attracted a great deal of researches, which propose multiple community models, including k-core \cite{DBLP:conf/icde/ChengKCO11} \cite{DBLP:journals/pvldb/LiQYM15}, clique \cite{pardalos1994maximum}\cite{wu2015review}, and k-ECC \cite{li2017effective}\cite{chang2013efficiently}. Subsequently, many researches, such as $k$-clique \cite{DBLP:conf/www/Tsourakakis15a} and $\gamma$-quasi-clique \cite{DBLP:journals/tkde/YuanQZCY18}  extend the models in order to describe community more comprehensively. Goldberg \cite{goldberg1984finding} defined the densest subgraph which takes the average node degree as the metric to evaluate its structure characteristic. Andersen and Chellapilla \cite{andersen2009finding} developed algorithms to find the densest subgraphs with two different size restriction. Han et al. \cite{zhongming2016ncss} proposed an association classification algorithm based on node centrality and the strength of structural relationships between nodes and associations. Hong et al. \cite{GraphReformCD_WWW22} proposed a model named GraphReformCD to find the strongly connected nodes in community. Gao et al. \cite{PCCD_sigir_20} proposed PCCD model to mine communities in sparse graphs. However, these studies focus on the lower-order community which is defined based on nodes and edges, ignoring the higher-order connectivity. Therefore, the higher-order model $k$-truss \cite{cohen2008trusses,DBLP:conf/sigmod/HuangCQTY14} takes the triangles as basic blocks to build communities. Akbas and Zhao \cite{akbas2017truss} designed an EquiTruss index to speed up the search for truss communities. Besides, Liu et al. \cite{liu2020truss} extended the truss model to directed graphs and developed a $D$-truss model. Yin et al. \cite{yin2017local} proposed MAPPR model which combine motifs with classical approximate personalized PageRank. Unfortunately, these existing higher-order methods cannot be used directly in the temporal networks as they ignore the temporal information.

\stitle{Temporal Community Mining.}
There are many researches extend the static models to temporal networks to capture both their structural and temporal information. For example, Li et al  \cite{DBLP:conf/icde/LiSQYD18} defined the persistent community in temporal networks based on $k$-core model \cite{DBLP:conf/icde/ChengKCO11,DBLP:journals/pvldb/LiQYM15}. Zhu et al. \cite{dccs:zhu2018diversified} and Hashemi et al. \cite{DBLP:conf/www/HashemiBL22} search for the core-based dense subgraphs in discontinuous time layer. Yu et al \cite{YUVLDB2021}  propose an index-based method with pruning strategies to speed up the search of historical $k$-cores. Similarly, Yang et al \cite{DBLP:conf/kdd/YangYWCZL16} and Lin et al  \cite{DBLP:journals/tsmc/LinYLWLJ22} combined the temporal information with the structure properties of clique to explore temporal quasi-cliques. Chu et al  \cite{OL:chu2019online} and Rozenshtein et al  \cite{rozenshtein2020finding} extended the basic concept of density to temporal networks, and proposed the corresponding algorithms to solve these NP-hard problems. Besides, Ma et al \cite{ma2017fast} proposed a series algorithms to solve the problem of finding heavy temporal subgraph in a special temporal networks where entity relationships keep fixed while the weight of edges change. Qin et al. \cite{qin2020periodic} developed a novel model to explore the periodicity of the community in temporal networks. Zhang et al. \cite{zhang2022significant} adopted two different strategies, top-down and bottom-up, to search their significant community. Besides,
Wang et al. \cite{TOIS_2021_2} used HR2vec method to capture user behavior from individual users to communities and entire networks.  Qin et al. \cite{Qintkde2020} propose periodic community models in temporal networks, including $\sigma$-periodic $k$-core and $\sigma$ -periodic $k$-clique. Tang et al. \cite{TangCRC22} propose a ($\theta, k$)-core reliable community (CRC) model in the weighted dynamic networks. However, these explorations in temporal networks mainly focus on the low-order structure of communities. So, few studies have attempted to explore the higher-order properties of communities in temporal networks. 
Paranjape et al. \cite{DBLP:conf/wsdm/ParanjapeBL17} counted the higher-order temporal motifs to explore the structure and function of temporal networks. Wang et al. \cite{cikm2020} provided an approximate temporal motif counting via random sampling. Porter et al. \cite{WWW2022_2} also developed a fast and accurate model-based method for counting motifs in temporal networks. However these methods only count the number but not promote the community. \textit{L-MEGA} \cite{LMega:fu2020local} mines  higher-order communities in time-evolving networks, starting with an query node and clustering the nodes with motif sub-structure. Unfortunately, capturing temporal information complicates the search for \textit{L-MEGA} in massive temporal networks. Thus it is necessary to develop a novel model for finding higher-order communities.

\section{Pseudocode for Local Exploration Strategies}\label{appendix-code-LS}
For clarity, we divide the local search algorithm (\textit{LS}) into two parts: local exploration strategy for finding MDT (Algorithm \ref{alg:local}) and expanding (Algorithm \ref{alg:expand}).

\begin{algorithm}[b]
\caption{Local exploration strategy for finding \textit{MDT} (LS)} \label{alg:local}
\begin{flushleft}
		\hspace*{0.02in} {\bf Input:}
Temporal graph $\mathcal{G}$, integer $\delta$ and node $q$\\
		\hspace*{0.02in} {\bf Output:}  The \textit{q-MDT} solution
	\end{flushleft}
	\begin{algorithmic}[1]
\State $k_l=INT\_MAX, k_h=INT\_MIN$;
\State $k^*=0$, $\mathcal{S}=\emptyset$, $\mathcal{H}=\emptyset$, $Q_1=\emptyset$;
\For{$v$ in $\mathcal{D}(q)$}
\State compute $TSup_{\mathcal G}(\bar e(q,v))$ with \textit{Sliding Window};
\State $k_l=min(k_l,TSup(\bar e))$;
$k_h=max(k_h,TSup(\bar e))$;
\EndFor
\While{$k_l\leq{k_h}$}
\State $k_m=(k_l+k_h)/2$;
\State $\mathcal B, Q_1$= Expanding$(\mathcal{G},\delta, Q_1, k_m, q)$;
\For{$\bar e(u,v) \in \mathcal{B} $}
\State	$\mathcal H=\mathcal H \cup \{e(u,v,t)|e(u,v,t)\in \mathcal G\}$;
\State	update $TSup_{\mathcal H}$ for related edges in $\mathcal H$;
\EndFor
\State $\mathcal S$, $k_{\mathcal H}^*$=decompose($\mathcal{H},q$);       
\If{$k_{\mathcal H}^*=0$}
\State $k_h=k_m$;
\EndIf
\If{$k_{\mathcal H}^*\neq 0$ and $k_{\mathcal H}^*<k_m$}
\State $k_l=k_{\mathcal H}^*;k_h=k_m$;
\EndIf
\If{$k_{\mathcal H}^*\neq 0$ and $k_{\mathcal H}^*\geq k_m$}
\State check connective property from $q$ and return;
\EndIf
\EndWhile
	\end{algorithmic}
\end{algorithm}	

Algorithm \ref{alg:local} begins by recording both the smallest $k_l$ and largest $k_h$ temporal support values (Line 5). It then selects $k_m = (k_l + k_h)/2$ as the threshold and calls Algorithm \ref{alg:expand} to extract an expanded subgraph $\mathcal{H}$ where all edges satisfy $TSup_{\mathcal G}(\bar e) \geq k_m$ (Lines 7-11). After that, the algorithm calls \textit{decompose} to get the current $k_{\mathcal H}^*$ of the \textit{q-MDT} in $\mathcal H$ (Line 12). 
Note that $k_{\mathcal H}^*$ can indicate what Algorithm \ref{alg:local} will do next:

Case 1 (Lines 13-14): If $k_{\mathcal H}^*=0$, the threshold $k_m$ is too strict, causing the expanded temporal subgraph may be much smaller than the exact solution. The algorithm updates $k_h=k_m$ and generates a new expansion.

Case 2 (Lines 15-16): If $0<k_{\mathcal H}^*<k_m$, a $(k_{\mathcal H}^*,\delta)$-truss can be mined on $\mathcal{H}$. According to Proposition \ref{pro:g_l_sup}, there must be a $(k^*,\delta)$-truss in $\mathcal{G}$, where $k^*\geq k_{\mathcal H}^*$. However, the $(k_{\mathcal H}^*,\delta)$-truss found on $\mathcal{H}$ may not be the maximal solution. 
Hence, the algorithm adjusts both $k_l=k_{\mathcal H}^*$ and $k_h=k_m$, and continues the search to get the exact solution.

Case 3 (Lines 17-18): If $k_{\mathcal H}^*\geq{k_m}$, the edges outside this subgraph can be safely pruned. The remaining subgraph satisfies the conditions (1) and (3) in Definition \ref{def:k-delta-truss}. The algorithm then checks the connectivity property to return the final exact \textit{q-MDT}

Algorithm \ref{alg:expand} defines one state variable for each edge. Initially, since all edges in $\mathcal{G}$ have not been visited, so Algorithm \ref{alg:expand} marks their state as $0$. When the edge $\bar e$ is accessed during the expansion and can be added into $\mathcal{H}$, Algorithm \ref{alg:expand} marks its state as $1$ ("visited"). Besides, if an edge was visited earlier but not added to $\mathcal{H}$, the algorithm set its state as -1. 
The algorithm also defines another queue $Q_1$ to track the process of expansion. For each expansion with $k$, queue $Q_1$ stores the edges $\bar{e}_1(u,w)$ that are adjacent to the "expandable" edge $\bar{e}_0(u,v)$, and satisfies  $TSup_{\mathcal{G}}(\bar e_1)<k$ and the triangle $\triangle_{uvw}$ satisfies $N(\triangle_{uvw},\delta)>0$. 
Especially, all expansions begin with the edges in the set $Q$ which initially consists of the valid edges induced by the query node $q$ at the first iteration and is updated to $Q_1$ in next round (Lines 2-5). For each edge $\bar{e}\in Q$, Algorithm \ref{alg:expand} first checks whether it is "expandable". If the state of an edge $\bar e$ is 1 or -1 with $TSup_{\mathcal{G}}(\bar{e})\geq{k}$, the edge $\bar e$ is regarded as "expandable". For an "expandable" edge, algorithm adds it into $\mathcal{H}$ and iteratively explores the edges which are in the same triangle with it (Lines 10-22).  Otherwise, algorithm adds it into the queue $Q_1$ to prepare for next round expansion (Lines 19-20). When the algorithm accesses two edges $\bar{e}_1(u,w)$ and $\bar{e}_2(v,w)$ through "expandable" edge $\bar{e}(u,v)$, where $N(\triangle_{uvw})>0$, $TSup_{\mathcal{G}}(\bar{e}_1)<{k}$ or $TSup_{\mathcal{G}}(\bar{e}_2)<{k}$, the algorithm only adds the edge with a smaller temporal support into $Q_1$ (Lines 14-18) in order to avoid duplicate visits. 

\begin{algorithm}[t]
    \caption{Expanding} 	
    \label{alg:expand}

    \begin{flushleft}
		\hspace*{0.02in} {\bf Input:}
$\mathcal{G}$, $\delta$, $Q_1$, $k$, $q$\\
		\hspace*{0.02in} {\bf Output:}  $\mathcal{B}$ and $Q_1$
	\end{flushleft}
	\begin{algorithmic}[1]
\State  $\mathcal{B}=\emptyset$, $Q_1=\emptyset$;

\If{this is the first round expansion}
\State $Q=\{\bar{e}(q,u)|TSup(\bar{e}(q,u))\geq{k}\}$;
\Else
\State $Q=Q_1$;
\EndIf
\While{$!Q.empty()$}
\State $\bar{e}(u,v)=Q.front()$;
\If{$\bar{e}.vis=1$ or ($\bar{e}.vis=-1 \& \bar{e}.TSup\geq{k}$)}
	\State $\mathcal{B}.insert(\bar e)$;
	\For{$w$ in $\mathcal D(u) \cap \mathcal D(v)$}
	
		\If{$TSup(\triangle_{uvw})=0$}
  \State \textbf {Continue};
  \EndIf
	\State	compute $TSup_{\mathcal G}(\bar e_1(u,w)), TSup_{\mathcal G}(\bar e_2(v,w))$
		\If{$TSup(\bar{e}_1)<k$ or $TSup(\bar{e}_2)<k$}
			\If{$TSup(\bar{e}_1)<TSup(\bar{e}_2)$}
   \State $Q_1.push(\bar{e}_1);\bar{e}_1.vis\gets{-1}$;
			\Else
   \State $Q_1.push(\bar{e}_2);\bar{e}_2.vis\gets{-1}$;
		\EndIf
  \EndIf
		\If{$\bar{e}_1.vis=0$}\State $Q.push(\bar{e}_1); \bar{e}_1.vis\gets{1}$;
  \EndIf
		\If{$\bar{e}_2.vis=0$}\State $Q.push(\bar{e}_2); \bar{e}_2.vis\gets{1}$;
  \EndIf
  \EndFor
\Else
\State $Q_1.push(\bar e)$;
\EndIf
\EndWhile
\State \Return{$\mathcal{B}, Q_1$};
\end{algorithmic}
\end{algorithm}	

\section{Examples}
\subsection{An Example of the Local Strategy}\label{appendix-example-local}
\begin{figure*}
    \centering
    \subfigure[$\mathcal H_1$ with $k=14$]{ 
        \label{fig:local_example:a} 
        \includegraphics[width=0.23\linewidth]{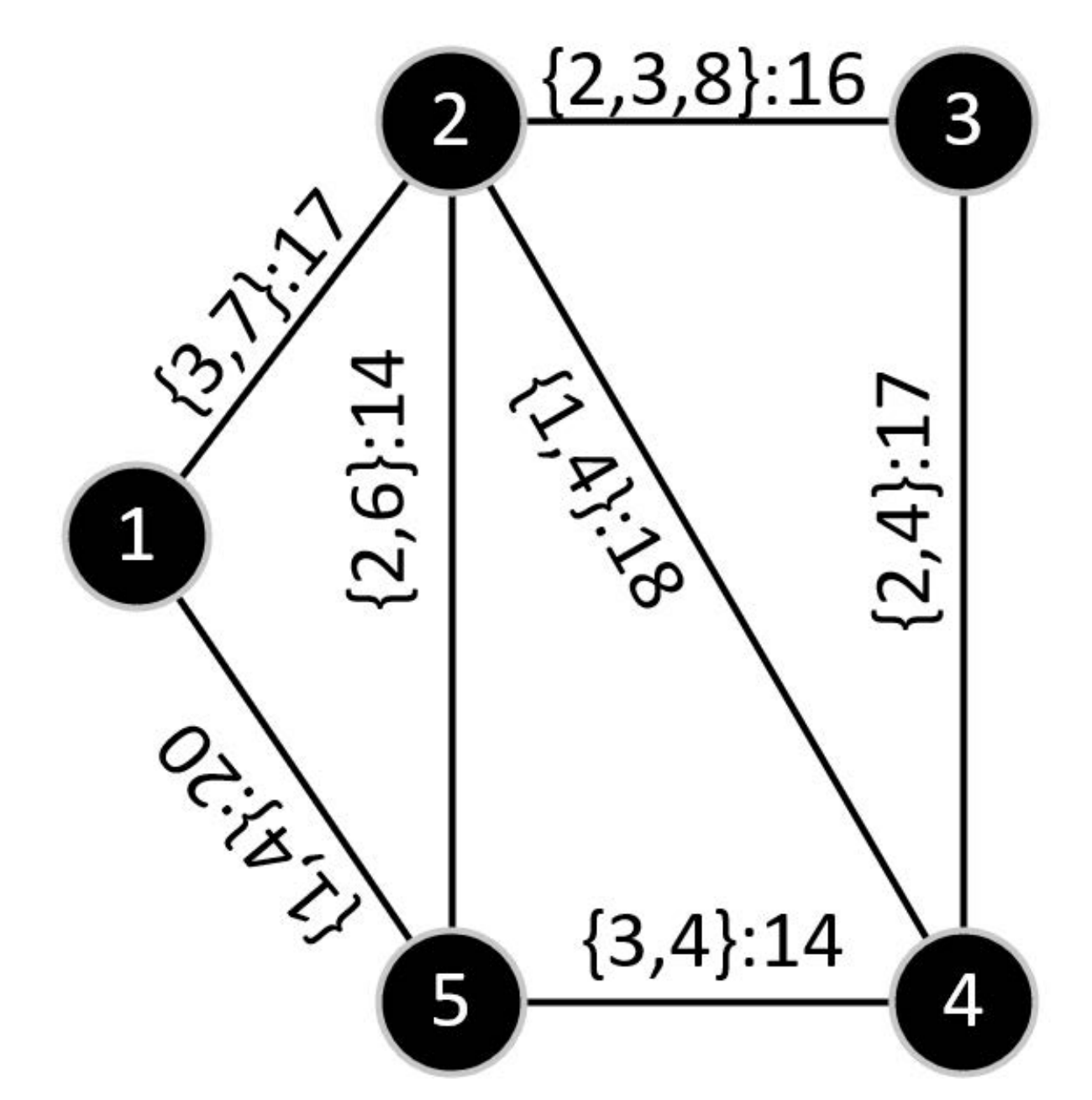} 
    } 
    \subfigure[$(4,3)$-truss in $\mathcal H_1$]{ 
        \label{fig:local_example:b} 
        \includegraphics[width=0.23\linewidth]{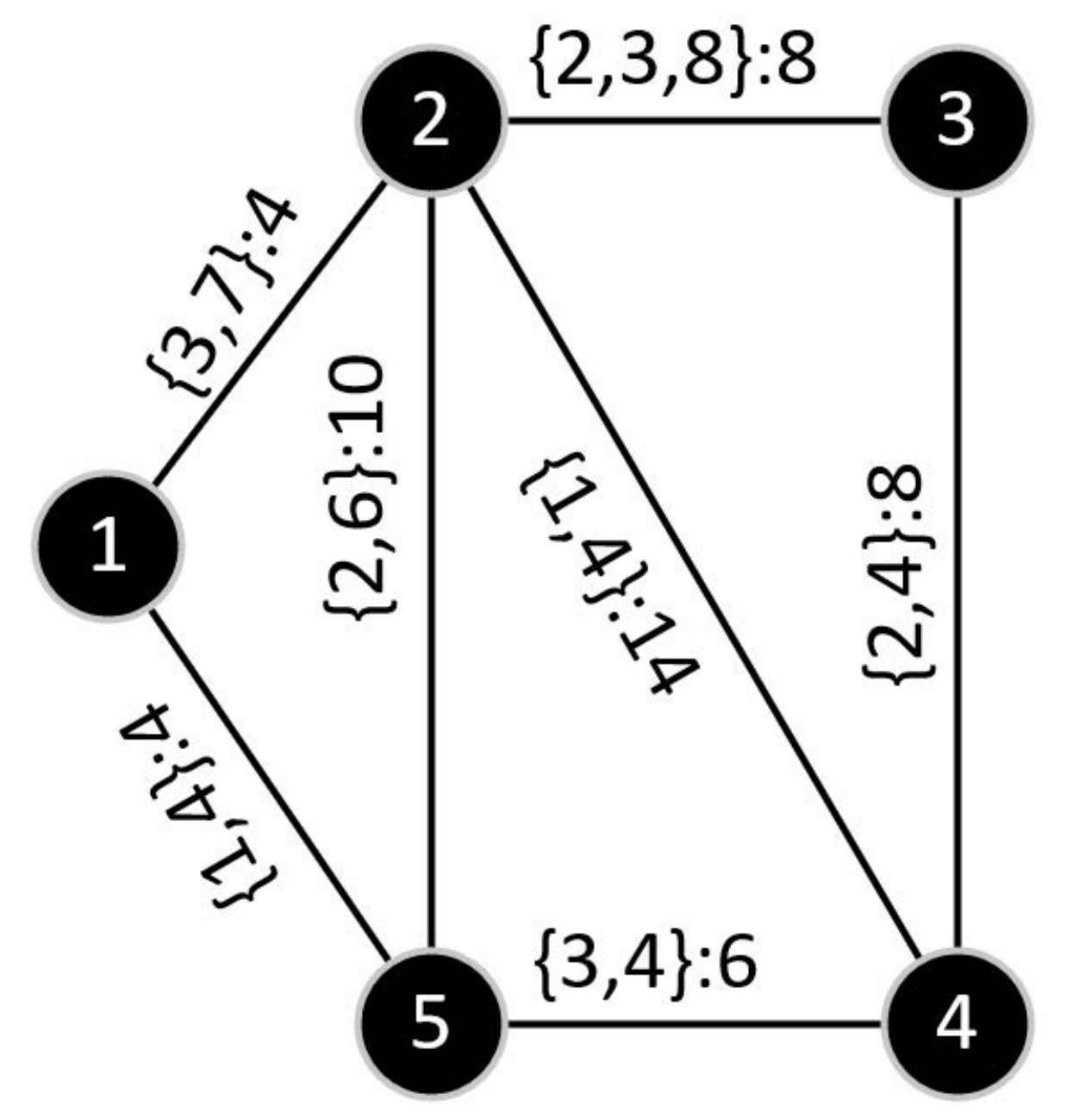} 
    } 
    \subfigure[$\mathcal H_2$ with $k=9$]{ 
        \label{fig:local_example:c}
        \includegraphics[width=0.23\linewidth]{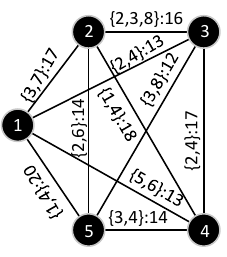} 
    } 
    \subfigure[$(12,3)$-truss in $\mathcal H_2$]{ 
        \label{fig:local_example:d} 
        \includegraphics[width=0.23\linewidth]{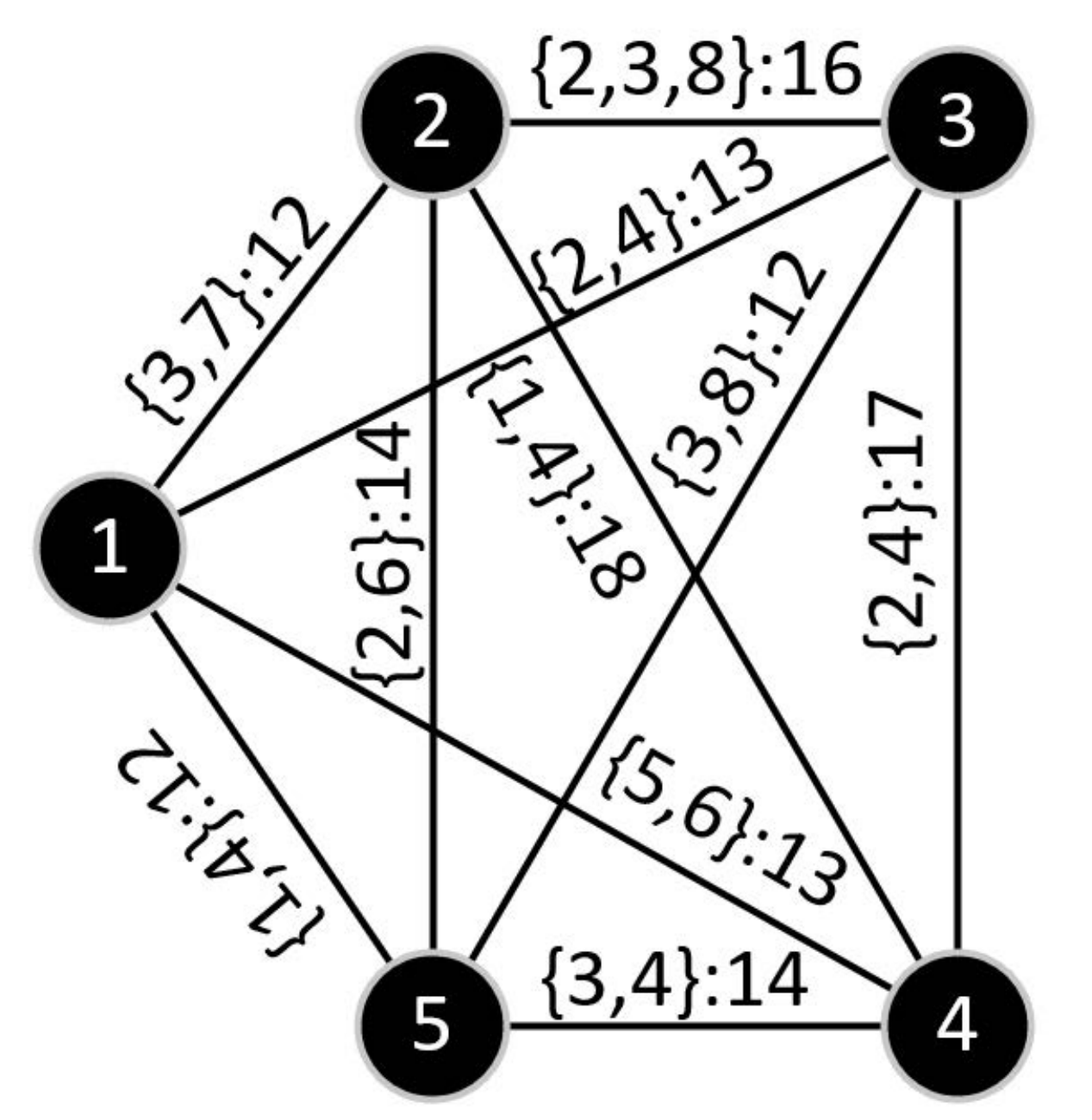} 
    } 
    \caption{The search process of local  strategy for \figurename\ref{fig:basic_example:a}}
    \label{fig:local_example}\vspace{-0.3cm}
\end{figure*}

Given the $\mathcal G$ in \figurename\ref{fig:basic_example:a}, and set $\delta=3$. \figurename\ref{fig:local_example} illustrates the search process of the local strategy. Specifically, \figurename\ref{fig:local_example:a} and \figurename\ref{fig:local_example:c} show the $TSup_{\mathcal G}(\bar e)$, \figurename\ref{fig:local_example:b} and \figurename\ref{fig:local_example:d} display the $TSup_{\mathcal H}(\bar e)$, which are exhibited after the timestamp.
As described earlier, the algorithm initializes $k_m=(TSup_{\mathcal G}(\bar e(1,6))+TSup_{\mathcal G}(\bar e(1,5)))/2=14$. The expanding algorithm starts with edge $\bar e(1,2)$ and checks the edges in triangle $\triangle_{123}, \triangle_{124}, \triangle_{125}$ and $\triangle_{126} $. As $TSup_{\mathcal G}(\bar e(1,5))>14$ and $TSup_{\mathcal{G}}(\bar e(2,5))>14$, they are added into  $\mathcal{H}_1$ and $Q$ to continue expanding. However, the edges $\bar e(2,6)$, $\bar e(1,4)$ and $\bar e(1,3)$ are recorded in $Q_1$ to prepare for next round expansion. Iteratively updating and checking edges in $Q$ until it is empty and the expanded temporal subgraph is shown in \figurename\ref{fig:local_example:a}. Next, the \textit{decompose} procedure is called in $\mathcal H_1$, and obtain a $(4,3)$-truss in $\mathcal{H}_1$ (\figurename\ref{fig:local_example:b}). As $4<k_m=14$, which indicates that setting $k_m=14$ as a threshold is too strict, so the temporal subgraph in \figurename\ref{fig:local_example:b} may not be the solution. Therefore, according to the expanding strategy, the algorithm sets $k_m=(k_l+k_h)/2=(4+14)/2=9$. Next, the second round expansion starts with the edges in $Q=\{\bar e(1,4),\bar e(1,3), \bar e(2,6),\bar e(3,5),\bar e(5,7)\}$, which are recorded in $Q_1$ at the first round , and gets the expanded subgraph $\mathcal H_2$ shown in \figurename\ref{fig:local_example:c}. In $\mathcal H_2$, the algorithm finds a $(12,3)$-truss as the final solution, and the Algorithm \ref{alg:local} stops as $12\geq k_m$.

\subsection{The example of TTS algorithm}\label{appendix-example-TTS}

\begin{figure}
\centering
\subfigure[Access \textit{TT} to initial Q]{ 
\label{fig:index_example:a} 
\includegraphics[width=0.45\linewidth]{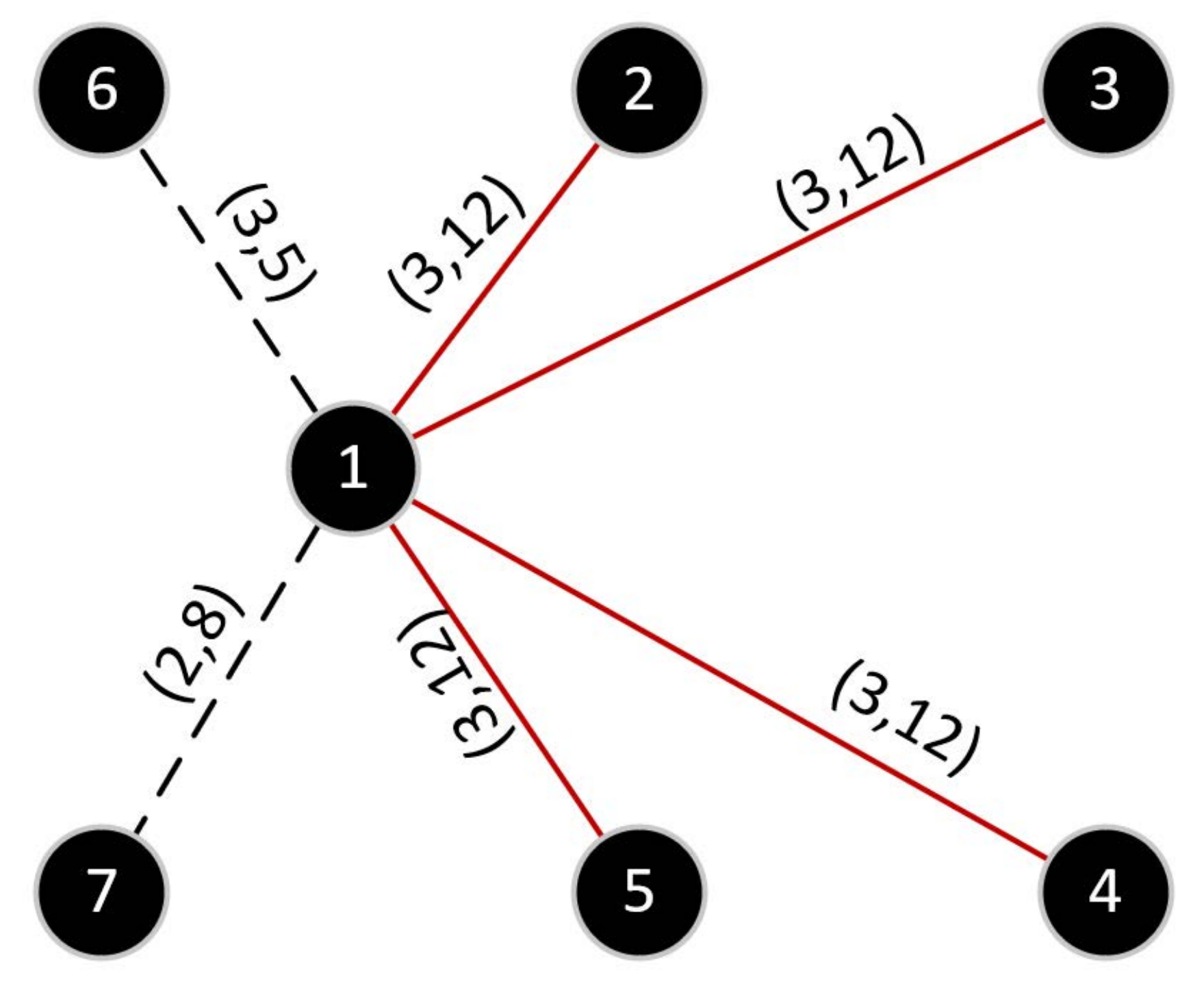} 
} 
\subfigure[Subgraph searched with $\bar e(1,2)$]{ 
\label{fig:index_example:b} 
\includegraphics[width=0.45\linewidth]{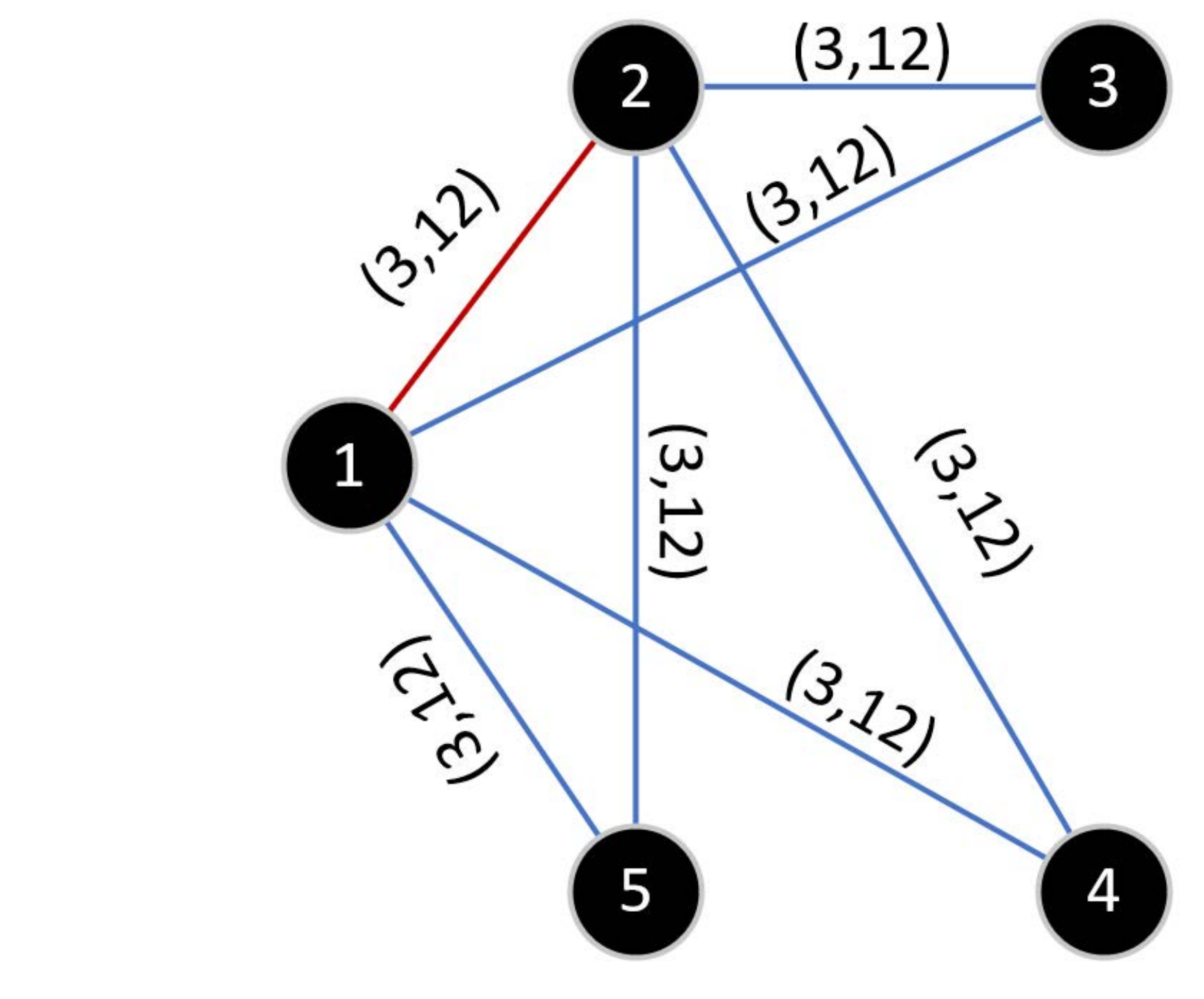} 
} 

\subfigure[Subgraph searched with $\bar e(1,3)$]{ 
\label{fig:index_example:c}
\includegraphics[width=0.45\linewidth]{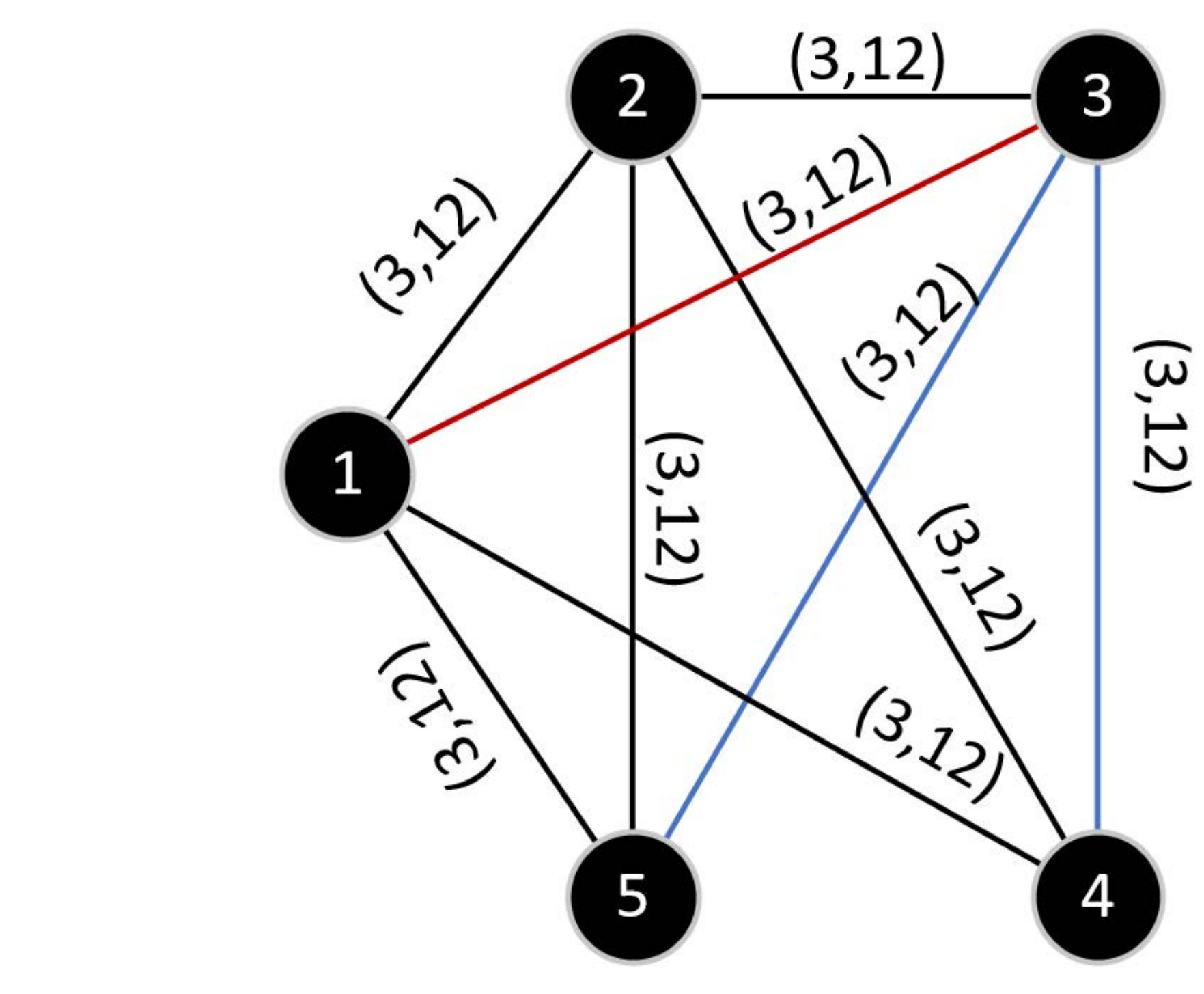} 
} 
\subfigure[\textit{q-MDT} searched with $\bar e(1,4)$]{ 
\label{fig:index_example:d} 
\includegraphics[width=0.45\linewidth]{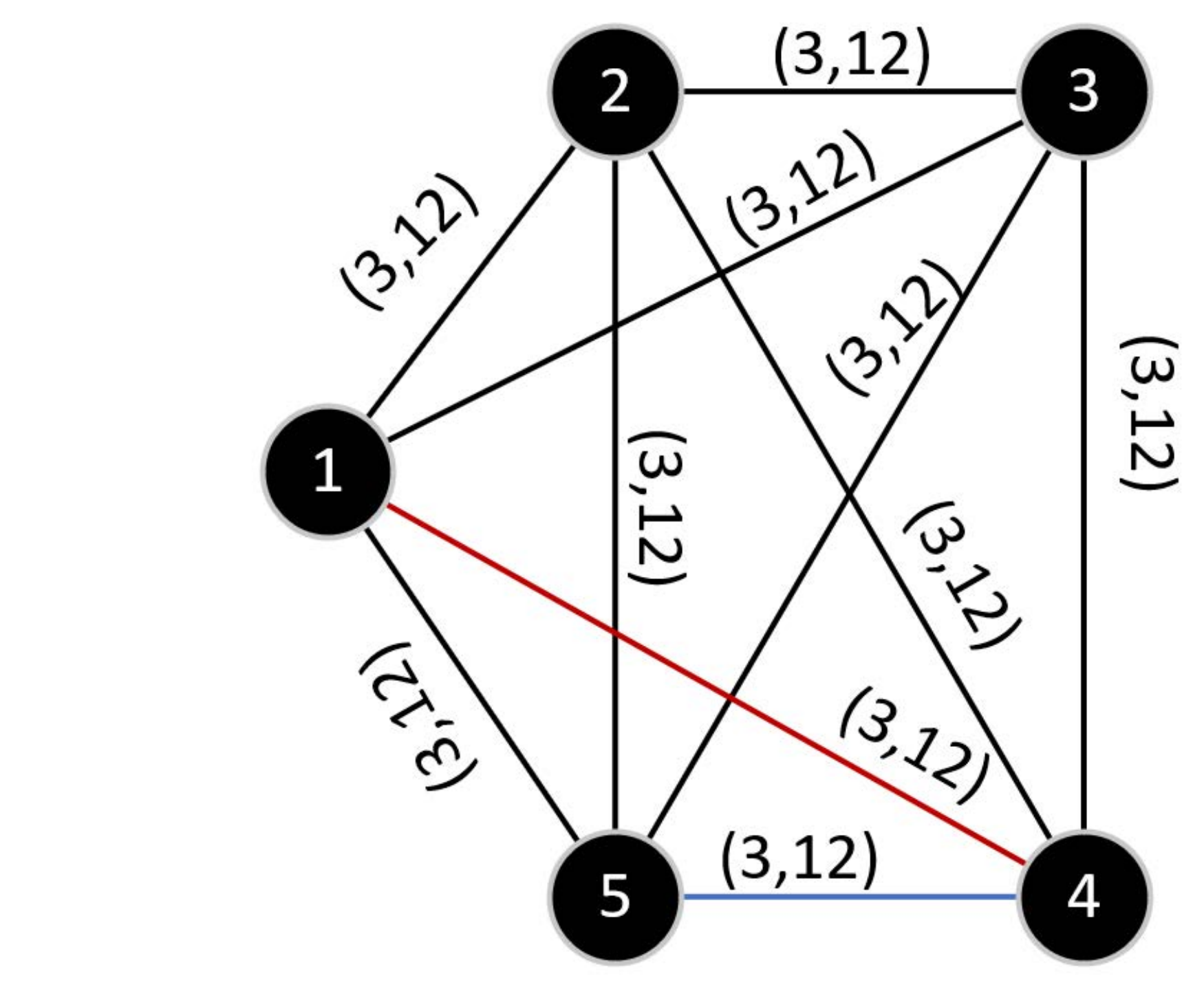} 
} 
\caption{The illustration of \textit{TTS} algorithm}
\label{fig:index_example} \vspace{-0.5cm}
\end{figure}

Table \ref{tab:index_example} shows the temporal trussness recorded in \textit{TT}-index for the triangles and edges in the temporal graph $\mathcal G$ (\figurename \ref{fig:basic_example:a}).
Consider temporal graph $\mathcal G$ in Figure \ref{fig:basic_example:a}, and set $\delta=3$. Algorithm \ref{alg:index_search} first accesses \textit{TT}-index to obtain the temporal trussness of the edges induced by query node (e.g., node 1). 
So $k^*=\max$ $\{\tau(\bar e(1,2)),\tau(\bar e(1,3)),\tau(\bar e(1,4)),\tau(\bar e(1,5)), \tau(\bar e(1,6)),\tau(\bar e(1,7))\}=12$ and $Q$ is initialized as $\{\bar e(1,2),\bar e(1,3),\bar e(1,4),\bar e(1,5)\}$ (i.e. the solid edges in \figurename \ref{fig:index_example:a}). Then, for each edge (e.g., $\bar e(1,2)$) in $Q$, the algorithm accesses the edges that are in the same triangle with $\bar e(1,2)$, and then checks whether the temporal trussness of another two edges is no less than $k^*$. As a result, edges $\{\bar e(1,3),\bar e(2,3),\bar e(1,5),\bar e(2,5), \bar e(1,4),\bar e(2,4) \}$ are added into the solution $Q$ because of $\bar e(1,2)$ (\figurename \ref{fig:index_example:b}). \figurename \ref{fig:index_example:c} and \figurename \ref{fig:index_example:d} show the temporal subgraph searched by $\bar e(1,3)$ and $\bar e(1,4)$ step by step, respectively. Finally, this search process terminates when $Q$ is empty and returns \figurename \ref{fig:index_example:d} as the final solution.

\begin{table}[!hb]
    \centering
    \caption{\textit{TT} for $\mathcal G$ in \figurename\ref{fig:basic_example:a}}
    \label{tab:index_example}
    \vspace{-0.2cm}
    \resizebox{0.48\textwidth}{!}{
	\begin{tabular}{c|c|c|c}
	\toprule
	$\bar e$ & \ \textit{TT}[e] & $\triangle$ &\textit{TT}[$\triangle$]\\
	\midrule
	$\bar e(1,2)$ & (1,3),(2,6),(3,12),(4,15),(5,18),(6,22),(7,24)&$\triangle_{123}$ &$\delta$=1\\
	$\bar e(1,3)$ & (1,3),(2,6),(3,12),(4,15),(5,18),(6,22),(7,24)&$\triangle_{124}$ &$\delta$=2\\
	$\bar e(1,4)$ & (1,2),(2,6),(3,12),(4,15),(5,18),(6,22),(7,24)&$\triangle_{125}$ &$\delta$=2\\
	$\bar e(1,5)$ & (1,2),(2,6),(3,12),(4,15),(5,18),(6,22),(7,24)&$\triangle_{126}$ &$\delta$=1\\
	$\bar e(1,6)$ & (1,2),(2,3),(3,5),(4,6),(5,8) &$\triangle_{134}$ &$\delta$=1\\
	$\bar e(1,7)$ & (1,2),(2,8) &$\triangle_{135}$ &$\delta$=1\\
        $\bar e(2,3)$ & (1,3),(2,7),(3,12),(4,15),(5,18),(6,22),(7,24)&$\triangle_{145}$ &$\delta$=1\\
        $\bar e(2,4)$ & (1,2),(2,7),(3,12),(4,15),(5,18),(6,22),(7,24)&$\triangle_{157}$ &$\delta$=1\\
	$\bar e(2,5)$ & (1,2),(2,7),(3,12),(4,15),(5,18),(6,22),(7,24)&$\triangle_{167}$ &$\delta$=2\\
        $\bar e(2,6)$ & (1,2),(2,3),(3,5),(4,6),(6,8)&$\triangle_{234}$ &$\delta$=1\\
	$\bar e(3,4)$ & (1,3),(2,7),(3,12),(4,15),(5,18),(6,22),(7,24)&$\triangle_{235}$ &$\delta$=1\\
	$\bar e(3,5)$ & (1,3),(2,7),(3,12),(4,15),(5,18),(6,22),(7,24)& $\triangle_{245}$ &$\delta$=2\\	
	$\bar e(4,5)$ & (1,3),(2,7),(3,12),(4,15),(5,18),(6,22),(7,24)& $\triangle_{345}$ &$\delta$=1\\	 
	$\bar e(5,7)$ & (1,2),(2,8)&$\triangle_{678}$ &$\delta$=2\\
	$\bar e(6,7)$ & (2,2),(3,4),(4,7),(5,10),(6,12)& &\\
	$\bar e(6,8)$ & (2,1),(3,4),(4,7),(5,10),(6,12)& &\\
	$\bar e(7,8)$ & (2,1),(3,4),(4,7),(5,10),(6,12)& &\\
	\bottomrule
	\end{tabular} 
    }
    \vspace{-0.3cm}
\end{table}

\section{Theorem Analysis}
\subsection{Proofs}\label{appendix-proofs}
Here, we provide the proofs of Proposition \ref{pro:trian_tsup}, \ref{pro:trian_up}, and \ref{pro:edge_tsup} in sequence.

\textbf{Proof of Proposition \ref{pro:trian_tsup}:} Time interval $[i, i+\delta+1]$ can be split into two $\delta$-length time intervals: $[i, i+\delta]$, $[i+1, i+\delta+1]$ which correspond to two adjacent temporal subgraphs of $\mathcal{G}^{\delta}$.
To compute the number of temporal triangles induced by $\triangle$ in $[i, i+\delta+1]$, we can compute the number of $\tilde \triangle$ appearing in $\mathcal{G}_{i}$ and  $\mathcal{G}_{i+1}$ separately and then sum them. However, temporal subgraphs $\mathcal{G}_i, \mathcal{G}_{i+1} \in \mathcal{G}^\delta$ overlap in $[i+1,i+\delta]$. So, they have an overlapping temporal subgraph $\mathcal{G}_{[i+1,i+\delta]}$, which is a temporal subgrpah in $(\delta-1)$-slice. Hence, the sum of them will count  $N_{\mathcal{G}_{[i+1,i+\delta]}}(\triangle,\delta-1)$ twice. So, 
\vspace{0.1cm}
\begin{equation}
\begin{aligned}
N_{\mathcal{G}_{[i, i+\delta+1]}^{\delta}}=N_{\mathcal{G}_i^\delta}(\triangle,\delta)+N_{\mathcal{G}_{i+1}^\delta}(\triangle,\delta)- N_{\mathcal{G}_{i+1}^{\delta-1}}(\triangle,\delta-1)\\
\end{aligned}
\end{equation}
\vspace{0.1cm}

For brevity, let $\Phi_{N_{\mathcal{G}_{i}^\delta}(\triangle,\delta)}$=$N_{\mathcal{G}_{i}^\delta}(\triangle,\delta)$-$N_{\mathcal{G}_{i}^{\delta-1}}(\triangle,\delta-1)$. Proposition \ref{pro:trian_tsup} shows that we can know the number of temporal triangles satisfy $\Delta(\tilde \triangle) \leq \delta$ in next time interval by summing the number of triangles in the current time interval and the difference value between the temporal triangle numbers in corresponding subgraphs of ($\delta$-1)-slice and $\delta$-slice. 
So, we can extend the proposition to the entire temporal graph and  compute the total number of temporal triangle satisfy $\Delta(\tilde \triangle)\leq \delta$ in the temporal graph $\mathcal G$.  

\textbf{Proof of Proposition \ref{pro:trian_up}:} For concise presentation, we let $N_{\mathcal{G}}$ denote $N_{\mathcal{G}}(\triangle,\delta)$. Similarly, $N_{\mathcal{G}_{i}^{\delta-1}}$ denotes $N_{\mathcal{G}_{i}^{\delta-1}}(\triangle,\delta-1)$. According to Proposition \ref{pro:trian_tsup}, $\forall \triangle \subseteq \mathcal{G}$ and an integer $\delta$, there exists an overlap between two adjacent temporal subgraphs of $\mathcal{G}^\delta$ when we compute $N(\triangle)$ of a triangle. So, 

\begin{equation} \label{eqa:tsup_in_delta}
  \begin{aligned}
N_{\mathcal{G}}(\triangle,\delta)&=N_{\mathcal{G}_{[1,\delta+1]}}+N_{\mathcal{G}_{[2,\delta+2]}}-N_{\mathcal{G}_{[2,\delta+1]}} + ...\\
&+ N_{\mathcal{G}_{[t_{max}-\delta,t_{max}]}}-N_{\mathcal{G}_{[t_{max}-\delta,t_{max}-1]}}\\
&=N_{\mathcal{G}_{1}^\delta}+\sum_{i=2}^{t_{max}-\delta}(N_{\mathcal{G}_{i}^\delta}
				-N_{\mathcal{G}_{i}^{\delta-1}}(\triangle,\delta-1))\\
&=N_{\mathcal{G}_{1}^\delta}+\sum_{i=2}^{t_{max}-\delta}\Phi_{N_{\mathcal{G}_{i}^\delta}(\triangle,\delta)}\\   
			\end{aligned}   
\end{equation}
Similarly,		
\begin{equation} \label{eqa:tsup_in_delta-1}
\begin{aligned}
 N_{\mathcal{G}}(\triangle,\delta-1)&=N_{\mathcal{G}_{[1,\delta-1]}}(\triangle,\delta)+ \sum_{i=2}^{t_{max}-\delta+1}(N_{\mathcal{G}_{i}^{\delta-1}}
				-N_{\mathcal{G}_{i}^{\delta-2}}(\triangle,\delta-2))\\
    &=N_{\mathcal{G}_{1}^{\delta-1}}+\sum_{i=2}^{t_{max}-\delta+1}\Phi_{N_{\mathcal{G}_{i}^{\delta-1}}(\triangle,\delta-1)}\\
\end{aligned}   
\end{equation}
Therefore, 
\begin{equation}    \label{eqa:get_train_up}
\begin{aligned} 
&N_{\mathcal{G}}(\triangle,\delta)-N_{\mathcal{G}}(\triangle,\delta-1) =\sum_{i=1}^{t_{max}-\delta}(\Phi_{N_{\mathcal{G}_{i}^\delta}(\triangle,\delta)}-\Phi_{N_{\mathcal{G}_{i+1}^{\delta-1}}(\triangle,\delta-1)}) ~~~~~~~~~~~~
			\end{aligned}   
		\end{equation}
		So, the Proposition \ref{pro:trian_up} holds.
		
\textbf{Proof of Proposition \ref{pro:edge_tsup}:} After knowing $N_{\mathcal{G}}(\triangle,\delta)$ of the triangles, we can compute the temporal support of a non-temporal edge as follows.

\begin{Proof}
For an edge $\bar{e}\in \mathcal{G}$, the temporal support $TSup_{\mathcal{G}}(\bar{e})=\sum_{j=1..n}N_{\mathcal{G}}(\triangle_j)$, then we have Equation \ref{eq:eq4}. And then the Proposition \ref{pro:edge_tsup} is proved.
\	
\begin{equation}
\label{eq:eq4}
\begin{aligned}
TSup_{\mathcal{G}}(\bar{e},\delta)&=\sum_{j=1}^{n} N_{\mathcal{G}}(\triangle_j,\delta) \\ &=\sum_{j=1}^{n}(N_{\mathcal{G}}(\triangle_j,\delta-1)+\sum_{i=1}^{t_{max}-\delta}(\Phi_{N_{\mathcal{G}_{i}^\delta}(\triangle_j,\delta)}-\Phi_{N_{\mathcal{G}_{i+1}^{\delta-1}}(\triangle_j,\delta-1)}))\\
			&=TSup_{\mathcal{G}}(\bar{e},\delta-1)+\sum_{j=1}^{n}\sum_{i=1}^{t_{max}-\delta}(\Phi_{N_{\mathcal{G}_{i}^\delta}(\triangle_j,\delta)}-\Phi_{N_{\mathcal{G}_{i+1}^{\delta-1}}(\triangle_j,\delta-1)})
		\end{aligned}
	\end{equation}

\end{Proof}	

\subsection{Complexity Analysis}\label{appendix-complexity}
Here, we present the complexity of the previous algorithm.

\textbf{Time complexity of \textit{Sliding Window}:}
There are at most $t_{max}$ timestamps, so the window in this edge at most slides $t_{max}$ times. For each temporal edge of $e_1$, there are two $\delta$-windows on $e_2$ and $e_3$, which at most access the edges $\delta^2$ times. So, assume there exist at most $t_{max}$ timestamps in $\mathcal G$, the time complexity of \textit{Sliding Window} is ${O}(t_{max}\delta^2)$.

\textbf{Time complexity of \textit{LS}:}
It is obvious that for each query node, Algorithm \ref{alg:expand} accesses at most $|{\{\bar e(u,v)|TSup_{{\mathcal G}}(\bar e)>k^*\}}|$ edges to compute their temporal support. 
Besides, counting temporal support requires calling \textit{Sliding Window} procedure for the triangles who include the edge satisfies that $TSup_{{\mathcal G}}(\bar e)\geq k^*$. And checking connection from the \textit{decompose}'s residual temporal suabgrpah requires accessing fewer edges than the expanded temporal subgraphs. So The time complexity of Algorithm \ref{alg:local} is $O(\sum_{\{\bar e(u,v)|TSup_{{\mathcal G}}(\bar e)>k^*\}}\min(\mathcal D(u),\mathcal D(v))+$ $|\{\triangle_{\bar e_1 \bar e_2 \bar e_2}|TSup_{ {\mathcal G}}\\(\bar e_i)>k^*, 1\leq i\leq 3\}|)$.

\textbf{Complexity of \textit{TT}-index:}
For each temporal subgraph $\mathcal G_{t}^{\delta} \in \delta$-slice, Algorithm \ref{alg:count_all_sup} needs to update the $N_{\mathcal G_{t}^{\delta} }(\triangle)$ and $TSup_{\mathcal G_{t}^{\delta}}(\bar e)$ for the triangles and edges influenced by the new inserted edges, respectively. Besides, accumulate the increment of the temporal support to gets the final temporal support needs to access at most all the triangles in $\mathcal G$. And then, employing the Algorithm \textit{decompose} takes $\sum_{\bar e(u,v)\in \mathcal G}\min (\mathcal D(u),\mathcal D(v))$ time. Therefore, given $\delta$, the update operation spends $\sum_{t=0}^{t_{max}-\delta_i}$ $\sum_{\bar e(u,v)\in  { G}_{[t+\delta,t+\delta]}}$ $\min(\mathcal D_{\mathcal G_{t}^\delta}(u), \mathcal D_{\mathcal G_{t}^\delta}(v))+|\triangle_{\mathcal{G}}|+\sum_{\bar e(u,v)\in \mathcal G}\min (\mathcal D(u),\mathcal D(v))$. Since $\delta $ varies from $0$ to $t_{max}$, thus the complexity for constructing \textit{TT} is
$O(\sum_{\delta=0}^{t_{max}}(\sum_{t=1}^{t_{max}-\delta}\sum_{\bar e(u,v)\in \mathcal {G}_{[t+\delta,t+\delta]}}\min(\mathcal D_{\mathcal G_{t}^\delta}(u),  \mathcal D_{\mathcal G_{t}^\delta}(v))+|\triangle_{\mathcal{G}}|+\sum_{\bar e(u,v)\in \mathcal G}\min (\mathcal D(u),\mathcal D(v))))$.

\section{Details of Experiments}
\subsection{Experimental Setting}\label{appendix-settings}

\textbf{Datasets.} The nine real-world datasets used in our experiments are outlined in Table \ref{tab:data}. These datasets are sourced from four distinct domains\footnote{http://snap.stanford.edu/, http://www.sociopatterns.org/, http://konect.cc/}. Specifically, Rmin, Primary, Lyon and Thiers datasets record face-to-face interactions between students and teachers at different schools (college, high school and primary school). Twitter and Facebook datasets represent popular social networks where users are vertices and interactions are edges. Enron is a dataset which consists of email communications from the Enron company. Lkml is a dataset that captures Linux kernel patchwork and related discussion recorded by the Linux kernel mailing list. DBLP documents scientific collaboration among scholars, providing a comprehensive record of co-authored publications.

\begin{table}[H]
	\begin{center}
		\caption{Dataset Statistics. $TS$ is the time scale of the timestamp.}
		\vspace{-0.2cm}
		\scalebox{0.94}{
			\begin{tabular}{crrrrr}
				\toprule
				Dataset & $|V|$ & $|\mathcal {G}|$ & $|\bar{\mathcal {G}}|$  &   $|T|$ &  $TS$ \\
				\midrule
				Rmin & 96  &76,551  & 2,539  & 5,576  &Hour\\ 
				Primary & 242 &26,351 &8,317 &20 &Hour\\
				Lyon & 242  &218,503  & 26,594   & 20 &Hour\\
				Thiers & 328  &352,374  & 43,496    & 50 &Hour\\ 	 
				Twitter & 304,198 &464,653 & 452,202   & 7  &Day\\
				Facebook & 45,813  &461,179  & 183,412    &223 &Week \\
				
				Enron & 86,978 & 697,956   &297,456   &177 &Week \\ 
				Lkml & 26,885 & 328,092  &159,996   & 98 & Month\\ 
				
				DBLP & 1,729,816  &12,007,380  & 8,546,306  &72 &Year\\ 
				\bottomrule
			\end{tabular}
		}\\
		\label{tab:data}
	\end{center}
	\vspace{-0.1cm}
\end{table}

\textbf{Baseline Models.}
\textit{MAPPR} \cite{yin2017local} and \textit{k-truss} \cite{DBLP:conf/sigmod/HuangCQTY14}  are static models which capture the higher-order cohesiveness of the subgraphs in non-temporal networks. \textit{OL} \cite{OL:chu2019online} models the dense temporal subgraphs which have the maximal burstness. \textit{PCore} explores the largest temporal subgraphs that satisfy persistence constraints \cite{DBLP:conf/icde/LiSQYD18}. \textit{DCCS} \cite{dccs:zhu2018diversified} and \textit{FirmCore} \cite{DBLP:conf/www/HashemiBL22} extend the static core model in multilayer networks to detect the cohesive temporal subgraphs. \textit{L-MEGA} \cite{LMega:fu2020local} aims to find the communities with higher-order motif on time-evolving graphs.  

\textbf{Settings.} All experiments were executed on a server with an Xeon 2.00GHz and 256GB memory running Ubuntu 18.04. All methods are implemented in C++.

\subsection{Scalability Evaluation}\label{appendix-scalability}
\begin{figure}[!h]
	\centering
	\subfigure[Running time with varying $|V|$]{ 
		\label{fig:sca:varyn} 
		\includegraphics[width=0.45\linewidth]{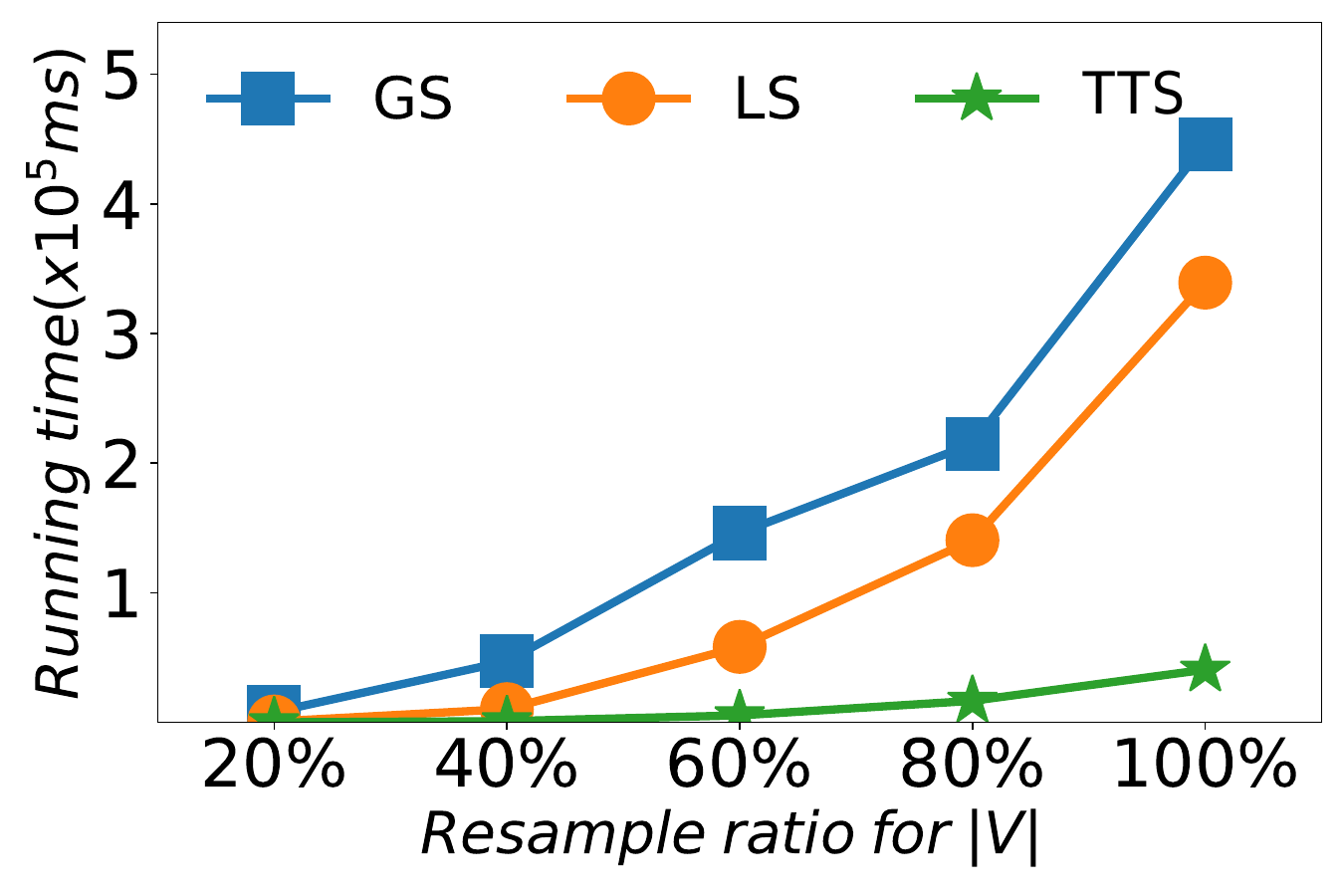} 
	} 
	\subfigure[Running time with varying $TS$]{ 
		\label{fig:sca:varyts} 
		\includegraphics[width=0.45\linewidth]{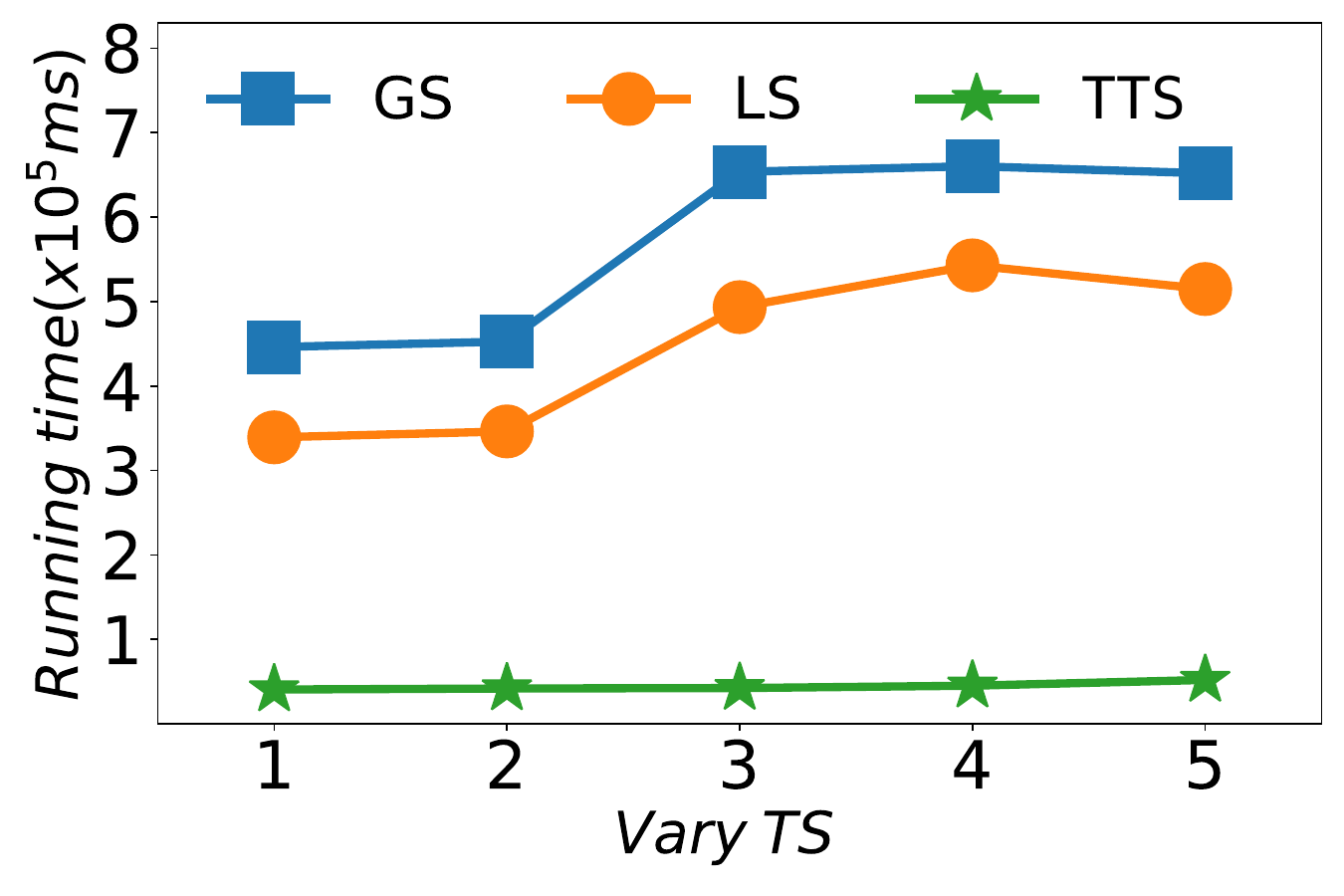} 
	} \vspace{-0.3cm}
	\caption{Scalability Testing}
    \vspace{-0.2cm}
	\label{fig:sca} 
\end{figure}

In order to evaluate the scalability of our  methods, we randomly extracted 20\%, 40\%, 60\%, and 80\% nodes from DBLP to generate synthetic networks. The query times on these synthetic networks are shown in \figurename\ref{fig:sca}. \figurename\ref{fig:sca:varyn} shows that the running time of \textit{GS} and \textit{LS} increases non-linearly as the number of nodes varies, consistent with their time complexity. In contrast, the running time of \textit{TTS} shows an approximately linear increase with the number of nodes. As shown in \figurename\ref{fig:sca:varyts}, the running time of \textit{GS} and \textit{LS} is sensitive to changes in time scale (\textit{TS}) ranging from 1 year to 5 years, whereas \textit{TTS} remains stable. Hence, \textit{TTS} demonstrates efficient handling of large-scale temporal graphs.  


\end{document}